\documentclass{aa}  

\usepackage{natbib}
\bibpunct{(}{)}{;}{a}{}{,}

\usepackage{graphicx}

\usepackage{natbib,twoopt}
\usepackage[breaklinks]{hyperref}      

\usepackage{soul}
\usepackage{orcidlink}

\definecolor{cobalt}{rgb}{0.06, 0.2, 0.65}
\hypersetup{
  colorlinks,
  citecolor=cobalt,
  linkcolor=[rgb]{0.8, 0.2, 1.0},
  urlcolor=cobalt,
}

\usepackage[labelfont=bf]{caption} 
\usepackage[varg]{txfonts}

\usepackage{tikz}
\usetikzlibrary{shapes.geometric, arrows}
\tikzstyle{startstop} = [rectangle, rounded corners, minimum width=3cm, minimum height=1cm,text centered, text width=6cm, draw=black, fill=red!20]
\tikzstyle{io} = [trapezium, trapezium left angle=70, trapezium right angle=110, minimum width=3cm, minimum height=1cm, text centered, text width=3cm, draw=black, fill=blue!10]
\tikzstyle{process} = [rectangle, minimum width=3cm, minimum height=1cm, text centered, text width=5cm, draw=black, fill=orange!20]
\tikzstyle{process2} = [rectangle, minimum width=3cm, minimum height=1cm, text centered, text width=3cm, draw=black, fill=orange!20]
\tikzstyle{decision} = [diamond, minimum width=3cm, minimum height=1cm, text centered, text width=3cm, draw=black, fill=purple!10]
\tikzstyle{arrow} = [thick,->,>=stealth]

\begin{document}

\title{[CII] luminosity models and large-scale image cubes based on COSMOS 2020 and ALPINE-ALMA [CII] data back to the epoch of reionisation}

\author{J. Clarke\inst{1}\thanks{\email{jclarke@ph1.uni-koeln.de}}\orcidlink{0009-0006-6570-5804} \and C. Karoumpis\inst{2}\orcidlink{0000-0003-3259-7457} \and D. Riechers\inst{1}\orcidlink{0000-0001-9585-1462} \and B. Magnelli\inst{3}\orcidlink{0000-0002-6777-6490} \and Y. Okada\inst{1}\orcidlink{0000-0002-6838-6435} \and A. Dev\inst{2}\orcidlink{0009-0008-6563-3681} \and T. Nikola\inst{4} \and F. Bertoldi \inst{2}\orcidlink{0000-0002-1707-1775}}\institute{I. Physikalisches Institut, Universität zu Köln, Zülpicher Straße 77, D-50937 Köln, Germany\and Argelander-Institut für Astronomie, Universität Bonn, Auf dem Hügel 71, 53121 Bonn, Germany\and Université Paris-Saclay, Université Paris Cité, CEA, CNRS, AIM, 91191 Gif-sur-Yvette, France\and Cornell Center for Astrophysics and Planetary Sciences, Cornell University, Ithaca, NY 14853}
 
\abstract 
{}
{We have implemented a novel method to create simulated [CII] emission line intensity mapping (LIM) data cubes using COSMOS 2020 galaxy catalogue data. It allows us to provide solid lower limits for previous simulation-based model predictions and the expected signal strength of upcoming surveys.}
{We applied [CII]158$\mu$m luminosity models to COSMOS 2020 to create LIM cubes  covering a $1.2\times1.2\deg^2$ sky area. These models were derived using galaxy bulk property data from the ALPINE-ALMA survey over the redshift range of $4.4<z<5.9$, while additional models were taken from the literature. The LIM cubes cover $3.42<z<3.87$, $4.14<z<4.76$, $5.34<z<6.31$, and $6.75<z<8.27$, matched to planned observations from the EoR-Spec module of the Prime-Cam instrument in the Fred Young Submillimeter Telescope (FYST). We also created predictions including additional galaxies below current detection limits by `extrapolating' from the faint end of the COSMOS 2020 luminosity function, comparing these to predictions from the literature. In addition, we computed the signal-to-noise (S/N) ratios for the power spectra, using parameters from the planned FYST survey with predicted instrumental noise levels.}
{We find lower limits for the expected power spectrum using the likely incomplete empirical data: when normalised by $2\pi^2$, the amplitudes at $k=1\,\textrm{Mpc}^{-1}$ are $3.06\times10^7, 1.43\times10^7, 9.80\times10^5, 2.77\times10^5\,(\textrm{Jy\,sr}^{-1})^2$ for the aforementioned redshift ranges. For the extrapolated sample, the power spectra are consistent with prior predictions, indicating that extrapolation is a viable method for creating mock LIM cubes. In this case, we expect a result of S/N$>$1 when using FYST parameters. However, our high-redshift results remain inconclusive because of the poor completeness of COSMOS 2020 at $z>6.3$. These predictions will be improved on the basis of future JWST data.}
{}


\titlerunning{[CII] image cubes based on COSMOS 2020 + ALPINE near epoch of reionisation}
\authorrunning{J. Clarke et al.}

\maketitle
%
\section{Introduction} \label{sec:intro}

The mechanisms underlying the Epoch of Reionisation (EoR) are not yet fully understood. This period marks the phase when high-energy photons emitted by the first stars and galaxies ionised the gas in the intergalactic medium, occurring at redshifts of $6<z<20,$ subject to specific cosmic conditions (e.g. \citealt{Haiman_1997}; \citealt{Barkana_2001}). These early galaxies trace the evolution of the Universe's dark matter structure, commonly referred to as the cosmic web. There is an ongoing debate on the intricacies of reionisation, including the relative contributions of stars versus active galactic nuclei (e.g. \citealt{Zaroubi_2013}; \citealt{Kulkarni_2017}; \citealt{Madau_2017}). Additionally, galaxies from these early times are distant and faint, with only the brightest being detectable. This limitation leaves the luminosity function of dim galaxies largely undetermined, and their impact on reionisation uncertain. As a result, constraining the specifics of the EoR remains a significant technical challenge.

Only the most sensitive pencil-beam surveys, such as those conducted by the \textit{Hubble} Space Telescope (HST) or \textit{James Webb} Space Telescope (JWST) (e.g. \citealt{Alvarez_2019}; \citealt{Treu_2023}), are capable of resolving the dimmest galaxies at the EoR. However, the area coverage of these surveys is significantly smaller than 1 square degree ($\deg^2$), which precludes us from gathering a more comprehensive galaxy sample or examining larger structures. To overcome this limitation, we can use a promising technique known as line intensity mapping (LIM, \citealt{Bernal_2022}). Unlike traditional observation methods, which focus on individual galaxies, LIM measures the aggregate integrated signal over a wide survey area that typically exceeds 1\,$\deg^2$, using a beam up to 1\,arcmin wide which covers a fractional bandwidth. This technique maps the tomography of a given region across different redshifts, resulting in a LIM cube (or intensity cube) — a sky map extended to three dimensions by each low-resolution frequency channel of the instrument. Therefore, LIM surveys extensive areas more efficiently than conventional techniques while also capturing the integrated intensity from elusive, low-luminosity galaxies within the aggregate signal. This approach provides valuable insights into galaxy clusters, the corresponding host dark matter halos and large-scale structure, and the luminosity function of the field. Consequently, LIM observations would allow us to constrain the early stages of the EoR. The primary analytical tool we use to find this information is the 3D spherically averaged power spectrum derived from the LIM cube. This two-point statistic connects first-order spatial correlations to each other, enabling the calculation of signal variations across different spatial scales (e.g. \citealt{Gelabert_1989}; \citealt{Barkana_2005}). The investigation of these power spectra has been central to previous work on LIM.

Line intensity mapping enables the investigation of various epochs of the universe by detecting different spectral lines, depending on the spectral range of the detector used. Of the most important lines in the EoR, including CO, HI, and [OIII], several experiments have specifically focused on the [CII] fine structure emission line ($^2P_{3/2}$--$^2P_{1/2}$ at rest frequency 1900.537\,GHz, \citealt{Harwit_1984}; \citealt{Watson_1985}). Previous observations have found a tight correlation between [CII] emission and the star formation rate (SFR) of galaxies (e.g. \mbox{\citealt{Boselli_2002}}; \mbox{\citealt{Vallini_2015}}; \mbox{\citealt{Lagache_2018}}), which is potentially related to how [CII] is a key cooling line for photo-dissociation regions (PDRs) in nearby galaxies. In this way [CII] can be used to trace massive star-forming regions, star-forming galaxies (SFGs) and, therefore, the ionising radiation that drives the EoR.

Numerous surveys are in preparation with an aim to probe the EoR. This study focusses on the EoR-Spec Deep Spectroscopic Survey (DSS), a major program for the Cerro Chajnantor Atacama Telescope (CCAT) Collaboration's second year of observations. The EoR-Spec DSS will utilise the EoR-Spec module within Prime-Cam, an instrument to be mounted on the Fred Young Submillimeter Telescope (FYST, \citealt{CCAT_Prime_Collaboration_2022}). Scheduled for its initial observations in 2026, EoR-Spec aims to span frequencies between 420 and 210 GHz, corresponding to $3.5<z<8$ for [CII] observations. Other similar projects include the Arizona Radio Observatory's Tomographic Intensity Mapping Experiment (TIME, \citealt{Crites_2014}) and the Atacama Pathfinder EXperiment's CarbON CII line in post-reionization and Reionization epoch (CONCERTO, \citealt{Lagache_2017}; \citealt{Dumitru_2019}; \citealt{CONCERTO_Collaboration_2020}; \citealt{Bethermin_2022}).

To prepare for work with observational data, prior studies have simulated galaxy samples to create [CII] intensity cubes and mock power spectra. Our work builds upon those data (e.g. \citealt{Silva_2015}; \citealt{Serra_2016}; \citealt{Chung_2020}; \citealt{Karoumpis_2022}; \citealt{Roy_2023}). This preparatory work relied on a variety of simulation tools, each underpinned by different assumptions for their simulated catalogues. Most models have found [CII] luminosity for their mock catalogues by assuming a strong, linear correlation between [CII] emission and SFR, using existing observational data for calibration, although a number of models were more complex, while others have used simulation work to determine their [CII] models (e.g. \mbox{\cite{Vallini_2015}, \citealt{Lagache_2018}, \citealt{Padmanabhan_2019})}. However, the scarcity of known [CII] emitters at high redshifts (\citealt{Schaerer_2010}; \citealt{Lagache_2018}) limits the robustness of these models. Therefore, this variety and uncertainty  result in predicted power spectra that can vary by more than 1\,dex (order of magnitude), even when focusing on identical frequency ranges. Such variability underscores the challenges faced in understanding the EoR and the role forthcoming surveys such as EoR-Spec DSS will play in advancing our knowledge.

In this work, we find robust lower bounds for the power spectra at this epoch, constraining the potential outcomes for simulations of LIM and future observations. Our findings validate prior projections and provide a foundation for error calibration of future observations. We constructed these limits by using the COSMOS 2020 galaxy catalogue (from now on referred to as COSMOS 2020, \citealt{Weaver_2022a}), the newest iteration of the COSMOS catalogues, which cover the well-documented COSMOS field. By combining galaxy data from this newest data release with several different [CII] emission models, we can create mock LIM cubes that are appropriate for the COSMOS field. Despite COSMOS 2020 not matching the depth of JWST surveys and so having lower completeness for faint galaxies, it is the most comprehensive catalogue available that combines the necessary large sky coverage and depth for creating mock LIM cubes. This data set only includes galaxies up to $z=9$ which are confirmed to exist, so consequently the constructed maps produce power spectra of the minimum possible magnitude. The COSMOS 2020 catalogue is also highly relevant because of its overlap with the area covered by the Extended-Cosmic Evolution Survey (E-COSMOS, \citealt{Scoville_2007}; \citealt{Aihara_2017}), a target for future observations by the EoR-Spec DSS. The [CII] emission models we used are sourced from existing literature and are complemented by models derived using empirical data from the ALMA Large Program to Investigate [CII] at Early Times (ALPINE, \citealt{Bethermin_2020}; \citealt{Faisst_2020}; \citealt{Le_F_vre_2020}), similar to the approaches of \cite{Schaerer_2020} and \cite{Romano_2022}. As ALPINE lies within the COSMOS field, these [CII] models should reflect the wider field. Beyond setting these lower limits, we validated our methodology by generating samples that account for the assumed incompleteness in COSMOS 2020. By extrapolating the original catalogue to include the fainter end of the luminosity function and to produce power spectra as a result, then comparing the results to previous simulations that assume this faint end does indeed exist, we aim to confirm the reliability of our sample and methodology. Aside from demonstrating the practicality of employing this strategy for creating realistic simulated LIM maps, we also scrutinised the analytical errors associated with these power spectra. In this way, we were able to assess the detectability of the power spectra. While our analysis is tailored to the specifications of EoR-Spec, the results may be relevant for other instruments and LIM studies, offering a versatile framework for future research in this area.

Section \ref{sec:method} of this paper presents a detailed overview of our methodology, including a procedural flowchart. This section describes the EoR-Spec DSS, the data sets from ALPINE and COSMOS 2020, and the process involved in generating initial line intensity mapping (LIM) cubes and power spectra. Section \ref{sec:results} examines the characteristics of these mock cubes and power spectra, making a comparison with prior studies and identifying the lower bounds for power spectra values. Section \ref{sec:interpolation} expands on these lower limits, along with a discussion of the methodology of our extrapolation technique to address COSMOS 2020's incompleteness. This section also outlines the enhanced samples and power spectra derived from this procedure. Lastly, Section \ref{sec:discussion} discusses the significance of our findings, the constraints they provide, and suggests ways for refining the simulated LIM cubes in future research.

We use a flat $\Lambda$CDM dark matter universe, taking $H_0=70 \textrm{kms}^{-1}\textrm{Mpc}^{-1}$, $\Omega_\textrm{M}=0.3$, and $\Omega_\Lambda=0.7$. However, we did not find meaningful differences when using other cosmologies (e.g. \citealt{Planck_2020}). We use magnitudes in AB format \citep{Oke_1983}.
\section{Method} \label{sec:method}
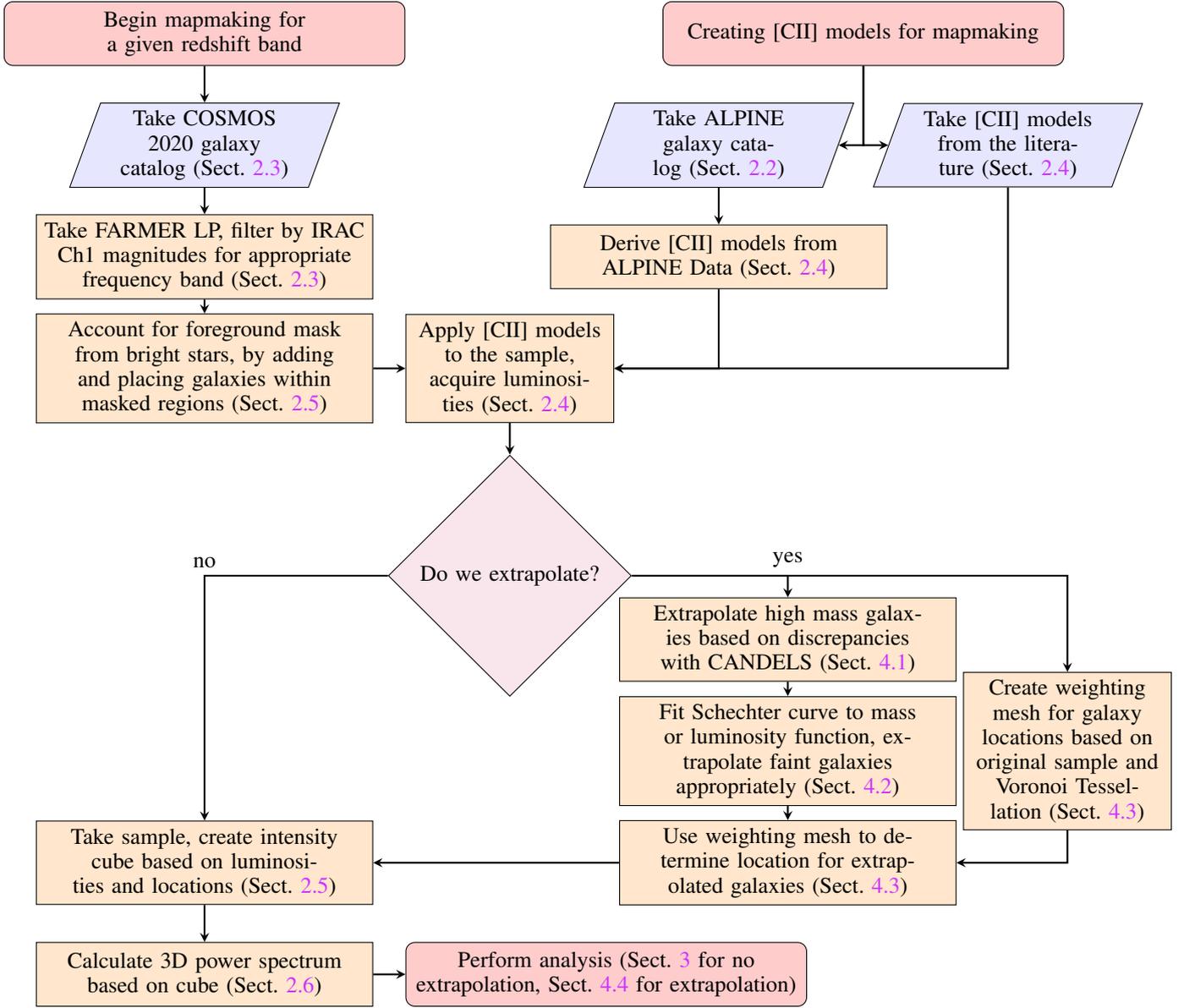
\begin{figure*}[t]
\centering
\begin{tikzpicture}[node distance=1.75cm]
\node (start) [startstop] {Begin map-making for a given redshift band};
\node (inCOSMOS) [io, below of=start] {Take COSMOS 2020 galaxy catalogue (Sect. \ref{sec:methodCOSMOS})};
\node (inALPINE) [io, below of=start, xshift=8cm] {Take ALPINE galaxy catalogue (Sect. \ref{sec:methodALPINE})}; 
\node (startModel) [startstop, above of=inALPINE, xshift=2.25cm] {Create [CII] models for map-making};
\node (inOLDLIT) [io, below of=start, xshift=12.5cm] {Take [CII] models from the literature (Sect. \ref{sec:methodCII})}; 

\draw [arrow] (start) -- (inCOSMOS);
\draw [arrow] (startModel) |- (inALPINE);
\draw [arrow] (startModel) |- (inOLDLIT);

\node (proGETSAMPLE) [process, below of=inCOSMOS] {Take FARMER LP, filter by IRAC Ch1 magnitudes for appropriate frequency band (Sect. \ref{sec:methodCOSMOS})};
\node (proADDMASK) [process, below of=proGETSAMPLE] {Account for foreground mask from bright stars, by adding and placing galaxies within masked regions (Sect. \ref{sec:methodMAP})};
\node (proAPPLYCII) [process2, right of=proADDMASK, xshift=3cm] {Apply [CII] models to the sample, acquire luminosities (Sect. \ref{sec:methodCII})};
\node (proALPINEMODEL) [process, below of=inALPINE] {Derive [CII] models from ALPINE Data (Sect. \ref{sec:methodCII})};

\draw [arrow] (inCOSMOS) -- (proGETSAMPLE);
\draw [arrow] (proGETSAMPLE) -- (proADDMASK);
\draw [arrow] (proADDMASK) -- (proAPPLYCII);
\draw [arrow] (inALPINE) -- (proALPINEMODEL);
\draw [arrow] (proALPINEMODEL) |- (proAPPLYCII);
\draw [arrow] (inOLDLIT) |- (proAPPLYCII);

\node (decINTERPOLATE) [decision, below of=proAPPLYCII, yshift=-1.5cm] {Do we extrapolate?};
\node (proMAP) [process, left of=decINTERPOLATE,yshift=-4.5cm,xshift=-3cm] {Take sample, create intensity cube based on luminosities and locations (Sect. \ref{sec:methodMAP})};
\node (proPS) [process, below of=proMAP] {Calculate 3D power spectrum based on cube (Sect. \ref{sec:methodPS})};
\node (stop) [startstop,right of=proPS,xshift=4.5cm] {Perform analysis (Sect. \ref{sec:results} for no extrapolation, Sect. \ref{sec:interpolationanalysis} for extrapolation)};

\draw [arrow] (proAPPLYCII) -- (decINTERPOLATE);
\draw [arrow] (decINTERPOLATE) -| node[anchor=south] {no} (proMAP);
\draw [arrow] (proMAP) -- (proPS);
\draw [arrow] (proPS) -- (stop);

\node (proINTERPCANDELS) [process, right of=decINTERPOLATE,xshift=2.575cm,yshift=-1cm] {Extrapolate high-mass galaxies based on discrepancies with CANDELS (Sect. \ref{sec:interpolationCANDELS})};
\node (proINTERPMLF) [process, below of=proINTERPCANDELS] {Fit Schechter curve to mass or luminosity function, extrapolate faint galaxies appropriately (Sect. \ref{sec:interpolationMLF})};
\node (proINTERPLOCS) [process, below of=proINTERPMLF] {Use weighting mesh to determine location for extrapolated galaxies (Sect. \ref{sec:interpolationVT})};
\node (proINTERPVT) [process2, right of=proINTERPMLF,xshift=2.6cm] {Create weighting mesh for galaxy locations based on original sample and Voronoi Tessellation (Sect. \ref{sec:interpolationVT})};

\draw [arrow] (decINTERPOLATE) -| node[anchor=south] {yes} (proINTERPCANDELS);
\draw [arrow] (decINTERPOLATE) -| (proINTERPVT);
\draw [arrow] (proINTERPCANDELS) -- (proINTERPMLF);
\draw [arrow] (proINTERPMLF) -- (proINTERPLOCS);
\draw [arrow] (proINTERPVT) |- (proINTERPLOCS);
\draw [arrow] (proINTERPLOCS) -- (proMAP);

\end{tikzpicture}
\caption{Flowchart showing the steps of intensity cube creation, with or without extrapolation. This forms an overview for the rest of the paper, therefore section labels are included for reference.}
\label{fig:FigFlowchart}
\end{figure*}
Here, we outline the methodology for creating a mock LIM cube, as visualised in Figure \ref{fig:FigFlowchart}. First, we created a sub-sample for a given frequency band from COSMOS 2020 (Sect. \ref{sec:methodCOSMOS}) and then account for the stellar mask within this sample (Sect. \ref{sec:methodMAP}). We applied the [CII] models to the bulk property data (i.e. stellar masses, SFRs) of the sub-sample to estimate [CII] luminosity data (Sect. \ref{sec:methodCII}), using models from the literature or those derived from ALPINE (Sect. \ref{sec:methodALPINE}) to create intensity cubes. Subsequently, we calculated the power spectra for these sub-samples (Sect. \ref{sec:methodPS}) and performed our analysis on them (Sect. \ref{sec:results}), which we expect to provide lower limits by their construction. We can also choose to extrapolate additional galaxies from the sub-sample, adjusting the bright galaxy population based on known incompleteness (Sect. \ref{sec:interpolationCANDELS}), assuming a significant population of faint galaxies exist (Sect. \ref{sec:interpolationMLF}). As a result, we produced power spectra that we expect to be concordant with existing simulation work. Galaxies are placed within the sample using Voronoi Tessellation techniques (Sect. \ref{sec:interpolationVT}), with the analysis of the resulting power spectra and their errors discussed in Sect. \ref{sec:interpolationanalysis}. We kept the procedure of extrapolation and the results it provides separate, as an extension of the default procedure we applied to the sub-sample of COSMOS 2020 and its corresponding lower limits.
\subsection{EoR-Spec DSS} \label{sec:methodFYST}
We characterised the survey we assumed for this work, EoR-Spec DSS \citep{CCAT_Prime_Collaboration_2022}, as follows. This survey offers a total of 4000 hours of coverage split between the Extended-Cosmic Evolution Survey (E-COSMOS, \citealt{Scoville_2007}; \citealt{Aihara_2017}) and Extended-Chandra Deep Field South (E-CDFS, \citealt{Giacconi_2002}) fields over the initial survey period. Each field covers a survey area of $4\deg^2$, notably larger than the areas covered by CONCERTO and TIME, thereby allowing EoR-Spec DSS to reach a larger scale clustering signal. 

The frequency range of EoR-Spec DSS, 420--210\,GHz, is to be split into four frequency bands. We took the specific bands used in past work (e.g. \citealt{Chung_2020}; \citealt{Karoumpis_2022}; \citealt{Roy_2023}), with the corresponding [CII] redshifts ranges and instrument properties listed in Table {\ref{table:EoRSpecProp}}. The frequency resolutions, approximately 2--4\,GHz, are wider than the typical spectral velocity widths of [CII] lines from the galaxies ($\sim$200\,kms$^-$, e.g. \mbox{\citealt{Wagg_2010}; \citealt{Bethermin_2020}}), because we expect LIM experiments to capture large numbers of galaxies within a single beam. As we are focussed on [CII] emission, we use the frequency and redshift bands interchangeably throughout this paper. 
\begin{table*}[t]
\caption{Instrument properties of EoR-Spec for the frequency bands of EoR-Spec DSS, with corresponding [CII] redshift ranges.}
\centering
\begin{tabular}{c c c c c}
 \hline\hline
 Frequency Band (GHz)& 410$\pm$20 & 350$\pm$20 & 280$\pm$20 & 225$\pm$20  \\ [0.5ex]
 \hline
 [CII] redshift range & 3.42-3.87 & 4.14-4.76 & 5.34-6.31 & 6.75-8.27 \\
 Frequency Resolution (GHz) & 4.1 & 3.5 & 2.8 & 2.2 \\
 Angular Resolution (arcsecs) & 37.2 & 39 & 46.2 & 52.8 \\ [1ex]
 \hline
\end{tabular}
\label{table:EoRSpecProp}
\tablebib{\cite{CCAT_Prime_Collaboration_2022}; \cite{Nikola_2022, Nikola_2023}}
\end{table*}
\subsection{Data: ALPINE} \label{sec:methodALPINE}
The ALPINE survey was undertaken between mid-2018 and early 2019 as an Atacama Large Millimeter/submillimeter Array (ALMA) Large Program, in order to investigate [CII] data at high redshift ($4.4<z<5.9$) in isolated star-forming regions within the COSMOS and CDFS fields (\citealt{Bethermin_2020}; \citealt{Faisst_2020}; \citealt{Le_F_vre_2020}). The authors targeted known galaxies that had strong UV emission, mostly detected from the Keck/DEIMOS campaigns (\citealt{Capak_2004}; \citealt{Mallery_2012}) and the VUDS survey \citep{Le_F_vre_2015}, in what is described as a pan-chromatic approach. These galaxies had UV and optical properties which were largely consistent with the first `normal' SFGs at $z>5$ detected with ALMA (\citealt{Riechers_2014}; \citealt{Capak_2015}), and so ALPINE also included a sub-set of the ALMA galaxies for re-observation. This resulted in a sample of 118 spectroscopically confirmed galaxies with redshifts between $4.4<z<4.65$ or $5.05<z<5.9$, falling within ALMA band 7 (275--373\,GHz) \mbox{(\citealt{Bethermin_2020}).} The gap in redshift exists due to an H$_2$O atmospheric feature around 325\,GHz.

\begin{figure}[t]
 \centering
 \includegraphics[width=\linewidth]{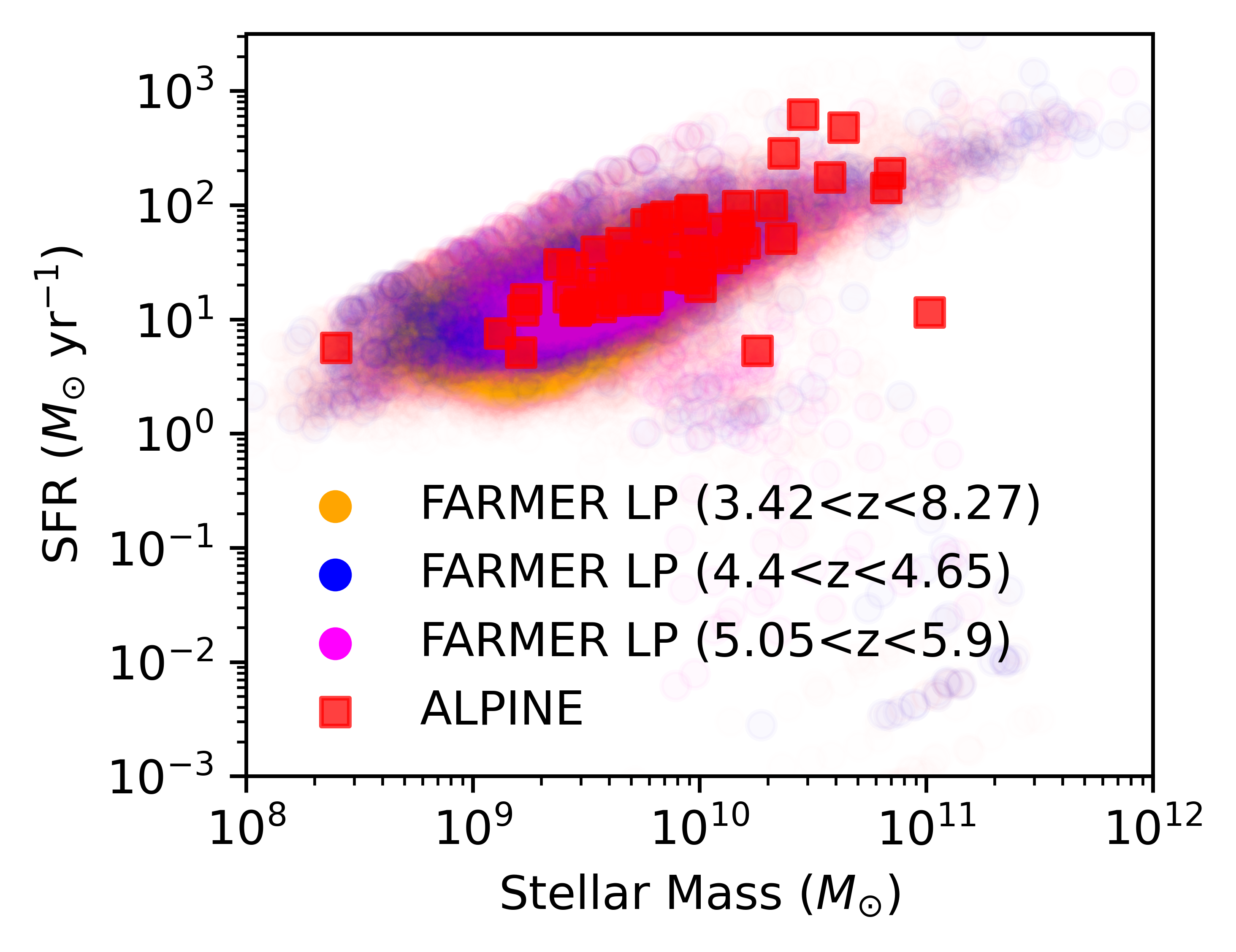}
 \captionof{figure}{Distributions of galaxy SFRs and stellar masses (galaxy main sequences) for FARMER LP and ALPINE. We also take main sequences from FARMER LP in the two separate redshift bands covered by ALPINE ($4.4<z<4.65$ and $5.05<z<5.9$) for comparison. The majority of ALPINE galaxies lie in this lower redshift band. All ALPINE galaxies, and most FARMER LP galaxies, lie in the galaxy main sequence area associated with SFGs (SFR\,$>10^0\,M_\odot\,\textrm{yr}^{-1}$ for all masses), although FARMER LP has a small number of quiescents which have low SFR and high stellar mass.}
 \label{fig:FigMainSeq}
\end{figure}

This sample is ideal for creating models of [CII] emission for the wider COSMOS field, as ALPINE galaxies are also present in the COSMOS 2020 galaxy catalogue in EoR-Spec DSS's redshift range (see Sect. \ref{sec:methodFYST}). We therefore took the 65 galaxies that have successful [CII] detection above 3.5\,$\sigma$, lie within the COSMOS field, and have the empirical bulk property data (stellar masses and SFRs) required to create [CII] models. Of these galaxies, 45 lie within $4.4<z<4.65$ and 20 lie within $5.05<z<5.9$. The bulk properties were determined using the LePhare code, as done in COSMOS 2015 (\citealt{Arnouts_2002}; \citealt{Davidzon_2017}), by matching ALPINE’s photometric sources to COSMOS 2015, HST and UltraVISTA $K_s$ images. Due to the selection methods used, this sample is biased towards SFGs over quiescent galaxies (non-star-forming galaxies) as shown in Fig. {\ref{fig:FigMainSeq}}; therefore, any models created from it may be poorly equipped to cover non-SFG mechanisms for [CII] emission.
\subsection{Data: COSMOS 2020} \label{sec:methodCOSMOS}
We created our intensity cubes using data from the COSMOS 2020 galaxy catalogue, release version 4.1.1 (March 5, 2023, \citealt{Weaver_2022a, Weaver_2022b}). These galaxies are observed in the UV through IR wavelengths, including data from GALEX, MegaCam, HST, Subaru, UltraVISTA DR4, and Spitzer. This sample is an evolution of previous work (\citealt{Scoville_2007}; \citealt{Davidzon_2017}) with additional inclusions from Hyper Suprime-Cam (HSC) Subaru Strategic Program (SSP) PDR2 \citep{Aihara_2019}, new Visible Infrared Survey Telescope for Astronomy (VISTA) data from DR4, and additional Spitzer Infrared Array Camera (IRAC) data. This catalogue includes $\sim$1.7 million galaxies with $z<9$ over a $1.7\times2 \deg^2$ area (149-151\,$\deg$ right ascension (RA), 1.4-3.1\,$\deg$ declination (Dec)), primarily selected using combined $izYJHK_s$ spectral band images. In practise this area is $1.279\deg^2$ (149.35-150.8\,$\deg$\,RA, 1.6-2.8\,$\deg$\,Dec) because of an outer mask region and a stellar mask to account for bright foreground stars. This is reduced further to $0.7\deg^2$ at $z>6.3$, the area of the four Ultra-Deep stripes of UltraVISTA (\citealt{McCracken_2012}). The COSMOS 2020 catalogue is organised in two separate samples: CLASSIC, which was created using Source Extractor and IRACLEAN \citep{Laigle_2016}; and THE FARMER, which was created using The Tractor photometry code \citep{Lang_2016}. photometric properties were computed for each sample using LePhare \citep{Arnouts_2002} and EAZY \citep{Brammer_2008}, leading to four sub-samples with different galaxy measurements and properties. As \cite{Weaver_2022b} used THE FARMER with the LePhare photometric code, we adopted that sub-sample and refer to it as FARMER LP throughout the paper, using this term synonymously with the COSMOS 2020 galaxy catalogue data. 

From FARMER LP, we took galaxies with photometric redshifts within a square area ($1.2\times1.2 \deg^2$, 149.6-150.8\,$\deg$\,RA, 1.6-1.8\,$\deg$\,Dec) in the redshift ranges described in Sect. \ref{sec:methodFYST}, $3.42<z<3.87$, $4.14<z<4.76$, $5.34<z<6.31$, and $6.75<z<8.27$. \cite{Weaver_2022b} removed galaxies with IRAC Channel 1 AB magnitudes above 26 as these galaxies are likely to be artefacts, and when we did the same we obtained 14\,611, 9322, 1732, 510 galaxies respectively. We took LP\_zPDF as photometric redshift, MASS\_MED for stellar mass, SFR\_MED for SFR, sSFR\_MED for specific star formation rate (sSFR), and IRAC\_CH1 for IRAC channel 1 magnitudes from the Spitzer Space Telescope.

\cite{Weaver_2022a} noted that COSMOS 2020 is increasingly incomplete at high redshift because they were not able to select galaxies by mass above $z>3.5$, due to a lack of $K_s$ band detections. Correspondingly we expect to miss many low-mass galaxies in the redshift ranges we cover, with the 75\% mass completeness limit of $10^{9.3}M_\odot$ for $z=3.5$. The ramifications of this are discussed extensively in Sect. \ref{sec:interpolation}. We also expect to miss many quiescent galaxies in FARMER LP (Weaver, priv. comm). However there is little evidence that they provide a significant fraction of the total [CII] emission at any redshift, as their [CII] emission is dominated by diffuse regions instead of PDRs (e.g. \citealt{Pierini_1999}; \citealt{Boselli_2002}). In addition, evidence from the local universe shows that most massive quiescent galaxies have $L_{[\textrm{CII}]}<10^7\,L_\odot$ (e.g. \citealt{Pineda_2018}; \citealt{Temi_2022}), several dex below typical SFGs. Therefore, this quiescent galaxy incompleteness and subsequent missing [CII] signal is unlikely to be a significant source of uncertainty for our study.
\subsection{[CII] modelling} \label{sec:methodCII}
When creating mock LIM cubes, we used several [CII] luminosity relations from the literature as well as relations derived from ALPINE, which are then applied to FARMER LP. While applying one model to all categories of galaxies is overly simplistic and fails to account for the nuances between different galaxy types, using a broad range of models should cover many potential scenarios including dependencies on different bulk properties. We took models from previous literature that used empirical data instead of those that used PDR modelling and other simulations, in order to reduce the number of assumptions made when creating our intensity cubes and power spectra. We also excluded models which were based on data that are irrelevant to our sample, such as models based solely on dwarf galaxies. All previous models assume a linear correlation in log-log space between [CII] and SFR as described in Eq. (\ref{eq:3}), with the fit parameters in Table \ref{table:data}:
\begin{equation}
\log_{10}\left(\frac{L_{[\textrm{CII}]}}{L_\odot}\right)=a+b\log_{10}\left(\frac{\textrm{SFR}}{M_\odot \textrm{yr}^{-1}}\right).
\label{eq:3}
\end{equation}

\cite{DeLooze_2014}: The authors determined linear log-log relations and the scatter between the SFR and IR line emission from dwarf galaxies in the Herschel Dwarf Galaxy Survey \citep{Madden_2013}. By using IR-[CII] proportionalities the authors then found linear log-log [CII]-SFR models, which were calculated separately for individual populations. We took the relations involving the entire sample or only starburst galaxies, as the other fits are specifically for dim dwarf galaxies which would be inappropriate to apply to the SFG-dominated FARMER LP. We refer to these models as DL14 Entire and Starburst.

\cite{Silva_2015}: The authors used a number of [CII] relations derived from high-redshift and local galaxies, as well as a SFR-$L_{\textrm{IR}}$-$L_{\textrm{FIR}}$-$L_{[\textrm{CII}]}$ relation that assumes a large proportion of [CII] luminosity comes from PDRs. We took one of their models, an empirical fit to \cite{DeLooze_2014}'s high-redshift galaxies, as it is the most appropriate for our purposes. We refer to it as Si15. 

\cite{Schaerer_2020} and \cite{Romano_2022}: The authors derived empirical linear log-log [CII]-SFR relations based on the ALPINE sample. \cite{Schaerer_2020} fitted the entire sample of ALPINE data combined with 36 other [CII] emission sources, while \cite{Romano_2022} added an artificial detection based on the non-detections within the sample. While the fundamental ideas behind their model creations are the same as ours, there are several notable differences such as how they included ALPINE data outside of the COSMOS field, and they did not find relations for properties outside of SFR. We refer to these models as Sc20 and Ro22.

\begin{table}[t]
\caption{Linear log-log [CII]-SFR fit parameters for Eq. (\ref{eq:3}).}
\centering
\begin{tabular}{c c c c }
 \hline\hline
 Model & a & b & References \\ [0.5ex]
 \hline
 DL14 Entire & 6.99 & 1.01 & 1\\
 DL14 Starburst & 7.06 & 1  & 1\\
 Si15 & 7.2204 & 0.8475 & 2\\
 Sc20 & 6.43 & 1.26  & 3\\
 Ro22 & 6.76 & 1.14 & 4\\ [1ex]
 \hline
\end{tabular}
\tablebib{(1) \cite{DeLooze_2014}; (2) \cite{Silva_2015}; (3) \cite{Schaerer_2020}; (4) \cite{Romano_2022}}
\label{table:data}
\end{table}
These models are all empirical and, thus, they were useful for our purposes; however, none of them are solely based on the ALPINE galaxies in the COSMOS field. Furthermore, as all of the models assume a [CII]-SFR relation, this results in a limited range of predictions. A number of other [CII] emission models were considered, notably those from \cite{Vallini_2015}, \cite{Lagache_2018}, and \cite{Padmanabhan_2019}, but they were inappropriate for our purposes, either being based on simulated galaxies or requiring bulk properties, such as halo mass, which were not recorded in FARMER LP. Nevertheless, these models were important in our considerations whilst creating models from ALPINE.

In order to create models from ALPINE data, we fitted [CII] results to bulk property data in a similar manner to \cite{Schaerer_2020} and \cite{Romano_2022}. However, unlike them we only used the 65 galaxies in ALPINE that lie within the COSMOS field with [CII] detections above the $3.5\,\sigma$ limit, and we fitted to more variables using Eq. (\ref{eq:5}):
\begin{gather}
\begin{aligned}
\log_{10}\left(\frac{L_{[\textrm{CII}]}}{L_\odot}\right) & =a+b\log_{10}(X)+c\log_{10}(Y)\\ & +d\log_{10}(X)\log_{10}(Y),
\end{aligned}
\label{eq:5}
\raisetag{8pt}   
\end{gather}
where $a$, $b$, $c$, $d$ are the fit coefficients, and $X$ and $Y$ are the bulk properties. We did attempt to fit to second-order terms, but no higher order models successfully converged. The bulk properties we primarily used were $\textrm{SFR}/M_\odot \textrm{yr}^{-1}$, stellar mass $M_\star/M_\odot$ (or in the format $M_\star/10^{10}M_\odot$), but we also included specific star formation rate $\textrm{sSFR}/\,\textrm{yr}^{-1}$, and metallicity in units of oxygen abundance $Z/12+\log\textrm{(O/H)}$. We used these bulk properties due to their intrinsic links to the physical properties of the galaxy itself, and therefore $L_{[\textrm{CII}]}$ emission. While these fits cannot replicate the intricacies of gas cloud emission within the ISM of a galaxy, we believe that a broad range of simple fits which cover more conceptual space than linear log-log fits with SFR should produce a variety of different intensity cubes and corresponding power spectra. For similar reasons we attempted fits with stellar mass and sSFR even though these properties are closely related to SFR, as we wanted the maximum amount of leeway for our fitting functions. Metallicity is not included in the ALPINE sample data, but we calculated it for each galaxy using \cite{Mannucci_2010}'s method (Eq. \ref{eq:6}):
\begin{multline}
\log_{10}\left(\frac{Z}{12+\log\textrm{(O/H)}}\right) =8.90 +0.37\log_{10}\left(\frac{M_\star}{10^{10} M_\odot}\right)\\ -0.14\log_{10}\left(\frac{\textrm{SFR}}{M_\odot \textrm{yr}^{-1}}\right) -0.19\log^2_{10}\left(\frac{M_\star}{10^{10} M_\odot}\right)\\ -0.054\log^2_{10}\left(\frac{\textrm{SFR}}{M_\odot \textrm{yr}^{-1}}\right) +0.12\log_{10}\left(\frac{M_\star}{10^{10} M_\odot}\right)\log_{10}\left(\frac{\textrm{SFR}}{M_\odot \textrm{yr}^{-1}}\right).
\label{eq:6}
\raisetag{47pt}
\end{multline}
We did not attempt to fit models with spectral lines (such as [OIII]), IR emission, or redshift, as these properties do not directly affect the physical environment that results in the [CII] emission. In this way, we only used [CII] models relating to the physical bulk properties of the galaxies.

These fits were applied to all 65 data points using standard least-squares fitting procedures, as well as to these data points sorted into 5 bins. We performed this binning using each bulk property. As this fitting was done for all combinations of bulk properties, we retrieved a large number of potential [CII] models. However, most of these models had poor fits or did not make physical sense, so we applied procedures to find the best models as described in Appendix \ref{appendix:sanitycheck}. This resulted in four best-fit models within the $4\,\sigma$ limit, shown in Eqs. (\ref{eq:em4}--\ref{eq:em7}). We calculated the errors in the fit parameters using Monte Carlo methods.
The equations are as follows:
\begin{gather}
\begin{aligned}
\log_{10}\left(\frac{L_{[\textrm{CII}]}}{L_\odot}\right)& =48.22(\pm12.07)-3.49(\pm2.24)\log_{10}\left(\frac{M_\star}{M_\odot}\right)\\& -5.616(\pm1.417)\log_{10}\left(\frac{Z}{12+\log\textrm{(O/H)}}\right)\\ &+0.508(\pm0.220)\log_{10}\left(\frac{M_\star}{M_\odot}\right)\log_{10}\left(\frac{Z}{12+\log\textrm{(O/H)}}\right),
\label{eq:em4}
\end{aligned}
\raisetag{47pt}
\end{gather}
\begin{align}
\log_{10}\left(\frac{L_{[\textrm{CII}]}}{L_\odot}\right) & =4.02(\pm0.35)+0.476(\pm0.035)\log_{10}\left(\frac{M_\star}{M_\odot}\right),
\label{eq:em5}
\end{align}
\begin{gather}
\begin{aligned}
\log_{10}\left(\frac{L_{[\textrm{CII}]}}{L_\odot}\right) & =3.31(\pm0.38)-0.33(\pm0.16)\log_{10}\left(\frac{\textrm{SFR}}{M_\odot \textrm{yr}^{-1}}\right)\\  & +0.601(\pm0.057)\log_{10}\left(\frac{M_\star}{M_\odot}\right),
\label{eq:em6}
\end{aligned}
\raisetag{21pt}
\end{gather}
\begin{gather}
\begin{aligned}
\log_{10}\left(\frac{L_{[\textrm{CII}]}}{L_\odot}\right) & =6.14(\pm1.29)-0.31(\pm0.15)\log_{10}\left(\frac{\textrm{sSFR}}{ \textrm{yr}^{-1}}\right)\\&+0.27(\pm0.12)\log_{10}\left(\frac{M_\star}{10^{10} M_\odot}\right).
\label{eq:em7}
\end{aligned}
\raisetag{21pt}
\end{gather}
These models are m1-m4 respectively. They show a greater variety of behaviour compared to the linear log-log [CII]-SFR models of the previous literature: m1 has a high level of complexity with cross terms in metallicity, m2 is a simplistic model with regards to stellar mass instead of SFR, and m3 and m4 have terms based on stellar mass and SFR. Despite these differences in construction and complexity, the resulting galaxy [CII] luminosities were relatively similar: when applied to any given galaxy, the [CII] luminosities always lay within 1\,dex. It initially seemed surprising that none of these models are solely based on SFR, in contrast to previous literature. However, m1-m4 still retain this core conceptual relation because there is a close correlation between SFR and stellar mass for SFGs, as shown in the galaxy main sequences of ALPINE and COSMOS 2020 (Fig. \ref{fig:FigMainSeq}).

We note that m1-m4 are likely to be more applicable to redshift bands $4.14<z<4.76$ and $5.34<z<6.31$, as the galaxy distribution of ALPINE is more representative of COSMOS 2020 at these redshifts. Consequently we treated results at $3.42<z<3.87$ and $6.75<z<8.27$ with caution, as discussed in Sect. \ref{sec:discussion}.

In conclusion, we took DL14, Si15, Sc20, and Ro22 from the previous literature (Eq. \ref{eq:3} with Table \ref{table:data}), and obtained m1-m4 from the ALPINE sample (Eqs. \ref{eq:em4}--\ref{eq:em7}). We group models together in upcoming figures to aid readability, typically using purple for DL14, Si15, Sc20, and Ro22, blue for m1-m4, orange for all models with no extrapolation, and red when using mass function extrapolation.

We did apply the models from \cite{Vallini_2015} and \cite{Lagache_2018} to the sample in order to compare specific statistics such as the luminosity function, as these models are commonly used throughout the literature. However, we do not include them in the power spectra or error discussions.
\subsection{Map creation}\label{sec:methodMAP}
\begin{figure}[t]
 \centering
 \includegraphics[width=\linewidth]{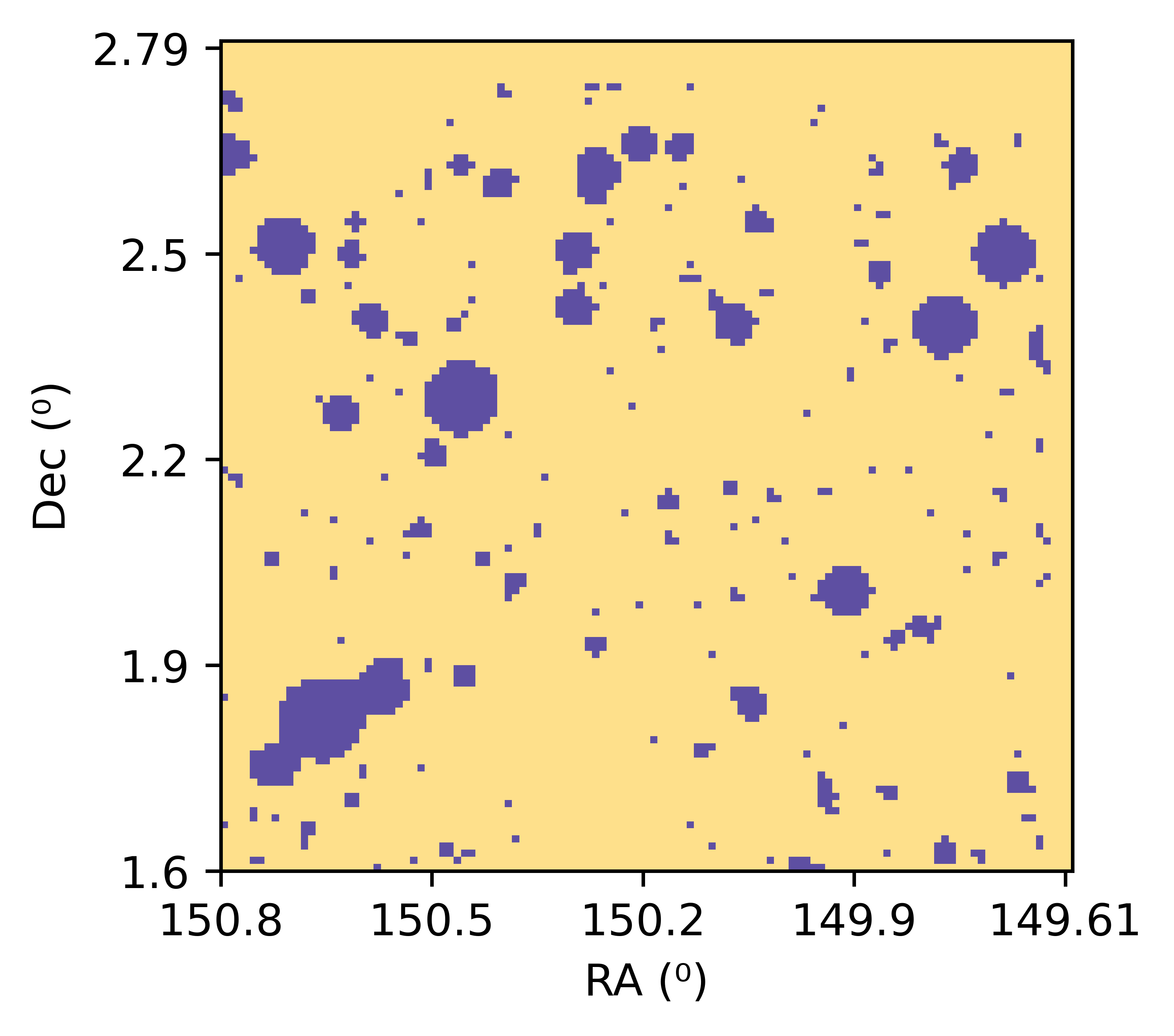}
 \captionof{figure}{Map of the area covered by COSMOS 2020 over $3.42<z<8.27$. The regions masked due to foreground stars (the stellar mask) are shown in blue.}
 \label{fig:Figz342StellarMask}
\end{figure}
When making our mock LIM cubes, we covered a $1.2\times1.2\deg^2$ area corresponding to 149.6--150.8\,$\deg$ RA and 1.6--2.8\,$\deg$ Dec, and an appropriate 40\,GHz wide frequency band. We determined the dimensions of the cube using the EoR-Spec specifications (\citealt{Nikola_2022, Nikola_2023}) and the data in the COSMOS 2020 catalogue paper \citep{Weaver_2022a}. In our cube, equivalent to a 3D array, each voxel has the dimensions relating to the angular and spectral resolution of EoR-Spec as discussed in Sect. \ref{sec:methodFYST} ($37.2\arcsec$, $39\arcsec$, $46.2\arcsec$, $52.8\arcsec$ and 4.1, 3.5, 2.8, 2.2\,GHz for $3.42<z<3.87$, $4.14<z<4.76$, $5.34<z<6.31$, $6.75<z<8.27$ respectively). Using the $1.2\times1.2 \deg^2$ square area this results in cube dimensions of $117\times117\times10$, $112\times112\times11$, $94\times94\times14$, and $82\times82\times18$ voxels, respectively. This third `depth' dimension spans a larger comoving distance than the 2D sky map width, despite having far fewer voxels, so the objects we refer to as `cubes' are actually cuboids.

In these 3D cubes we inserted each galaxy's luminosity at the corresponding voxel location, depending on its RA, Dec, and luminosity distance which we calculated using its redshift and the cosmology in Sect. \ref{sec:intro}. We only included galaxies with photometric redshifts corresponding to the relevant channels, and made sure to convert the galaxy co-ordinate data in FARMER LP to fit this new co-ordinate scheme, accounting for the outer mask surrounding COSMOS 2020 and the FARMER LP catalogue pixel size of $0.15\arcsec$. However, when creating the mock cubes, we found that the width of the spectral bins is narrower than the uncertainty in photometric redshift for many FARMER LP galaxies (typically $\Delta z\approx0.1$). This redshift scatter implies that galaxies from the sample could potentially be in different frequency slices of the intensity cube, which has implications for the structure of the mock cubes and therefore the resulting power spectrum. Upon investigation we found that the difference in power spectra magnitude from this scatter is less than 0.1\,dex, as discussed in Appendix \ref{appendix:mapmakingvar}. We note that this is an issue for our technique of constructing mock cubes and our results and this  will not be a problem for the actual instrument.

As we need intensity cubes, we used Eq. (\ref{eq:9}) to determine intensity ($I_{[\textrm{CII}]}$) from [CII] luminosity:
\begin{gather}
\frac{I_{[\textrm{CII}]}}{L_\odot/\textrm{Mpc}^2/\textrm{GHz}/\textrm{rad}^2}=\frac{\textrm{rad}^2}{\Delta\theta_\textrm{beam}^2}\sum_{i}\frac{\textrm{GHz}}{\Delta\nu_0}\frac{L_{[\textrm{CII}],i}}{L_\odot}\frac{\textrm{Mpc}^2}{4\pi r_i^2(1+z_i)^2},
\label{eq:9}
\raisetag{21pt}
\end{gather}
where $L_{[\textrm{CII}]}$ is the luminosity of each galaxy in the voxel (referred to by index $i$), $\Delta\nu_0$ is the frequency channel of a given redshift slice in the 3D data cube, $r(1+z)$ is the luminosity distance because $r$ is the comoving distance and $z$ is the redshift, and $\Delta\theta_\textrm{beam}$ is the instrument beam width (specifically the full width half maximum), which is equivalent to the angular size of the voxel on the sky. We then converted these units to $\textrm{Jy\,sr}^{-1}$, which are our preferred intensity units by convention. 

\begin{figure}[t]
 \centering
 \includegraphics[width=\linewidth]{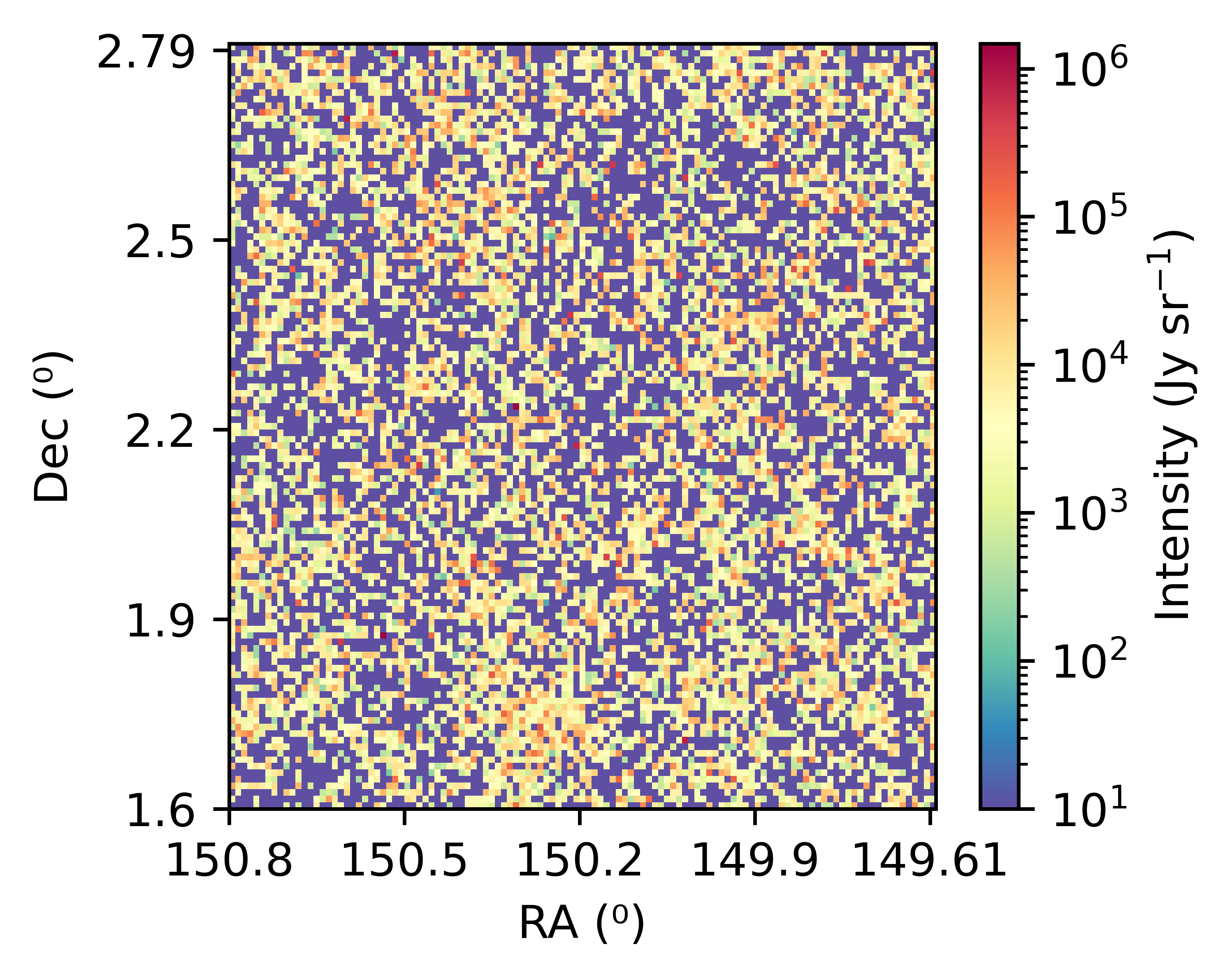}
 \captionof{figure}{2D cross-section of the 3D intensity cube at $3.42<z<3.87$, having applied m1 to FARMER LP. We include additional galaxies extrapolated in the stellar mask region as indicated by Fig. \ref{fig:Figz342StellarMask}.}
 \label{fig:Figz342CroppedMap}
\end{figure}

After carrying out this procedure for the original FARMER LP sample, we then accounted for masked regions of COSMOS 2020 due to foreground stars, namely, the `stellar mask'. This removal of unwanted signal creates areas of no intensity, as shown in Fig. \ref{fig:Figz342StellarMask}. As these regions span over a total area of $0.12\deg^2$, approximately 10\% of the area of FARMER LP, this could potentially impact the power spectra. The statistical averaging would be skewed by the voids, thereby resulting in an increase of power spectra magnitude by approximately 10\%, and compensating for the changed volume could add additional complications. We used extrapolation to counteract this: the process of adding mock galaxy data to FARMER LP sample based on projected incompleteness of the original sample. In the case of mask extrapolation, we added a number of galaxies to the sample in the masked regions to ensure the galaxy density of masked and unmasked regions was made equal, that is $\sim$10\% of the number of galaxies in the original sample. For example, for the 14\,611 galaxies in the $3.42<z<3.87$ region, we would add 1\,964 galaxies. The locations were randomly picked within the voids, and the properties of the galaxies were randomly selected from the existing galaxies in the redshift band, which on average reproduced the existing luminosity function and galaxy distribution of the sample. In this way we do not expect the [CII] luminosity function to be meaningfully affected by this extrapolation. To ensure we reproduced these existing statistics in these extrapolated data, we created ten random masks to use in intensity cube creation, and averaged the resulting power spectra of these cubes. We discuss this further in Appendix \ref{appendix:mapmakingvar}, where we demonstrated that these corrections are appropriate. An example of an extrapolated intensity cube is shown in Fig. \ref{fig:Figz342CroppedMap}.

\begin{figure}[t]
 \centering
 \includegraphics[width=\linewidth]{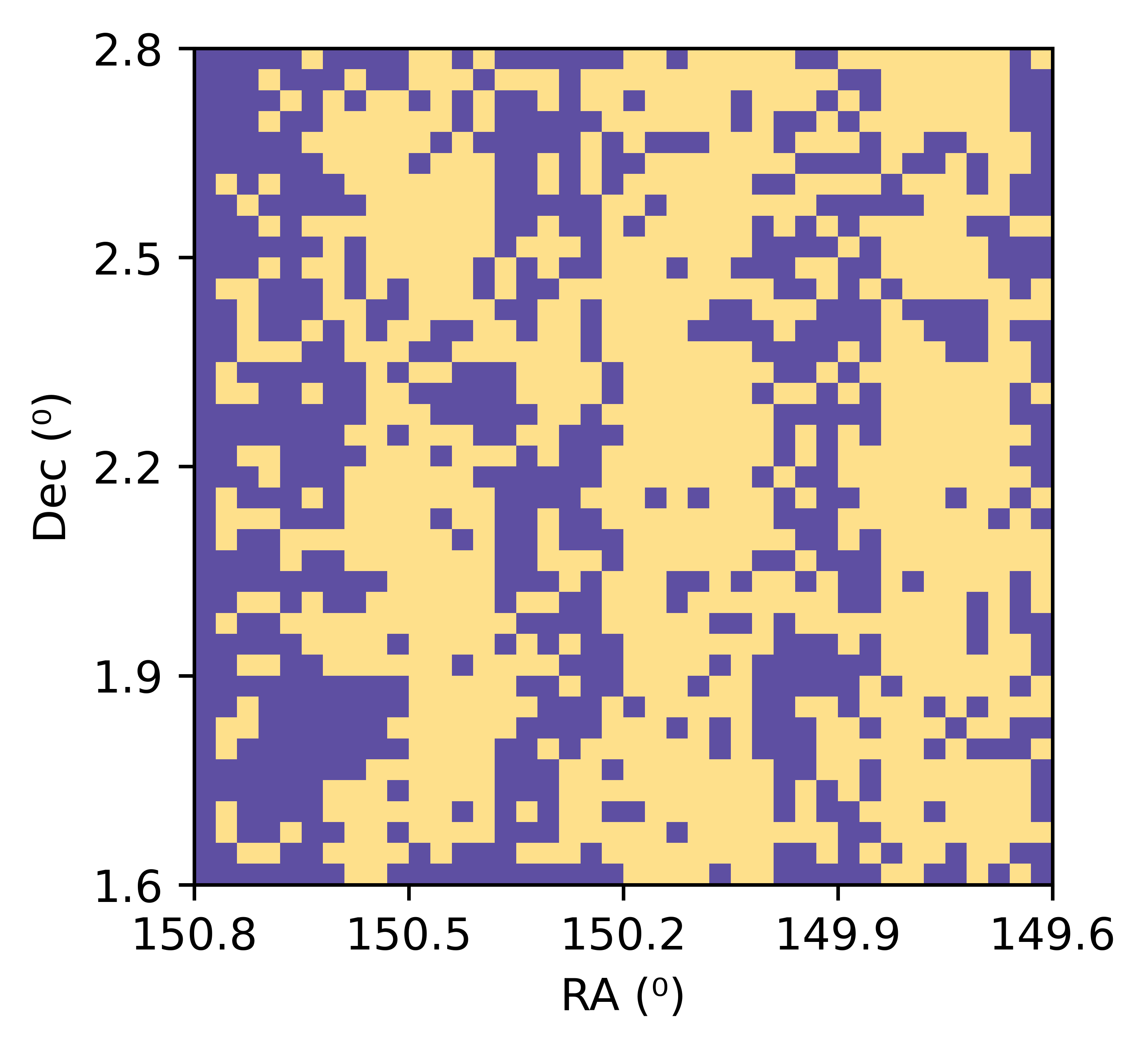}
 \captionof{figure}{As Fig. \ref{fig:Figz342StellarMask}, but only for FARMER LP galaxies with $z>6.3$. Even in the yellow stripes where we expect galaxies to be visible, there are blank pixels as we can only detect the brightest galaxies.}
 \label{fig:Figz675StellarMask}
\end{figure}

For $z>6.3$ most galaxy data are concentrated in the $0.7\deg^2$ covered by the four Ultra-Deep stripes of UltraVISTA \citep{McCracken_2012}, the areas defined by \cite{Weaver_2022a}. We therefore applied our mask extrapolation methods to the regions not covered by the stripes for the band $6.75<z<8.27$, as shown in Fig. \ref{fig:Figz675StellarMask}. This mask extrapolation approximately doubled the FARMER LP sample in that redshift range as the areas of no signal are significantly larger when compared to lower redshifts, and likely compromised any large-scale structure in that band. This is discussed further in Appendix \ref{appendix:mapmakingvar}.

\subsection{Power spectra} \label{sec:methodPS}
The 3D spherically averaged power spectrum is the primary statistic used to analyse intensity cubes in previous work. In the context of upcoming LIM experiments, power spectra will allow us to probe the luminosity function of the covered fields and to determine the existence and impact of regions with significant galaxy clustering. This statistic is therefore useful in characterising the EoR, so we applied it to our mock intensity cubes. When performing this for maps created from mock samples, as done in this work, we already know the `input' luminosity function and clustering. Therefore, the resulting power spectra can be viewed as predictions for if reality matches our simulated conditions.

Firstly we established the dimensions of the voxels and the intensity cube in comoving space, in units of Mpc. We then took the 3D intensity cubes and performed a Fast Fourier Transform (FFT) over all elements in the cube, producing a new 3D cube centred around the origin point in Fourier space, where each intensity element was transformed into corresponding Fourier amplitudes. These Fourier amplitudes are multiplied by the volume of a voxel divided by total number of voxels (i.e. volume of a voxel per voxel), a normalisation factor to arrive at the units of the Fourier transform, thereby preventing the physical scale of the intensity cube from influencing the power spectra amplitude. 

\begin{figure}[t]
 \centering
 \includegraphics[width=\linewidth]{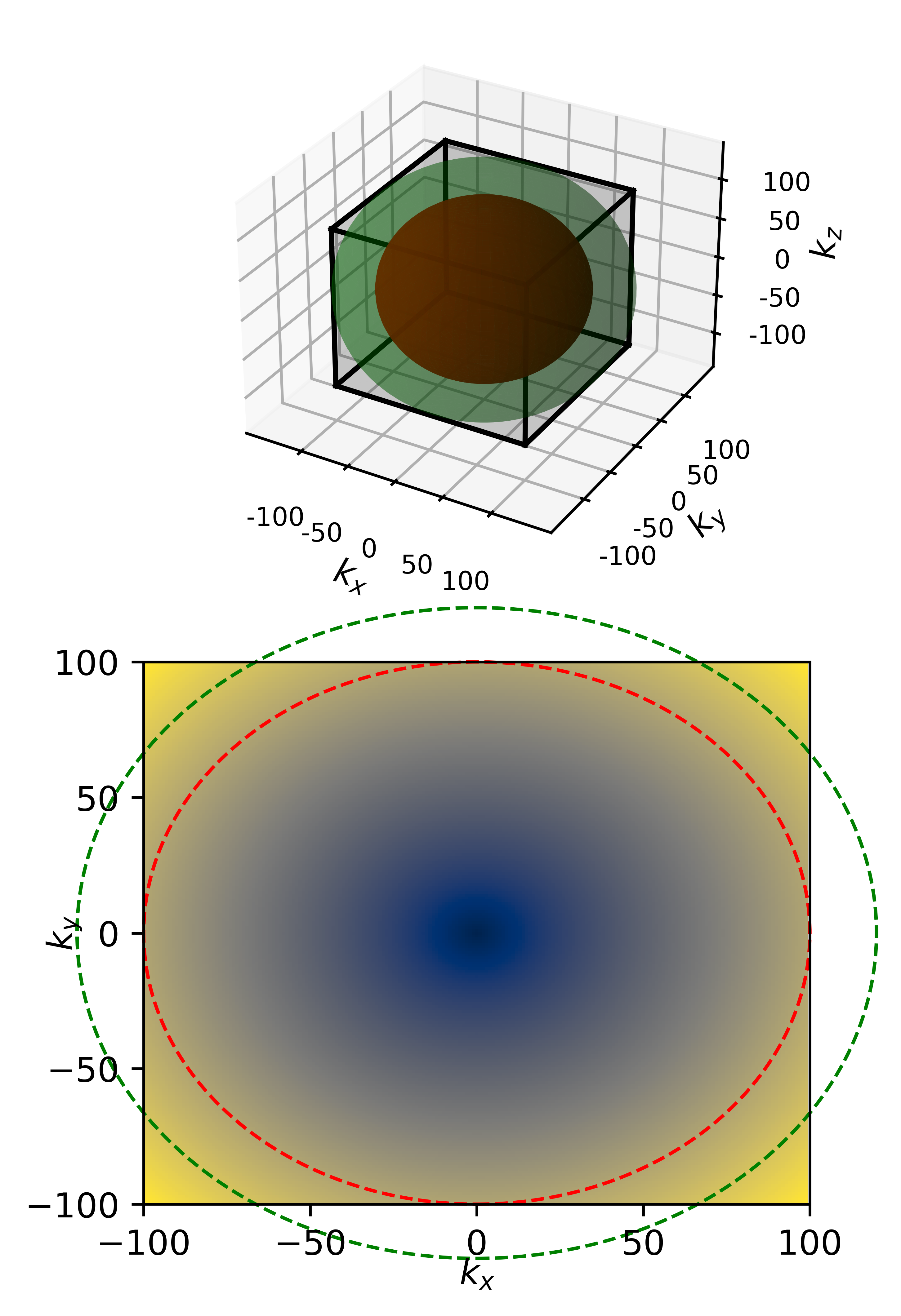}
 \captionof{figure}{Fourier modes of a 3D cube and a 2D map (upper and lower subplot respectively), in the manner of \cite{Ponthieu_2011}. The power spectrum $P(k)$ at a given $k$ value is the average of all spatial frequencies with the same magnitude, that is concentric shells or rings around the centre where $k=0$. The inner shell or circle in red, which touches the edges of the box, is the largest $k$ value where all $k$ modes can be included. All $k$ values with higher magnitude, such as the outer shell or circle in green, lie outside the box and thus miss $k$ modes, introducing greater uncertainties in the corresponding power spectrum value. In these examples $k$ is unit-less.}
 \label{fig:FigPonthieuDiagram}
\end{figure}

To get from this Fourier space cube to the power spectra $P(k)$, we had to spherically average the Fourier amplitudes for each spatial frequency bin. We determined the spatial frequency co-ordinates $k_x$, $k_y$, and $k_z$ corresponding to each Fourier amplitude, equivalent to $x$, $y$, $z$ in physical space. Following this we calculated the magnitude of the spatial frequency co-ordinates for each Fourier amplitude, which we refer to as $k$, using Pythagoras' theorem. Subsequently we averaged the Fourier amplitudes of all points with $k$ co-ordinates within a given $k$ bin, of width $\Delta k$. These averaged values correspond to $P(k)$. To visualise this concept we can use Fig. \ref{fig:FigPonthieuDiagram}, based on a diagram by \cite{Ponthieu_2011}. For our 3D cubes, we move in concentric shells of width $\Delta k$ away from the central value of $k=0$, finding the average of all values within each shell. This indicates that amplitudes corresponding to the highest $k$ values will have a greater degree of uncertainty as there will be missing $k$ modes in the cube, visualised by the green circles in the figure. The physical interpretation of these frequencies is that small $k$ represents larger physical scales, up to and including the whole intensity cube, while large $k$ represents small physical scales, including the variation of signal within individual beam widths. In the context of LIM the power spectrum at small $k$ is dominated by the impact of galaxy clusters and other large structures in `clustering signal', and the power spectrum at large $k$ represents the differences between individual galaxies in `shot-noise signal'. 

As our $k$ values are calculated by Fourier transforming the comoving length of the cube in Mpc, subdivided by the number of voxels in each dimension, $k$ is in units of Mpc$^{-1}$. Consequently, the spherically averaged power spectra $P(k)$ has units Mpc$^3(\textrm{Jy\,sr}^{-1})^2$. By convention we multiplied $P(k)$ through to find $k^3P(k)/2\pi^2$ in units of $(\textrm{Jy\,sr}^{-1})^2$, and all power spectra are to be shown in this format.

Due to the size and resolution of our cubes, the $k$ scales are restricted. For a spherically averaged power spectrum the largest $k$ scale is limited by the smallest physical scale, equivalent to the comoving distance covered by one voxel on the sky map (the beam width), as discussed in detail by \cite{Karoumpis_2022}.
\begin{align}
k_{\textrm{max}}&=\frac{\pi}{r_{\perp,\textrm{min}}}, \\
r_{\perp,\textrm{min}}&=D_\textrm{angular}(z_{\textrm{cen}}) \Delta\theta_\textrm{beam},
\label{eq:11}
\end{align}
where $D_\textrm{angular}(z_{\textrm{cen}})$ is the comoving angular distance at the mean redshift of the voxel in Mpc. The smallest $k$ scale is limited by the largest physical scale, equivalent to the diagonal between two opposite corners of the intensity cube, across the distance on the sky map and the full distance covered by the redshift range:
\begin{align}
k_{\textrm{min}}&=\frac{2\pi}{r_{\textrm{max}}},
\end{align}
\begin{align}
r_{\textrm{max}}&=\sqrt{r_{\perp,\textrm{max}}^2+r_{\parallel,\textrm{max}}^2},
\end{align}
\begin{align}
r_{\perp,\textrm{max}}&=D_\textrm{angular}(z_{\textrm{cen}}) \Delta\theta_\textrm{map},
\end{align}
\begin{align}
r_{\parallel,\textrm{max}}&=\frac{c}{H_0}\int_{z_{\textrm{min}}}^{z_{\textrm{max}}}\frac{\textrm{d}z}{\sqrt{\Omega_\textrm{M} (1+z)^3+\Omega_\Lambda}},
\end{align}

where $c$ is the speed of light, $\Delta\theta_\textrm{map}$ is the angular size of the whole sky map, $z_{\textrm{min}}$ is the low end of the redshift band, $z_{\textrm{max}}$ is the high end of the redshift band, and $\Omega_\textrm{M}$ and $\Omega_\Lambda$ are the cosmological parameters of our flat cosmology. No spatial frequencies exist outside this range. 

Throughout our analysis we used two different measures to find the width of the $k$ bins, that is the widths of the shells. We primarily used the narrowest possible interval, where there is a separate $k$ bin for each pixel in a ring around the origin in the 2D plane. This corresponds to using half the number of pixels along the map dimension$+1$ (60, 57, 50, 42 for $3.42<z<3.87$, $4.14<z<4.76$, $5.34<z<6.31$, $6.75<z<8.27$ respectively) giving a frequency width between $k$ modes of $\Delta k=0.043,0.040,0.037,0.035$\,Mpc$^{-1}$ respectively. The second method used far broader $k$ bins with $\Delta k=0.3$\,Mpc$^{-1}$, similar to those used in preliminary LIM work such as \mbox{\cite{Chung_2022Alt}}. All power spectra in subsequent figures feature narrower bins unless stated otherwise, due to their increased spatial frequency resolution as well as their use in most prior work.

We also determined a first estimate of the error in our power spectra in order to find the S/N. This error, $\sigma$, depends on instrumental thermal noise, the sample variance from binning $k$ modes, and instrumental beam smoothing. It was formulated by \cite{Li_2016}, and we followed the adapted version of their process as used by \cite{Chung_2020} and \cite{Karoumpis_2022}. The full details of its calculation is shown in Appendix \ref{appendix:psnoise}, with the most important takeaway being that $\sigma$ is inversely proportional to bin size. Correspondingly power spectra with narrower $k$ bins have greater relative errors.
\section{Results} \label{sec:results}
We now discuss the sample of empirical data, and the power spectra resulting from the generated 3D intensity cubes, comparing our models to each other and to previous simulated work. Due to the incompleteness of FARMER LP, the range covered by our power spectra form clear lower limits, the minimum estimate for a LIM cube from observational data. By performing error analysis we also discuss the challenges in determining the power spectra for EoR-Spec in this minimum case.
\subsection{Sample analysis} \label{sec:resultsIA}
In order to provide appropriate context to the upcoming power spectra, we discuss the $L_{[\textrm{CII}]}$ distribution of galaxies in FARMER LP when applying [CII] models, using intensity-frequency and luminosity function diagrams. In addition, we directly compare these statistics to the samples of previous literature, primarily \cite{Yue_2015}, \cite{Breyesse_2017}, \cite{Chung_2020}, and \cite{Karoumpis_2022}. We first examine the intensity-frequency (or intensity-redshift) graph and the [CII] luminosity function of FARMER LP and compare our results to those from \cite{Karoumpis_2022}. Here we averaged the intensity of each individual slice within the intensity cubes, equivalent to each individual frequency channel within the 40\,GHz bands, and plotted against frequency or redshift. We did this for all models applied to the FARMER LP galaxies for the frequency channels covered by our cubes.
\begin{figure}[t] 
 \centering
\includegraphics[width=\linewidth]{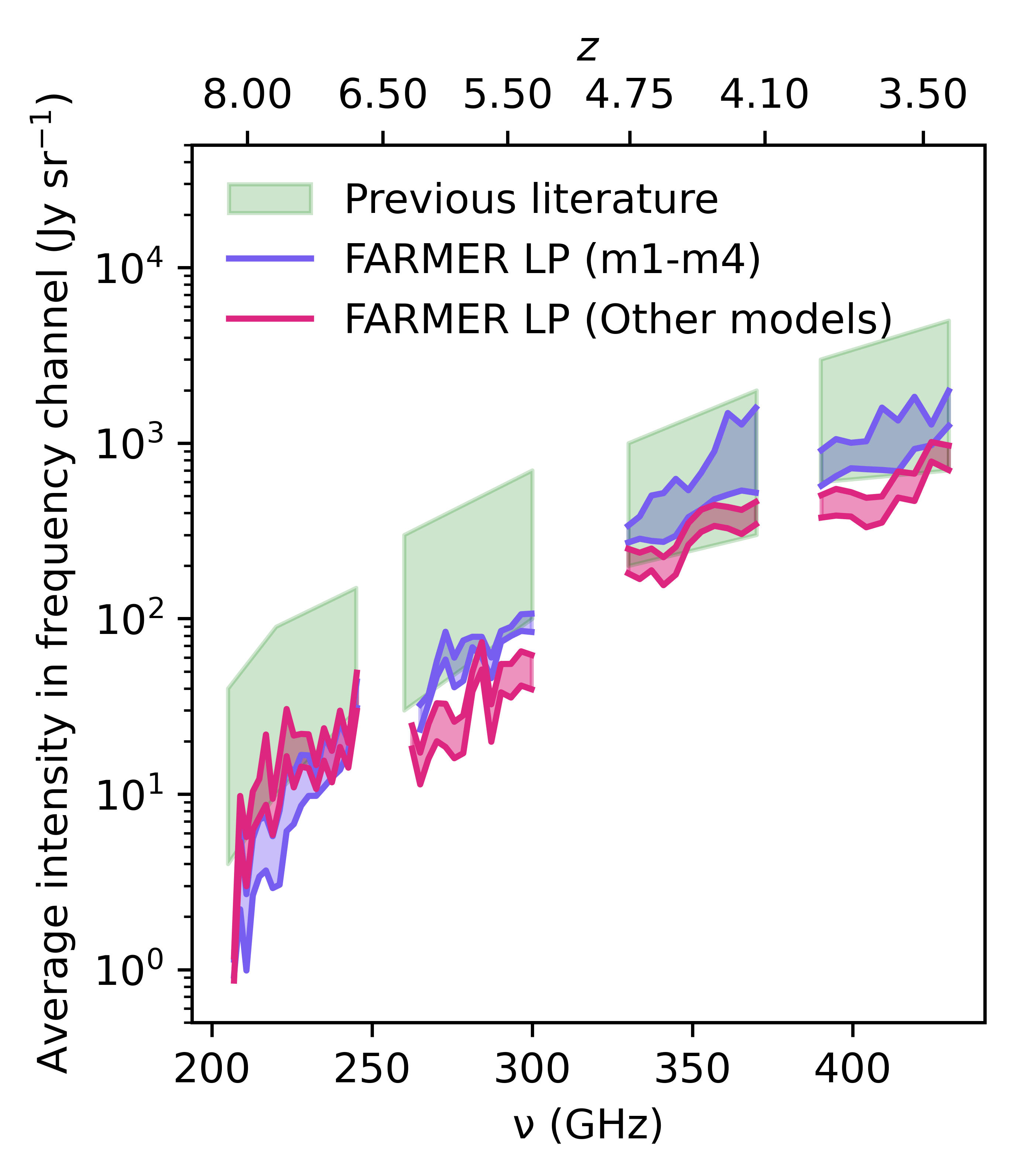}
 \captionof{figure}{Average [CII] intensity of slices of the 3D intensity cubes (in units of Jy\,sr$^{-1}$), plotted against the frequencies and redshifts of the respective slices. We separate the results from our cubes into the models from ALPINE and the models from the previous literature, in blue and purple respectively. We compared the results of our cubes to the equivalent statistic from previous simulated cubes in green (all taken from \citealt{Karoumpis_2022}, including itself and \citealt{Yue_2015}; \citealt{Breyesse_2017}; \citealt{Chung_2020}).}
 \label{fig:FigIvsFreq}
\end{figure} 

We only performed this analysis for the 40\,GHz frequency bands we cover to maintain consistency with other works, as well as to not misleadingly cover frequencies where there is weak transmission through the atmosphere. Figure \ref{fig:FigIvsFreq} shows that the previous literature is broadly in agreement with itself, within 1\,dex, and demonstrates the expected result: low mean intensity at low frequency, which rises with frequency in a curve shaped similarly to a logarithm. Their results follow the natural relation we expect as galaxies with the same luminosity at greater distances would give less flux, assuming [CII] emission does not significantly change with redshift. Our intensity cube channels agree with previous work at low redshift, albeit with slightly lower average intensity, indicating that FARMER LP is unlikely to be missing a significant amount of the galaxy luminosity function at lower redshift. Consequently, the significant decrease in average intensity at $5.34<z<6.31$ indicates that we have lost much of the dim end of the luminosity function at this distance. In contrast the lack of a sharp decrease at $6.75<z<8.27$ initially seems surprising in relation to the expected trend. However this behaviour is explained by the much greater mask extrapolation, drawing from a limited sample which is already biased towards the most luminous galaxies, potentially inserting more bright galaxies than we expect to exist. This indicates that our methodology is likely to be inaccurate for the $6.75<z<8.27$ band, at least when using data from the COSMOS 2020 catalogue. Finally, it is shown that m1-m4 produce higher intensity slice averages when applied to FARMER LP in comparison to the previous literature models, except for $6.75<z<8.27$ (a band that ALPINE data do not cover). This potentially indicates that ALPINE’s bias towards galaxies that are further along the main sequence could produce models that overestimate [CII] intensity, which we discuss in Sect. \ref{sec:discussion}.

It is important to note that it will be challenging to cleanly recover this intensity-frequency statistic from actual observations. This is because instrumental white noise and foreground signal prevent us from accurately measuring the average [CII] intensity specifically, as discussed by \cite{Breyesse_2017} and \cite{Karoumpis_2022}. We used this statistic here as it is a clear method to compare different samples of [CII] data to each other in this idealised scenario.

\begin{figure}[t]
 \centering
 \includegraphics[width=\linewidth]{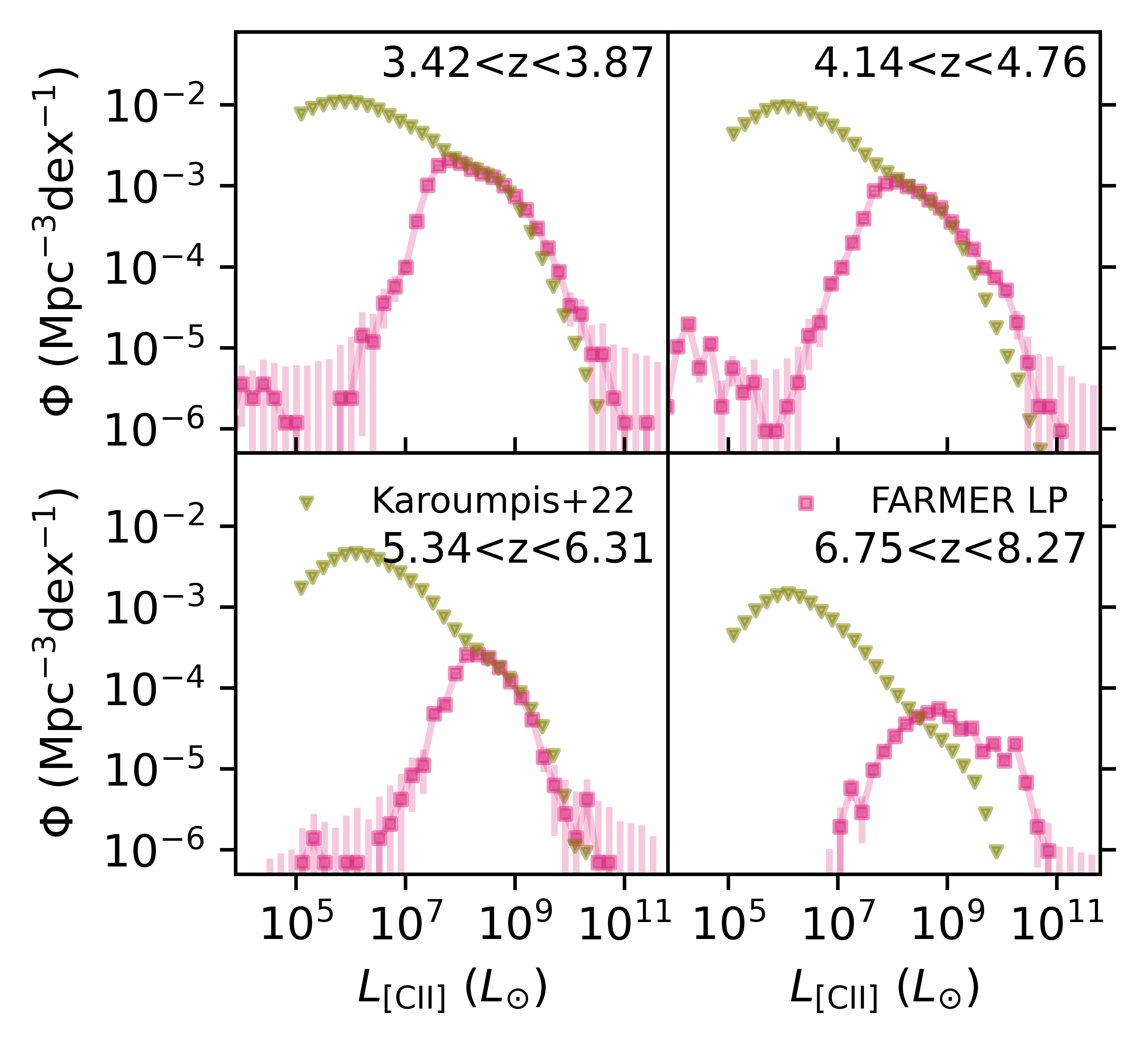}
 \captionof{figure}{Luminosity functions of FARMER LP in the four frequency bins within the range of EoR-Spec, including the extrapolated mask galaxies, shown down to $10^4L_\odot$. Comparison luminosity functions in the same redshift bands from \cite{Karoumpis_2022} are also shown. As they used the [CII] model from Eq. (8) of \cite{Vallini_2015}, we applied the same model to ensure consistency. We did not include this SFR- and metallicity-based [CII] model in our wider analysis because it was derived from radiative transfer cosmological simulations instead of from an existing sample. For all redshift bands there is a clear discrepancy below $10^8L_\odot$, as the models from \cite{Karoumpis_2022} assume a significant population exists below this point, which FARMER LP does not have. At lower redshift bands, especially visible for $4.14<z<4.76$, we see small upturns in the luminosity function below $10^6L_\odot$. In addition, the luminosity function of FARMER LP shows larger populations of high-luminosity galaxies for all bands except $5.34<z<6.31$. This is most noticeable for $6.75<z<8.27$, where our function is greater by 1\,dex. These deviations are likely caused or exacerbated by the propagated errors when calculating [CII] luminosity, as shown by the shaded error bars.}
 \label{fig:FigLFRedshiftComp}
\end{figure}

We also examined the [CII] luminosity function of FARMER LP for our different redshift bands in Fig. \ref{fig:FigLFRedshiftComp}, a standard measure that shows the relative density of galaxies with a given luminosity (number per unit volume per unit luminosity dex, where in this specific context dex is the magnitude interval which we take as 0.1). These luminosity functions verify that FARMER LP has a small population of [CII]-dim galaxies, because the expected Schechter curve shape \citep{Schechter_1976} of the luminosity function falls off at $\sim$10$^8L_\odot$ for each redshift band. For comparison, we note that the lowest [CII] luminosity from a local dwarf galaxy relevant to our work is $\sim$10$^4L_\odot$ \citep{Cormier_2015}, and that $10^6L_\odot$ is a typical [CII] luminosity of dwarf galaxies. This missing population for FARMER LP is in contrast to previous simulated work, which we show by overlaying the sample luminosity function used by \cite{Karoumpis_2022} when they used \cite{Vallini_2015}'s [CII] model. This discrepancy increases drastically below $L=10^8L_\odot$, indicating potential incompleteness in FARMER LP. However it is vital to state that we do not claim that the sample from \cite{Karoumpis_2022} is more accurate than FARMER LP, but instead that simulated predictions assume greater contributions to [CII] emission from the dim end of the luminosity function. Consequently, this means that our work in the form presented in this section is useful as a scenario where the expected low-luminosity galaxies do not exist, or contribute significantly less to the [CII] emission than previous simulations predict. This  becomes visible in the power spectrum when we make our comparisons to previous simulations.

Furthermore, FARMER LP has a slightly larger proportion of high-luminosity galaxies compared to some previous simulations (see $3.42<z<3.87$, $4.14<z<4.76$). For $6.75<z<8.27$, this difference exceeds 1\,dex above $10^{10}L\odot$, even eclipsing FARMER LP from $5.34<z<6.31$. While the discrepancies at low redshifts are most likely due to specific simulation parameters in previous simulations, the specifics of the \cite{Vallini_2015} model, or a slight sample bias towards high-luminosity galaxies in FARMER LP, the discrepancies at high redshift are far more significant. The high-redshift case indicates that previous simulations vastly underestimate the number of bright galaxies at high redshift, or that there is some sample or methodological error in COSMOS 2020 for galaxies with $z>6.3$. This is also shown in our results in Fig. \ref{fig:FigIvsFreq}. It is possible that this is a consequence of error propagation when calculating [CII] luminosity, as shown by the error bars in the luminosity function found by Monte Carlo simulations. We discuss this further with direct comparisons to the ALMA Reionization Era Bright Emission Line Survey (REBELS, \citealt{Bouwens_2022}) in Appendix \ref{appendix:mapmakingvar}. While we  show the results at this redshift band using this mask extrapolation to maintain consistency with the other redshift bands, we believe that results at $6.75<z<8.27$ should be viewed with caution.

\subsection{Power spectra, comparison to previous work, and lower limits}  \label{sec:resultsIPS}
\begin{figure*}[t]
 \centering
 \includegraphics[width=\linewidth]{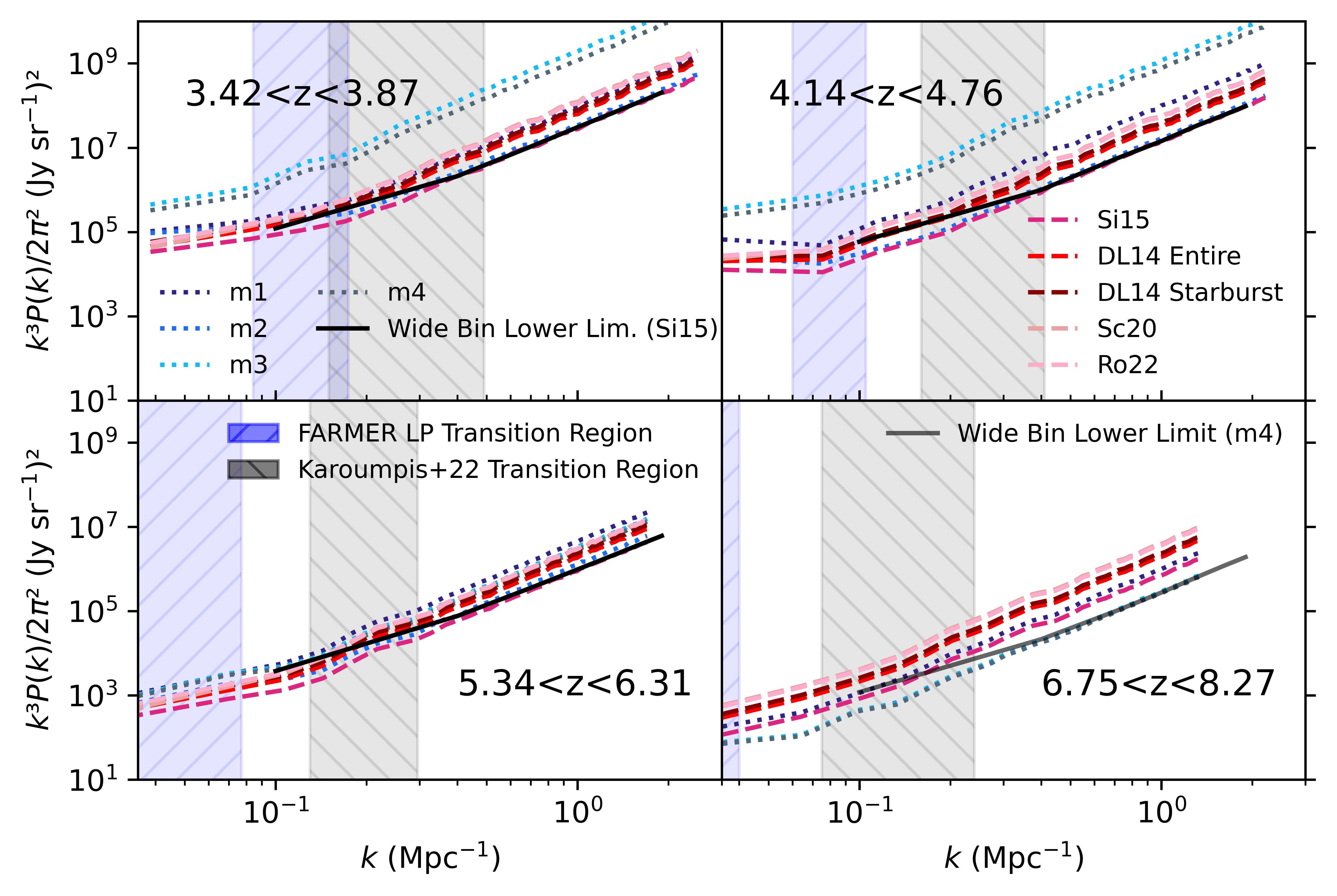}
 \captionof{figure}{Power spectra of models m1-m4, Si15, DL14, Sc20 and Ro22 for all redshift bands, demonstrating a spread of up to 2\,dex. We always show power spectra in the $k^3P(k)/2\pi^2$ format. Differences in maximum and minimum $k$ with regards to redshift are due to different beam and map size. The lines are kept consistent across all future graphs, however, the colours change to display the contrast with respect to previous simulation works. While most of these power spectra are shown with narrower $k$ bins as discussed in Sect. \ref{sec:methodPS}, we also overlay the wider $k$ bin versions ($\Delta k=0.3$\,Mpc$^{-1}$) of the absolute minimum power spectra (Si15 for most redshifts, m4 for $6.75<z<8.27$). These minimum values are quantified in Table \ref{table:k vals}. The upturn at small $k$ for these lower limits is due to using wider $k$ bins compared to the other power spectra. The transition region (see text) for our calculated models is shown in hatched blue, with a transition region from \cite{Karoumpis_2022} shown in hatched grey for contrast.}
 \label{fig:FigCompareModels}
\end{figure*}

Our power spectra, derived from the mock samples we produced, can be viewed as predictions for if the [CII] luminosity function and clustering of the FARMER LP sample matched that of actual observations. Therefore, in the context of Figs. \ref{fig:FigIvsFreq} and \ref{fig:FigLFRedshiftComp}, these power spectra provide lower limits for the results from EoR-Spec's observation of E-COSMOS. Before we compare our work and the constraints they provide to the previous literature, we first show the nuances between each individual model's power spectra for all redshift bands in Fig. \ref{fig:FigCompareModels}.
\begin{figure*}[t]
 \centering
 \includegraphics[width=\linewidth]{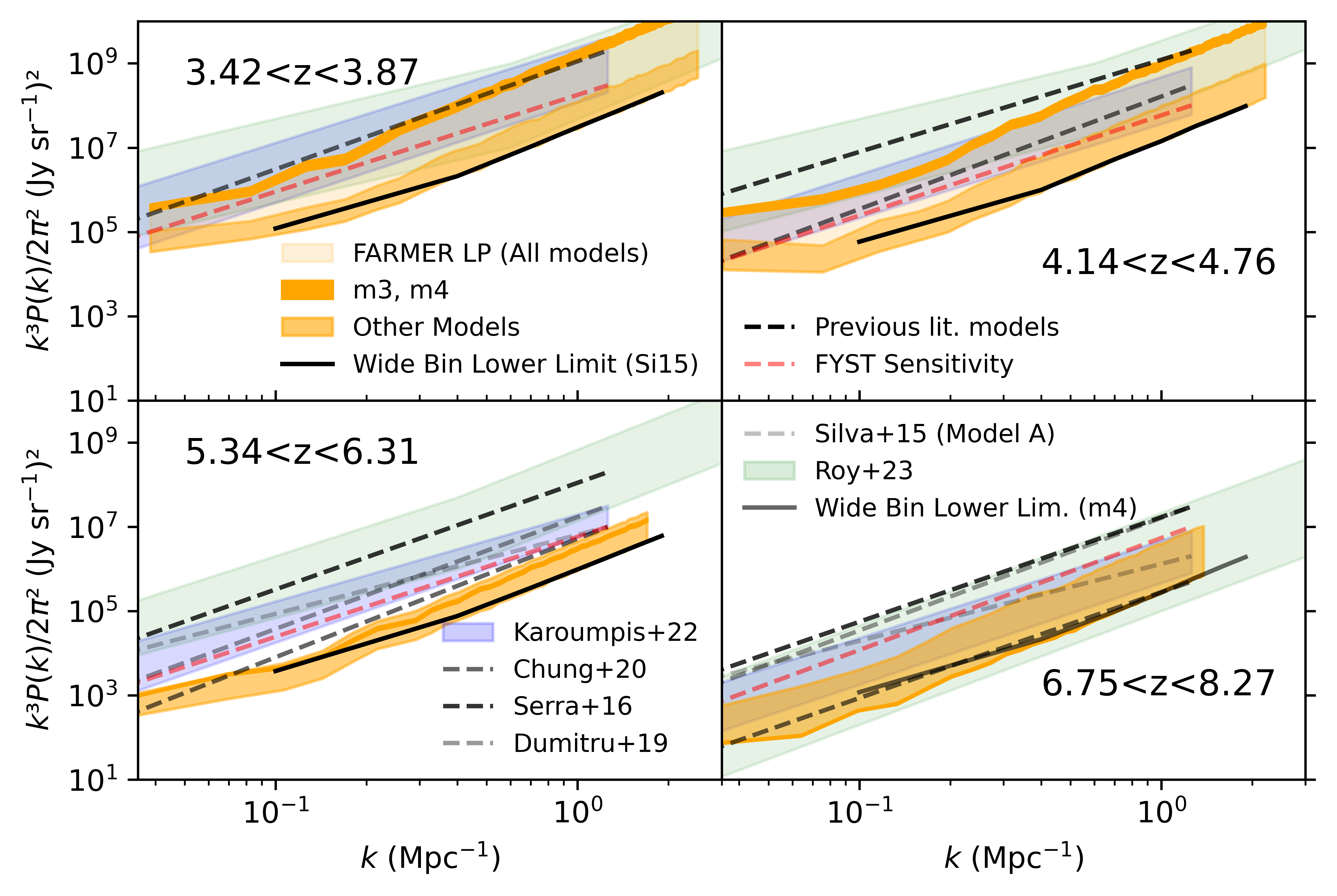}
 \captionof{figure}{Power spectra at all frequency channels comparing our work to previous literature power spectra. Our models are grouped together in orange, with different shadings used to highlight models m3 and m4 at $3.42<z<3.87$ and $4.14<z<4.76$. We show all models with narrow $k$ bins, equivalent to the best resolution possible, however the lowest models are also shown with bin width $\Delta k=0.3$\,Mpc$^{-1}$, these power spectra having worse resolution but being more accurate to expected observations. Previous simulation work is shown with dashed lines, with the exception of \mbox{\cite{Karoumpis_2022} and \cite{Roy_2023}}, where the range of predictions is shown in blue and green shaded regions respectively. We also include the power spectra of the expected EoR-Spec sensitivity taken from \cite{CCAT_Prime_Collaboration_2022}.}
 \label{fig:FigCompareDefaultPS}
\end{figure*}

We found trends amongst the magnitudes of these models, equivalent to their shot-noise and therefore their sample luminosity functions, which describe their fundamental behaviour. Overall, we saw that most models (m1, Si15, DL14, Sc20, Ro22) stay within 1\,dex over all redshift bands. These models, which are primarily from previous literature, exhibit the expected downwards shift in magnitude with increasing redshift as intensity decreases. This trend does not continue at $6.75<z<8.27$, with the magnitudes not meaningfully decreasing due to the unusual nature of that sub-sample as discussed earlier (e.g. Fig. \ref{fig:FigIvsFreq}). The m2 model is similar but experiences a decrease between the final bands, which indicates that the model assigns relatively low luminosities to the galaxies at $6.75<z<8.27$. The m3 and m4 models have significantly larger magnitudes compared to other models for $3.42<z<3.87$ and $4.14<z<4.76$, are roughly equivalent for $5.34<z<6.31$, and have significantly lower magnitudes for $6.75<z<8.27$. This indicates that m3 and m4 assigned a higher average luminosity than other [CII] models at low redshift. This includes dim galaxy detections and lower average luminosity than other models at high redshift, which only includes bright galaxy detections.
These deviations most likely arise from the construction of the ALPINE models (Eqs. {\ref{eq:em4}}--{\ref{eq:em7}}) and how they interact with the specific properties of the sample galaxies in each band. Overall, m1 is the most complex model we created, however it behaving as a typical model has precedence due to its structural similarity to [CII] models formed from simulations, such as from \mbox{\citealt{Vallini_2015}}. In contrast, m2 directly relates the [CII] emission to stellar mass, which is correlated with but not equivalent to SFR. For most redshift bands there is no significant difference, however in FARMER LP galaxy detections at $z>6.3$ are made by UltraVISTA in NIR. For those distant galaxies, this corresponds to rest-frame UV emission and so, those galaxies must have high UV luminosities to be detected. As there is a strong correlation between UV emission and SFR, because UV emission is dominated by the unobscured light from young massive stars (e.g. \mbox{\citealt{Kennicutt_1998}}), most of the detected galaxies in FARMER LP have high SFR but they do not necessarily have high stellar mass. Galaxies with high stellar mass at these distances would have high rest-frame optical emission, however at present we cannot probe this emission as we do not have access to the relevant bands from JWST. Therefore, m2 is likely to assign these galaxies lower [CII] emission when compared to models from the literature which relate to SFR, and so having a power spectrum with lower magnitude for $6.75<z<8.27$. For m3 and m4, these models have opposing linear terms in stellar mass and SFR or sSFR. Due to the correlation between these opposing terms, we see a smaller difference in assigned [CII] emission between bright and dim galaxies. This is a useful idea to explore within our models, but it deviates significantly from the established literature. When applying m3 and m4 to the low-redshift bands of FARMER LP, the many smaller galaxies are assigned relatively high [CII] emission, resulting in power spectra with higher magnitudes. We see the inverse effect for high-redshift bands which only have a few large galaxies. In this way, power spectra from m1-m4 show useful results for unconventional models of [CII] emission.

In addition to the narrow $k$ bin power spectra, we also superimpose a version of the lowest magnitude power spectra whilst using large $k$ bins, to more accurately visualise what upcoming observations would find in the minimum possible case. While these power spectra appear to have stronger clustering signal at small $k$ compared to the narrow bin power spectra, this is only because they average the power spectra over a wider $k$ range and so have less resolution on clustering scales. Even so, these power spectra cover the same intensity cubes as the narrow bin versions, and so can still be taken as lower limits. To quantify these lower limits, we show the power spectra values from Fig. \ref{fig:FigCompareModels} in Table \ref{table:k vals}.

\begingroup
\setlength{\tabcolsep}{3pt} 
\begin{table}
\caption{Lower limits of FARMER LP power spectra, when $\Delta k=0.3$\,Mpc$^{-1}$.}
\centering
\begin{tabular}{c c c c c}
\hline\hline
\, &\multicolumn{4}{c}{$k^3P(k)/2\pi^2$ ((Jy\,sr$^{-1}$)$^2$)}\\
 $k$\,(Mpc$^{-1}$) & 0.25-0.55 & 0.55-0.85& 0.85-1.15& 1.15-1.45\\
\hline
$3.42<z<3.87$ &2.11$\times10^6$& 1.06$\times10^7$& 3.06$\times10^7$& 6.56$\times10^7$\\
$4.14<z<4.76$ &9.93$\times10^5$& 5.29$\times10^6$& 1.43$\times10^7$& 3.26$\times10^7$\\
$5.34<z<6.31$ &7.66$\times10^4$& 3.58$\times10^5$& 9.80$\times10^5$& 2.07$\times10^6$\\
$6.75<z<8.27$ &2.16$\times10^4$& 9.80$\times10^4$& 2.77$\times10^5$& 6.26$\times10^5$\\
\hline
\end{tabular}
\tablefoot{For most redshift bands this minimum model comes from Si15, with the exception of $6.75<z<8.27$ where the minimum is m4 (Fig. \ref{fig:FigCompareModels}). Due to the averaging of power spectra within these wide bins, there are slight discrepancies on clustering scales when comparing these power spectra to those with narrower bin sizes (and so higher frequency resolution).}
\label{table:k vals}
\end{table}
\endgroup
Subsequently, we determined the impact of galaxy clustering on the power spectrum. As discussed in detail by \cite{Uzgil_2014}, the shot-noise (small scales, right hand side of each subplot) is effectively a Poisson noise effect which creates the power law shape in the $k^3P(k)/2\pi^2$ power spectra. The clustering component (large scales, left hand side of each subplot) of the power spectrum is shown by a `kick' at low $k$, a deviation from the shot-noise power law. The region where both contributions are approximately equal is called the transition region, a concept we take from \cite{Karoumpis_2022} to visualise the strength of clustering on different scales more clearly. We calculated this for FARMER LP (hatched blue region) by finding the $k$ modes where the magnitude of the power spectrum for each [CII] model is approximately twice the shot-noise power law, that is where the additional clustering signal component is equal to the shot-noise component, and then taking the median $k$ modes of these points for all the models of the given redshift band. By looking at these transition regions, we found that the kick is only noticeable at low redshifts for all models ($3.42<z<3.87$, $4.14<z<4.76$), with power spectra never leaving the transition region for higher redshifts. Weaker clustering at higher redshift is similar to the work of \cite{Roy_2023}. Furthermore, for low-redshift bands most models exhibit kicks at $k\approx10^{-1}\,\textrm{Mpc}^{-1}$, with notable deviations for m3 and m4. This is potentially due to m3 and m4 having low variation between galaxy [CII] emission as discussed earlier, reducing the impact of specific overdense regions. 

When comparing this clustering to that of previous simulations, their power spectra have far greater clustering signal, as shown by the transition region from \cite{Karoumpis_2022} for their power spectra in Fig. \ref{fig:FigCompareModels} (hatched grey region). This contrast is likely due to FARMER LP excluding many of the low-luminosity galaxies that surround more luminous galaxies in the sample. This therefore reduces the impact of any existing galaxy clusters on the power spectra, a problem which becomes far worse for the high-redshift bands with greater incompleteness. This idea can also be described by power spectra statistics as discussed by \cite{Uzgil_2014} and \cite{Karoumpis_2022}, as low-luminosity galaxies in the sample have a significantly greater impact on the clustering signal component than the shot-noise component. In addition, as $6.75<z<8.27$ had a far larger mask (and thus more mask extrapolation), this likely resulted in most large-scale structure being destroyed. Finally, when using large $k$ bins the kicks appear to occur at higher $k$, however this effect is due to the lower precision of these bins.

In this way, by making intensity cubes and the corresponding power spectra whilst only using galaxies from COSMOS 2020, that is galaxies which we know exist without any additional signal, we derived a lower limit for possible [CII] power spectra. It is therefore unlikely that the future observed power spectra of EoR-Spec will stray below the limits as quantified in Table \ref{table:k vals}. 

We then compared our power spectra to the simulated power spectra from the literature in detail, as shown in Fig. \ref{fig:FigCompareDefaultPS}. In  graphs below, we group our power spectra together to make them more readable, with specific smaller groupings being used to examine specific models. The methodology of the previous literature is described below.

\cite{Karoumpis_2022}: The authors used $4\times4 \deg^2$ mock scans created by the Illustris TNG300-1 simulation with frequency bands in the range covered by EoR-Spec. They assigned SFR and other bulk properties to their galaxies using abundance matching. From this, the authors selected a $2\times2 \deg^2$ region, apply models from \cite{Vallini_2015}, \cite{Lagache_2018}, and Sc20 and obtained a range of predictions. We show the range they covered using the blue shaded area in Fig. \ref{fig:FigCompareDefaultPS} and similar figures.

\cite{Chung_2020}: The authors used the \cite{Lagache_2018} model with an added scatter of $\sim$0.5\,dex to emulate the deviation within said model, creating cubes from the galaxy-halo model of UNIVERSE MACHINE with EoR-Spec's frequency bands. They acquired maps with an approximate size of $2\times2 \deg^2$. Their results indicate a relatively weak clustering signal and that detection past $z\approx6$ will be challenging.

\cite{Serra_2016}: Using galaxy data from $z<4$, the authors used $L_{\textrm{IR}}$ emission to infer $L_{[\textrm{CII}]}$ emission assisted by a halo model. Their power spectra have a relatively high magnitude, giving an upper bound when compared to other simulations. The authors aimed to cover CONCERTO (200--360\,GHz) with an area of $2 \deg^2$, which overlaps with all our bands except $3.42<z<3.87$.

\cite{Dumitru_2019}: The authors used \cite{Lagache_2018} without any added scatter, applied to a hydro-dynamical cosmological simulation made by the Sherwood simulation suite \citep{Bolton_2016}, which was combined with the G.A.S semi-analytical model \citep{Cousin_2016} to determine SFR. They covered snapshots at high redshifts ($z\approx6.3, 7.1, 8.2, 9$) with an area of $1.5\times1.5 \deg^2$.

\cite{Silva_2015}: The authors used simulations from the SImfast 2021 code (\citealt{Santos_2010}; \citealt{Silva_2013}) and galaxy data from \cite{De_Lucia_2007}, applied a halo mass-SFR relation, and created their cubes using their four separate [CII] models (Si15 and three others). They assumed strong clustering signal and weak shot-noise, leading to a large kick at relatively high $k$ modes. The authors covered a range of 200--300\,GHz ($5.33<z<8.5$), but we only include their work at $5.34<z<6.31$ as their specific frequency bands overlap poorly with $6.75<z<8.27$. This map has a $1.35\times1.35 \deg^2$ coverage, similar to our work.

\cite{Roy_2023}: The authors used the LIMpy package to generate power spectra with SFR provided by UNIVERSE MACHINE and Illustris TNG. They implemented a broad range of SFR models: \cite{Eli_Visbal_2010}, all models from Si15, \cite{Fonseca_2016}, \cite{Lagache_2018}, and Sc20. We  show the range they covered using the green shaded area. They covered the same redshift bands as us, with map sizes of $4\times4 \deg^2$.

As these simulated works typically cover larger sky map scales than the area covered by COSMOS 2020, their power spectra extend to small $k$. However as power spectra are normalised with cube volume we do not expect any meaningful discrepancies in the magnitudes, and as the previous literature typically used narrow $k$ bins we expect to see the kicks to be directly comparable to our models when using narrow $k$ bins. We also show the first estimate of FYST's sensitivity from \cite{CCAT_Prime_Collaboration_2022}, the approximate detection limit, for comparison purposes.

\begin{figure*}[t]
 \centering
 \includegraphics[width=\linewidth]{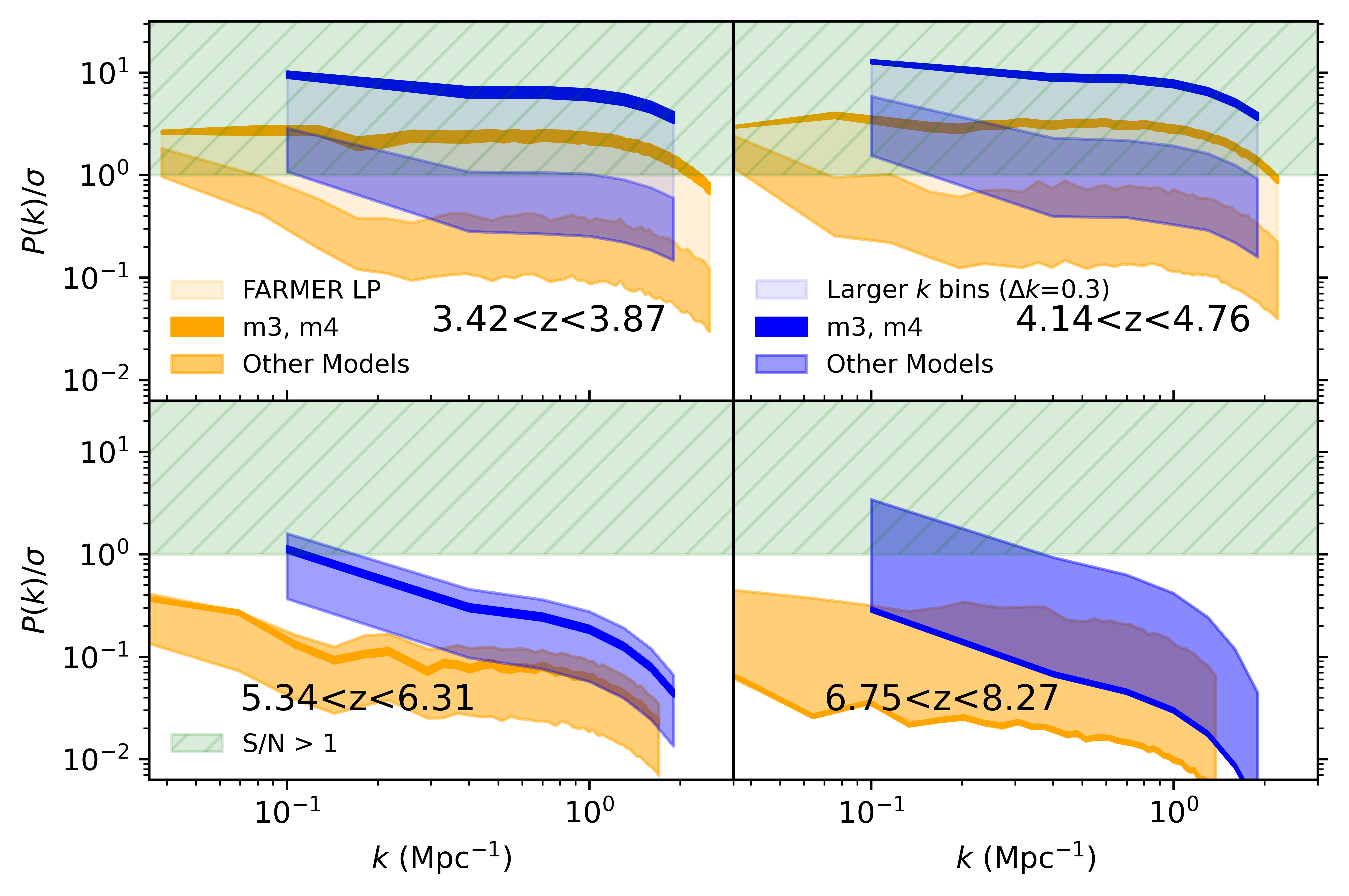}
 \captionof{figure}{Comparing $P(k)$ and its error, $\sigma$, by finding their ratio (equivalent to S/N). We show the ratio of $P(k)/\sigma$ for all models and redshifts. Orange is for narrower bins of width $\Delta k=0.043,0.040,0.037,0.035$\,Mpc$^{-1}$ respectively, whilst blue is for bins with width $\Delta k=0.3$\,Mpc$^{-1}$. If this ratio is greater than 1, denoted by the green hatched region, we have greater signal than noise. m3 and m4 are highlighted by shading.}
 \label{fig:FigDefaultPkDivSigma}
\end{figure*}

Our power spectra are almost always lower in magnitude when compared to the previous literature, including lying below the expected sensitivity of the instrument. At high $k$ most of our models partially overlap with some previous simulated work such as the lower ends of \cite{Karoumpis_2022} and \cite{Roy_2023}. This overlap shrinks as $k$ decreases, indicating that previous work has stronger clustering components, as indicated by the transition regions in Fig. \ref{fig:FigCompareModels}. While the absolute lower limits do not exclude any previous simulation work, models m3 and m4 eclipse the lower ends of other simulated models and the expected EoR-Spec sensitivity power spectrum for $3.42<z<3.87$ and $4.14<z<4.76$. However, other power spectra from the literature including \cite{Serra_2016}, \cite{Dumitru_2019}, and the upper end of \cite{Roy_2023} are significantly greater in magnitude than any of FARMER LP power spectra, which is to be expected as these works assume a much greater contribution from faint galaxies. $6.75<z<8.27$ is the exception to these trends with significantly greater overlap, however as noted in Sects. \ref{sec:methodMAP}, \ref{sec:resultsIA} and Appendix \ref{appendix:mapmakingvar} this band is likely limited in use.

In summary, Figs. \ref{fig:FigIvsFreq} and \ref{fig:FigLFRedshiftComp} show that COSMOS 2020 provides an empirical lower limit to simulation-based estimates for the luminosity function when we apply our luminosity models. This is also demonstrated in the power spectra of Fig. \ref{fig:FigCompareModels}. Our lower limits for power spectra based on existing data cannot definitively exclude the lower ends of previous simulated power spectra from the literature (Fig. \ref{fig:FigCompareDefaultPS}), and the FARMER LP power spectra have comparatively weaker clustering components due to the lack of faint galaxies in the sample.

\subsection{Error analysis}  \label{sec:resultsEA}
We found a first estimate of the error in the power spectra and visualised it in the form of the S/N $P(k)/\sigma$, where $\sigma$ is the analytically calculated instrumentation error of EoR-Spec (process described in Appendix \ref{appendix:psnoise}). While we cannot presently include all possible error contributions, such as errors from foreground contamination removal and sky noise, this does include instrumental thermal noise, the sample variance from binning $k$ modes, and instrumental beam smoothing. Measurements of $\sigma$ will be improved when we obtain the initial data, after the first light of EoR-Spec, but we consider this a good first estimate of noise in the power spectra.

When viewing the relative error for power spectra in Fig. \ref{fig:FigDefaultPkDivSigma}, we can determine useful information at $k$ scales where the S/N is above 1, that is when there is more signal than noise (denoted by the hatched green region). When using narrow $k$ bins this is possible for m3 and m4 at $3.42<z<3.87$ and $4.14<z<4.76$, but all other models fail to reach S/N$>1$ at these redshifts, and all models fail at higher redshifts. If we use wider bins we retrieve greater S/N as $\sigma$ is inversely proportional to the number of $k$ modes in a bin, a number which increases if we use wider bins (Appendix \ref{appendix:psnoise}). However, noise still dominates for lower magnitude models and models at higher redshifts. We recalculated these for greater observation time than our assumed 2000 hours, however even when using large bins, we only achieved S/N$>1$ for most models at $z<6.3$ after tripling our observation time. These calculations also assume that signal extraction is 100\% efficient, so in reality these ratios are likely to be lower. 

This metric indicates that observations cannot retrieve any useful information in the initial observation period for the most pessimistic cases, even when using wide $k$ bins. This issue is more pressing for shot-noise regions, which typically have greater errors compared to clustering regions, thereby leading to problems when constraining the luminosity function of observed fields. In this case higher redshift bands would be a significant technical challenge, as discussed by \cite{Chung_2020}.

\section{Extrapolation}  \label{sec:interpolation}
\begin{figure*}[t]
 \sidecaption
 \includegraphics[width=12cm]{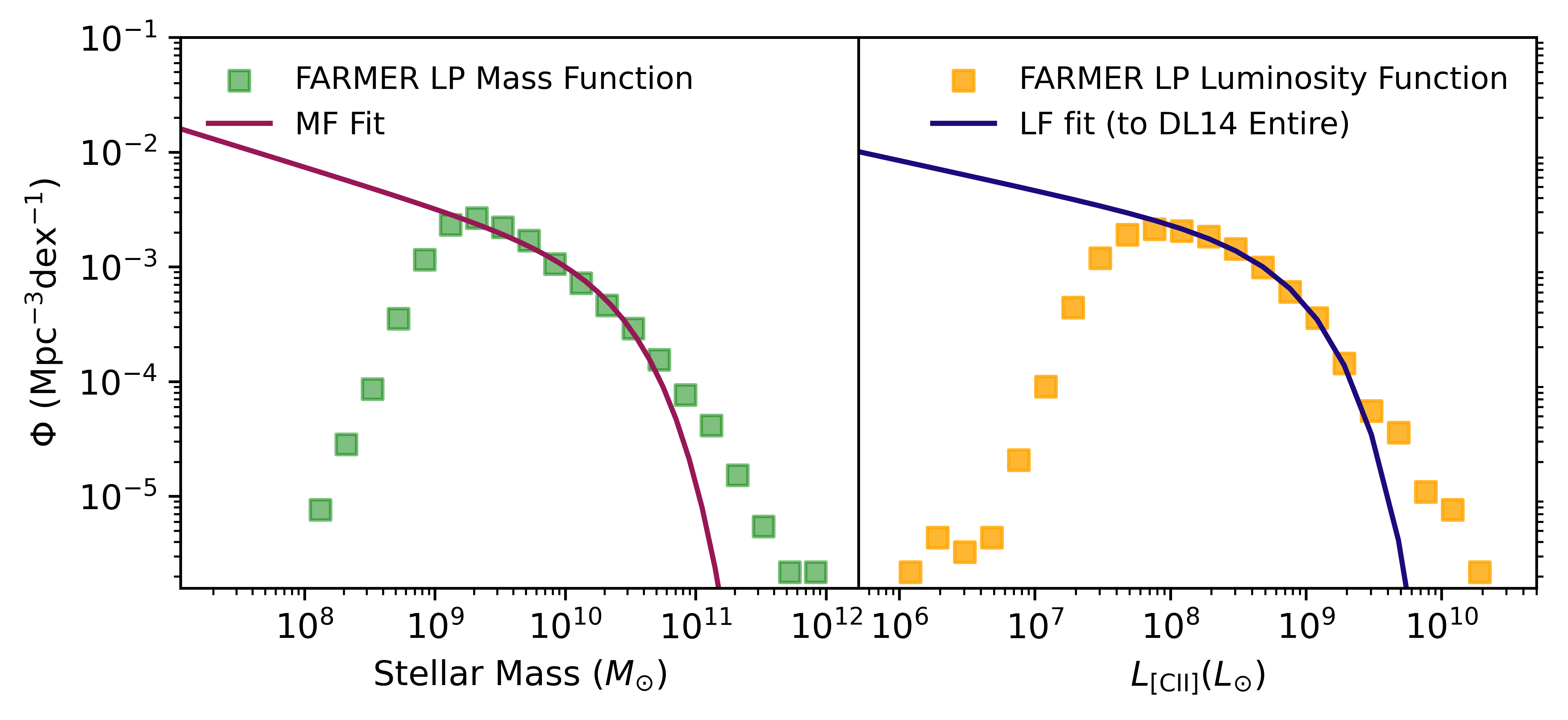}
 \caption{Examples of how Schechter fits are applied to mass functions (left) and luminosity functions (right) at $3.42<z<3.87$. The luminosity function in this case is when we apply DL14 Entire to FARMER LP. In both cases, we applied the CANDELS correction to the sample before fitting the Schechter curves. We extrapolate galaxies corresponding to the difference between the projected curves and the existing functions. While there are slight discrepancies at high mass or luminosity, this is acceptable as this extrapolation method focuses on dim galaxies, and there are higher errors for the bright end of these functions (as shown in Fig. \ref{fig:FigLFRedshiftComp}).}
 \label{fig:FigBothm2L18LF}
\end{figure*} 
The limited completeness of FARMER LP is useful for making our lower bound mock LIM cubes, however it is noticeably different compared to previous simulation work. This is due to their implicit assumption of fainter galaxies at high redshift contributing significantly to the [CII] luminosity function. In this context we decided to create a hypothetically more complete sample by extrapolating from FARMER LP, creating additional galaxies to fill in perceived missing gaps in the sample's mass or luminosity function, which are then added to the mock LIM cubes. By creating a variety of power spectra from physically plausible samples and checking their concordance with previous simulations, we could verify if the simulated works are consistent with existing empirical data from COSMOS 2020. In addition, as part of this extension we aimed to create reasonable upper limits for the observed power spectra, although these will not be as rigorous as the lower limits due to the inherent uncertainties in extrapolating from an existing mass or luminosity function.

When extrapolating we had to determine the number of additional galaxies with appropriate bulk properties, and devise a method to sensibly add these galaxies to the existing cubes. We explored three techniques of extrapolation: exploiting data from surveys which probed deeper than COSMOS 2020, extrapolating from the mass function of FARMER LP, and extrapolating from the luminosity functions we generated from applying [CII] models to FARMER LP. Each process was performed separately for each frequency band to account for the sample variation on different cosmological timescales. The galaxies were then added to the LIM cubes appropriately using a Voronoi Tessellation (VT) technique. After creating these samples and corresponding power spectra, we repeated our analysis with the power spectra and relative errors to show concordance with previous simulation work.
\subsection{Using CANDELS data} \label{sec:interpolationCANDELS}
\begin{figure*}[t]
 \sidecaption
 \includegraphics[width=12cm]{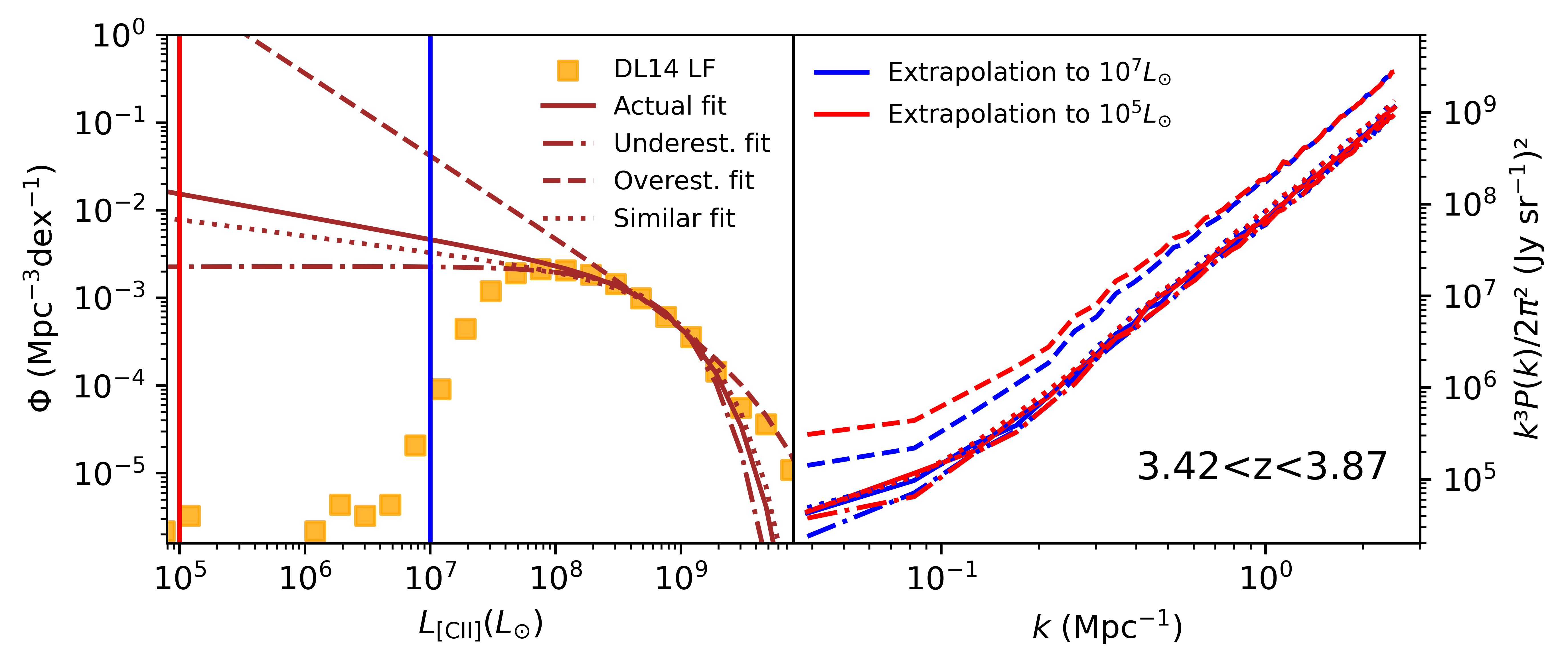}
 \caption{Comparison of different Schechter curve fits to the DL14 Entire [CII] luminosity function at $3.42<z<3.87$. This sample includes FARMER LP, the stellar mask and CANDELS factor adjustments. We include the actual fit we chose, as well as fits that are underestimates, overestimates, or were similar. Parameters for these fits are in Appendix \ref{appendix:fitparams}. The left subplot shows the actual Schechter curves themselves, including the two limits of extrapolation we investigate, and the right subplots show the resulting power spectra. The power spectra are mostly similar, except for the `overestimate' case.}
 \label{fig:FigDW}
\end{figure*}
We made a first estimate of the incompleteness of FARMER LP by comparing it to deeper surveys in the wider COSMOS field. We primarily used the Cosmic Assembly Near-infrared Deep Extragalactic Legacy Survey (CANDELS, \citealt{Nayyeri_2017}), which covers a $\sim$0.06$\deg^2$ region within one of UltraVISTA's Ultra-Deep stripes in the COSMOS field, using HST data. \cite{Weaver_2022a, Weaver_2022b} noted that the galaxy number ratio of FARMER LP to CANDELS is $\sim$75\% for $z>2.5$ within the region probed by CANDELS, implying that FARMER LP did not detect at least 25\% of galaxies in that region. Therefore we planned to use the ratio between FARMER LP and CANDELS to find an appropriate number of galaxies to add to the sample. In order to do this we calculated this ratio again for each frequency band, for galaxies past a certain stellar mass threshold where FARMER LP was determined to be almost `mass complete' (excluding discrepancies with CANDELS). We used this mass limit as we only intended to correct for the high-mass galaxies with this method, the fainter galaxies being accounted for by Schechter curve comparisons. This mass threshold was calculated in several ways to ensure the limit was accurate, first by using the following equations from \cite{Weaver_2022a, Weaver_2022b} for 95\% mass completeness:
\begin{eqnarray}
M_{\textrm{lim, `22}}= -1.51\times10^6 (1+z)+6.81\times10^7 (1+z)^2,\\
M_{\textrm{lim, `23}}= -3.23\times10^7 (1+z)+7.83\times10^7 (1+z)^2. 
\label{eq:mlim}
\end{eqnarray}
In addition to these methods we also manually calculated the $M_{\textrm{lim}}$ value by following the same procedure as \cite{Weaver_2022b}. However, instead of re-scaling the 30th mass percentile of galaxies by IRAC Ch1 luminosity, we took the 40th percentile of masses within FARMER LP because of our focus on high-redshift bright galaxies. All three methods return the same stellar mass thresholds within 0.05\,dex: $10^{9.2}$, $10^{9.3}$, $10^{9.5}$, $10^{9.7}M_\odot$ for $3.42<z<3.87$, $4.14<z<4.76$, $5.34<z<6.31$, and $6.75<z<8.27$ respectively. We then found the number ratio of galaxies in FARMER LP to CANDELS above these mass thresholds (after accounting for the stellar mask regions of FARMER LP within the CANDELS region), which were 78\%, 78\%, 51\%, and 58\% respectively. We always extrapolated additional galaxies above the mass thresholds in the FARMER LP sample following these ratios before applying the other extrapolation methods. For example, when we applied this correction to the $5.34<z<6.31$ sub-sample, we generated $(100/51)-1\approx1$ galaxy for each existing FARMER LP galaxy with stellar mass above $10^{9.5}M_\odot$, duplicating existing galaxies to determine the properties of these new galaxies. That is, we added an 1 extra galaxy for each original galaxies in FARMER LP with stellar mass above the mass completeness threshold, intuitively accounting for the 49\% assumed incompleteness in this FARMER LP band. In the $3.42<z<3.87$ band, we generated $(100/78)-1\approx0.28$ galaxies per original galaxy above $10^{9.2}M_\odot$, or approximately 1 extra galaxy for 4 originals, and so on. In this way, the CANDELS factor corrects for the high end of the galaxy main sequence, with Schechter fits correcting for the low end.
\subsection{Extrapolating from mass and luminosity functions using Schechter curves} \label{sec:interpolationMLF}
\begin{figure*}[t]
 \sidecaption
 \includegraphics[width=12cm]{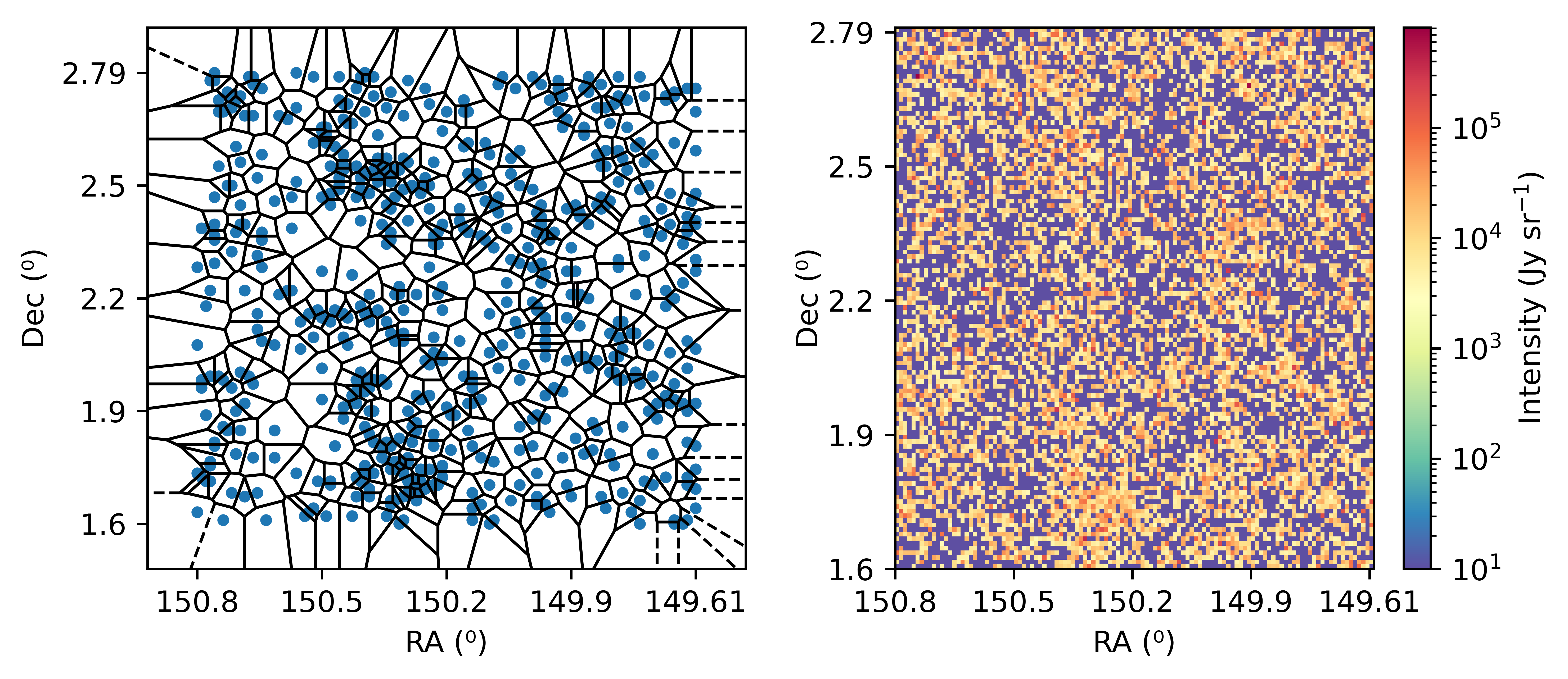}
 \caption{Example VT map when applied to the m4 FARMER LP map at $3.42<z<3.87$. We typically use 3D weighting maps, however we show this example in 2D for clarity. Here this example takes nodes (blue) as the top 3\% brightest pixels.}
 \label{fig:FigVoronoi}
\end{figure*}
The assumed missing dim end of the [CII] luminosity function for models applied to FARMER LP has been a key indicator of incompleteness, as shown in Fig. \ref{fig:FigLFRedshiftComp}. Correspondingly, we attempted to construct more complete samples by extrapolating out from this end of the function. As a way to verify this technique we also did the same with the mass function of FARMER LP. A basic visualisation of this is shown in Fig. \ref{fig:FigBothm2L18LF}: by computing the difference between our extrapolated function and the existing function for faint galaxies within each redshift band, we can estimate the number of `missing' galaxies in FARMER LP and add them to our sample. 

To find the function we used to add galaxies, we assumed that the mass and luminosity functions of galaxy populations are likely to follow a Schechter function \citep{Schechter_1976} as in the following equations:
\begin{align}
\Phi(M)\textrm{d}M&=\left(\frac{\Phi_0}{M_\textrm{c}/M_\odot}\right) \left(\frac{M}{M_\textrm{c}}\right)^\alpha e^{-\frac{M}{M_\textrm{c}}}\textrm{d}M,
\label{eq:19}
\end{align}
\begin{align}
\Phi(L)\textrm{d}L&=\left(\frac{\Phi_0}{L_\textrm{c}/L_\odot}\right) \left(\frac{L}{L_\textrm{c}}\right)^\alpha e^{-\frac{L}{L_\textrm{c}}}\textrm{d}L,
\label{eq:20}
\end{align}
where $\Phi$ is the number of galaxies per unit mass or luminosity per unit dex per unit volume and $L_\textrm{c}$, $M_\textrm{c}$, $\Phi_0$, and $\alpha$ are fit parameters. The number of galaxies in each mass or luminosity band of the function can be found by multiplying $\Phi$ by the dex and the volume covered by the frequency band we used. As before, we always used dex=mass or luminosity interval=0.1, with the comoving volume calculated using our given cosmology - $4\,630\,000, 5\,880\,000, 7\,900\,000, 10\,300\,000$\,Mpc$^3$ for $3.42<z<3.87$, $4.14<z<4.76$, $5.34<z<6.31$, and $6.75<z<8.27$ respectively. After calculating the existing mass or luminosity function for FARMER LP when including the added CANDELS correction, we attempted to fit Eqs. (\ref{eq:19}), (\ref{eq:20}) to the high-mass or  -luminosity points. We recorded the parameters for all of the fits we used in Appendix \ref{appendix:fitparams}.

Once we obtained our expected curves, we generated a number of galaxies equivalent to the difference between the Schechter fit and the actual FARMER LP function for each dex interval. When generating galaxies using a luminosity function, we assigned their bulk properties so that they reproduce the appropriate [CII] luminosity for the given luminosity band. For the mass function case, we assigned galaxy stellar masses according to the mass of the dex band and then used the galaxy main sequence of the given redshift band to calculate SFR, and therefore sSFR and metallicity (Eq. \ref{eq:22}). The main sequence parameters are in Table \ref{table:data3}). In addition to our own fits, we also used the mass function fit parameters found by \cite{Weaver_2022b} for SFGs in COSMOS 2020.
\begin{equation}
    \log_{10}\left(\frac{\textrm{SFR}}{M_\odot\textrm{yr}^{-1}}\right)=a+b\log_{10}\left(\frac{M_\star}{M_\odot}\right).
\label{eq:22}
\end{equation}
\begin{table}[t]
\caption{Galaxy main sequence fit parameters for redshift bands.}
\centering
\begin{tabular}{c c c}
 \hline\hline
 Model & a & b \\ [0.5ex]
 \hline
 $3.42<z<3.87$ & -6.42 & 0.789 \\
 $4.14<z<4.76$ & -5.81 & 0.740  \\
 $5.34<z<6.31$ & -6.79 & 0.834  \\
 $6.75<z<8.27$ & -7.68 & 0.942  \\ [1ex]
 \hline
\end{tabular}
\tablefoot{We found these by applying least-squares fitting between stellar mass and SFR using a standard linear log-log model, similar to Fig. \ref{fig:FigMainSeq}.}
\label{table:data3}
\end{table}
After completing the extrapolation, we took these enhanced samples and ran the same procedures as described in Sects. \ref{sec:methodMAP} and \ref{sec:methodPS}. When using the samples extrapolated using a [CII] luminosity function, we only applied the [CII] model which was used to create the corresponding luminosity function.

However, there are significant limitations with this technique. In many cases we could produce multiple valid Schechter fits for a given function as demonstrated by Figure \ref{fig:FigDW}, thereby adding different numbers of extrapolated galaxies and altering the resulting power spectra. In the `overestimate fit', a vastly unrealistic number of galaxies were extrapolated due to the steep gradient of the fit, in this case half a million galaxies in a $1.2\times1.2 \deg^2$ area per 0.1\,dex band. While we found negligible differences between the power spectra for the other fits, the shot-noise for the overestimate fit was inflated by at least 0.2\,dex. Furthermore, the minimum mass or luminosity point for extrapolation was unknown, as we do not know the properties of the faintest galaxies at high redshift. Nevertheless, we calibrated these end points by reviewing the literature on the smallest dwarf galaxies in the local universe, where we found [CII] luminosities as low as $10^4L_\odot$ and stellar masses as low as $10^6M_\odot$ (e.g. \citealt{Cormier_2015}; \citealt{Madden_2013}). As it is unlikely that the Schechter function accurately predicts galaxy numbers for the very dimmest galaxies, we only extrapolated down to $10^5L_\odot$ or $10^7M_\odot$. In addition we experimented with the end points $10^7L_\odot$ or $10^8M_\odot$, which are the end point of [CII]-metallicity relations (\citealt{DeLooze_2014}; \citealt{Lagache_2017}) and a typical dwarf galaxy mass respectively. We found minimal differences between the power spectra from these different extrapolation depths for most fits in Fig. \ref{fig:FigDW}, so it is likely that the contribution of signal from the smallest galaxies is negligible. The primary exception to this trend is the overestimate fit, where the clustering signal is greatly inflated at $k<10^0\,\textrm{Mpc}^{-1}$ when we extrapolate to $10^5\,L_\odot$. As discussed in Sect. \ref{sec:resultsIPS}, this is due to faint galaxies greatly contributing to the clustering signal. Therefore we only included fits that do not demonstrate this dramatic increase in shot-noise and clustering signal. However there were still multiple valid fits in some cases, and while this is not a problem for the final power spectrum magnitude, we discuss constraining these fits further with additional data in Sect. \ref{sec:discussion}.

\subsection{Voronoi tessellation for galaxy distribution} \label{sec:interpolationVT}

Once we generated the additional galaxies via extrapolation, we determined the new galaxies' map positions and redshifts. Instead of determining these co-ordinates randomly, which was appropriate for adding galaxies in the small stellar mask areas but would destroy any of the existing structure within the intensity cube, we implemented a weighting cube. This is an array that has the same dimensions of the intensity cube and stores relative weights in each voxel, which then calibrates the random selection of $x$, $y$ and $z$ co-ordinates for the extrapolated galaxies. The weights used in the weighting cube are the normalised stellar mass of FARMER LP's galaxies, and are stored within the voxels in the same way that [CII] intensity is stored within the intensity cube. We also added a proportion of a galaxy's weight to voxels on the same redshift slice within two spaces, to prevent over-weighting for specific pixels. The ratio of weights of `central pixel':`adjacent pixel':`pixel two spaces away' is 3:2:1. Furthermore, we added a baseline weight equivalent to adding an average galaxy to each voxel. This method was inspired by and produces similar maps to Voronoi tessellation (VT) (e.g. \citealt{Ramella_2001}; \citealt{Kim_2002}), which determines overdensities and voids by identifying bright nodes on a map and drawing equidistant lines between them. An example of this applied to FARMER LP is shown in Fig. \ref{fig:FigVoronoi}, where we treat high-weight voxels as nodes. In this way the highest mass galaxies of FARMER LP (primarily SFGs) become the bright centres of large galaxy clusters within the assumed dark matter halos. This process assumes the majority of extrapolated dim galaxies lie within these clusters, however does not meaningfully account for other structures such as cosmological filaments. Consequently we must therefore exercise caution when viewing the strength of clustering signal in power spectra from extrapolation, which we discuss in Sect. \ref{sec:discussion}. We also discuss the impact of VT on power spectra compared to randomly distributing galaxies in Appendix \ref{appendix:mapmakingvar}, to ensure that our initial assumptions are somewhat reasonable.

It would be ideal to use a weighting cube based on known overdensities in the COSMOS field, however at time of publication we only have a protocluster density map for $6<z<7.7$ \citep{Brinch_2023}. As this would only apply to the redshift band $6.75<z<8.27$, which is unlikely to give useful results as previously discussed, we did not explore this further.

\subsection{Power spectra and error analysis for extrapolated sample} \label{sec:interpolationanalysis}
\begin{figure*}[t]
 \centering
 \includegraphics[width=\linewidth]{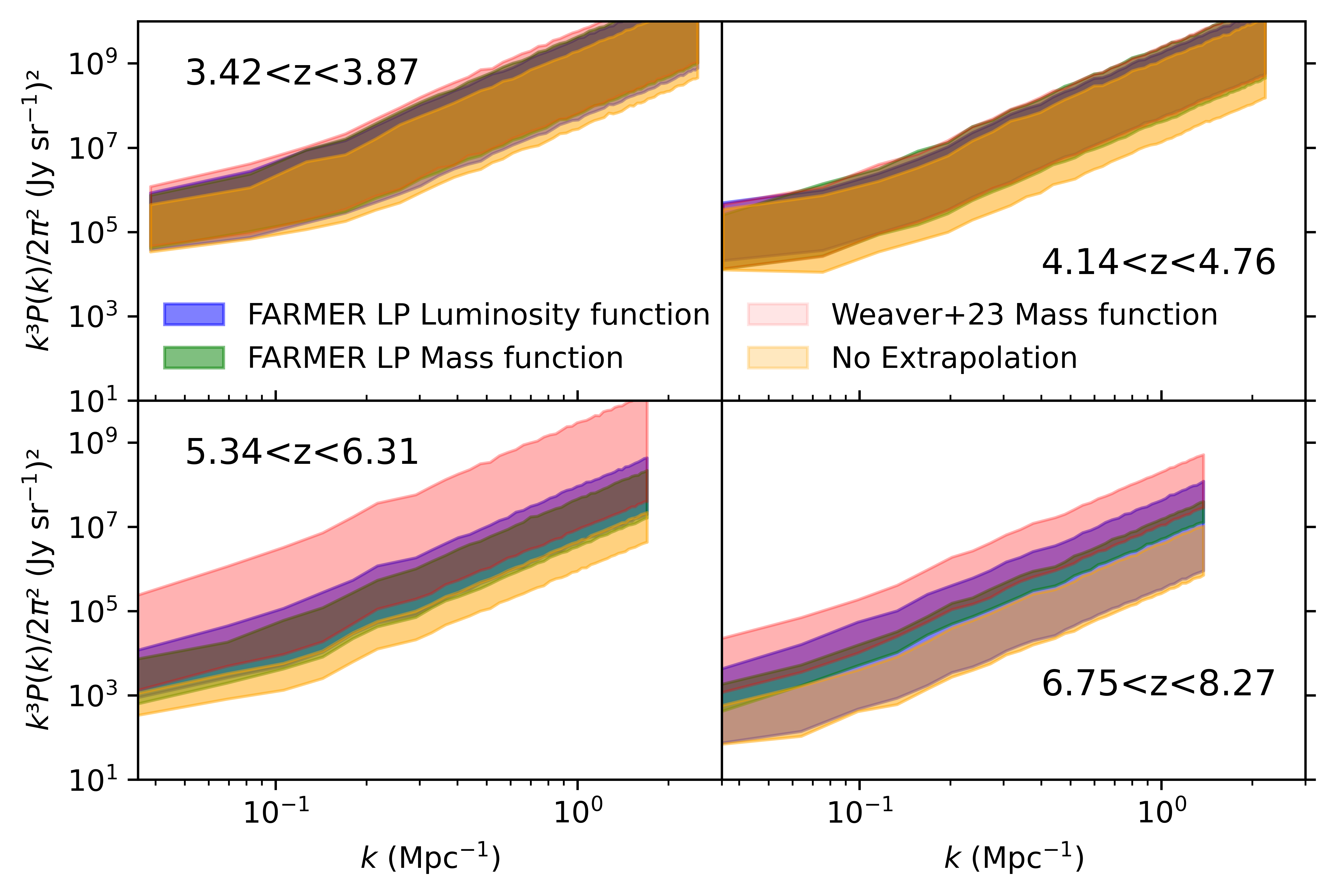}
 \captionof{figure}{Comparison of extrapolated power spectra from different extrapolation techniques. The models in orange are those from the original FARMER LP sample with no extrapolation, with blue representing FARMER LP luminosity function extrapolation, green representing FARMER LP mass function extrapolation, and red representing mass extrapolation using \cite{Weaver_2022b} parameters. The key point here is the difference between extrapolated models and the original sample in orange.}
 \label{fig:FigManyinterpPS}
\end{figure*}

After extrapolating from FARMER LP, we created power spectra from these expansions to the existing samples. When comparing the power spectra from different extrapolation techniques in Fig. \ref{fig:FigManyinterpPS}, we found a consistent increase in magnitude for all methods. This is to be expected as extrapolating the missing low-mass or low-luminosity galaxies produces fundamentally similar outcomes, as low-mass galaxies are likely to be dim. The power spectra magnitude increase is relatively small at low redshift ($0.2-0.4$\,dex at $3.42<z<3.87$) but is more significant at high redshift ($0.5-0.8$\,dex at $5.34<z<6.31$). This is consistent with the idea that FARMER LP is more complete at lower redshift bands, as the survey was able to find a higher proportion of dim galaxies, so extrapolation has more impact for higher-redshift sub-samples with worse completeness. When examining the clustering component we found a small increase due to the large number of dim galaxies added around assumed clusters using VT. For example, in $3.42<z<3.87$ we found that the transition region occurs around $k\approx0.25$ for all extrapolation methods, that is, where the clustering signal is approximately equal to the shot-noise component. However, these transition regions are still at smaller $k$ when compared to the simulation work of \cite{Karoumpis_2022}, a result that is somewhat surprising considering our attempts to mimic larger structure. This may be a consequence of the specific assumptions we made when applying VT because of the lack of frame of reference we have for high-redshift structures, as discussed in Sect. \ref{sec:discussion}. In addition, this could result from previous literature deliberately seeding their galaxies around specific dark matter halos.

\begin{figure*}[t]
 \centering
 \includegraphics[width=\linewidth]{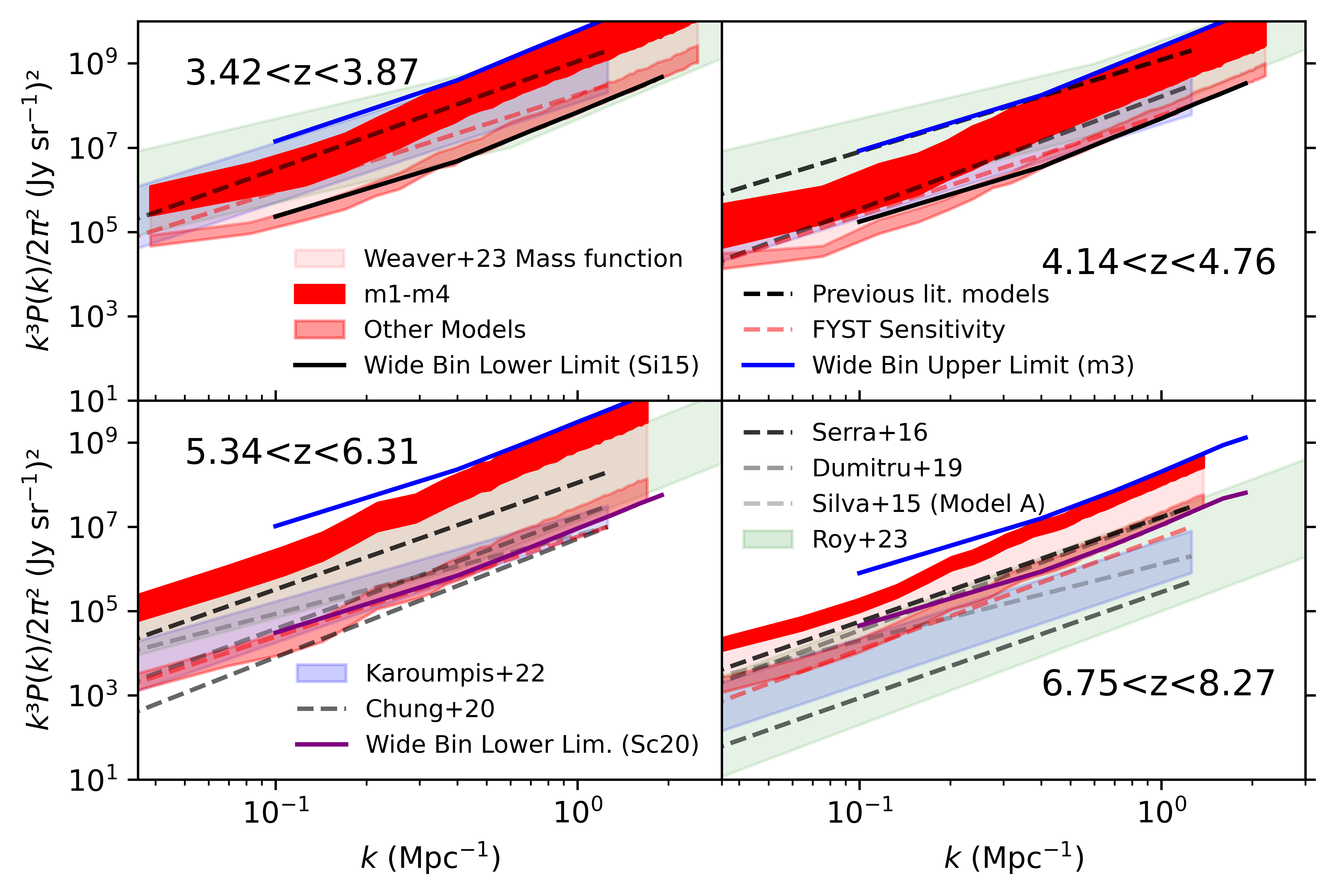}
 \captionof{figure}{As Fig. \ref{fig:FigCompareDefaultPS}, but for the extrapolated data based on the mass function fit with parameters given by \cite{Weaver_2022b}. We group the ALPINE models together and increase shade density for emphasis.}
 \label{fig:FigSpecificinterpPS}
\end{figure*}

To verify that this extrapolation method leads to results consistent with the literature, we must directly compare their power spectra. In our figures such as Fig. \ref{fig:FigSpecificinterpPS} we only used the mass function extrapolation sample from \cite{Weaver_2022b} for clarity, however the conclusions are the same for all extrapolation techniques. Overall we found significant overlap between our models and the previous literature at all redshifts. This is demonstrated by our models managing to reach the shot-noise component of all models from previous literature. We also visualise this by showing the non-absolute upper and lower limits from extrapolation when using wide bins, with results shown in Table \ref{table:compare k interp}. Furthermore, almost all of our models exceeded the expected FYST sensitivity limit, an alternate measure predicting detectability, which implies that if our extrapolated samples were accurate to reality we would be able to recover usable results. This concordance is therefore useful in demonstrating that extrapolating from existing samples (which are useful for absolute lower limits) can be used to replicate predictions where dim galaxies have a significant impact (therefore providing reasonable upper limits). Consequently, existing catalogues can model power spectra from LIM in a variety of test scenarios.

However, there are some caveats to the results from extrapolation. $6.75<z<8.27$ remains an outlier, with power spectra at that redshift being multiple dex above some predictions at the same redshift. This is because COSMOS 2020 only includes high-luminosity galaxies in this redshift range, so the fits and subsequent extrapolation from the existing sample produced a vast number of galaxies and thereby skewed the power spectra magnitude. We see this in the Schechter fit parameters in Table \ref{table:lumfitdata}, where the `knee' of the Schechter fit, $L_\textrm{c}$, is far higher for most models at this redshift. At lower redshifts, only the ALPINE models (m1-m4) reach the magnitudes of the high-magnitude models from the previous literature, with all our other models only overlapping with the lower estimates. This may be a consequence of these models being fitted using high-luminosity data, and thus producing high luminosities even when applied to smaller galaxies (as discussed in Sect. {\ref{sec:resultsIPS}}). Therefore it may be best to view these uppermost limits of our work with a level of caution. Despite these discrepancies, and the differences in clustering signal, it is reassuring to see this overall agreement between the sample extrapolated from FARMER LP and the previous literature. In this way, our extrapolation methods are a viable way to create extensions from existing samples.
\begingroup
\setlength{\tabcolsep}{3pt}
\begin{table}[t]
\caption{Soft lower and upper limits of FARMER LP power spectra from an extrapolated sample, using $\Delta k=0.3$\,Mpc$^{-1}$.}
\centering
\begin{tabular}{c c c c c}
\hline\hline
\, &\multicolumn{4}{c}{$k^3P(k)/2\pi^2$ ((Jy\,sr$^{-1}$)$^2$)}\\
 $k$\,(Mpc$^{-1}$) & 0.25-0.55 & 0.55-0.85& 0.85-1.15& 1.15-1.45\\
\hline
 \, &\multicolumn{4}{c}{Lower limits from extrapolation}\\
\hline
$3.42<z<3.87$ &4.79$\times10^6$& 2.49$\times10^7$& 6.93$\times10^7$& 1.52$\times10^8$\\
$4.14<z<4.76$ &3.48$\times10^6$& 1.81$\times10^7$& 4.91$\times10^7$& 1.11$\times10^8$\\
$5.34<z<6.31$ &6.98$\times10^5$& 3.29$\times10^6$& 9.19$\times10^6$& 1.91$\times10^7$\\
$6.75<z<8.27$ &8.68$\times10^5$& 3.87$\times10^6$& 1.11$\times10^7$& 2.50$\times10^7$\\
\hline
 \, &\multicolumn{4}{c}{Upper limits from extrapolation}\\
\hline
$3.42<z<3.87$ &3.98$\times10^8$& 2.13$\times10^9$& 6.02$\times10^9$& 1.31$\times10^{10}$\\
$4.14<z<4.76$ &1.75$\times10^8$& 8.92$\times10^8$& 2.51$\times10^9$& 5.38$\times10^9$\\
$5.34<z<6.31$ &2.30$\times10^8$& 1.10$\times10^9$& 3.08$\times10^9$& 6.47$\times10^9$\\
$6.75<z<8.27$ &1.59$\times10^7$& 7.15$\times10^7$& 2.04$\times10^8$& 4.58$\times10^8$\\
\hline
\end{tabular}
\tablefoot{These limits are for the power spectra from the \cite{Weaver_2022b} mass function extrapolation. The lower limits are Si15 for $3.42<z<3.87$ and $4.14<z<4.76$ and Sc20 for other bands, whilst the upper limits are m3 for all bands (Fig. \ref{fig:FigSpecificinterpPS}).}
\label{table:compare k interp}
\end{table}
\endgroup

\begin{figure*}[t]
 \centering
 \includegraphics[width=\linewidth]{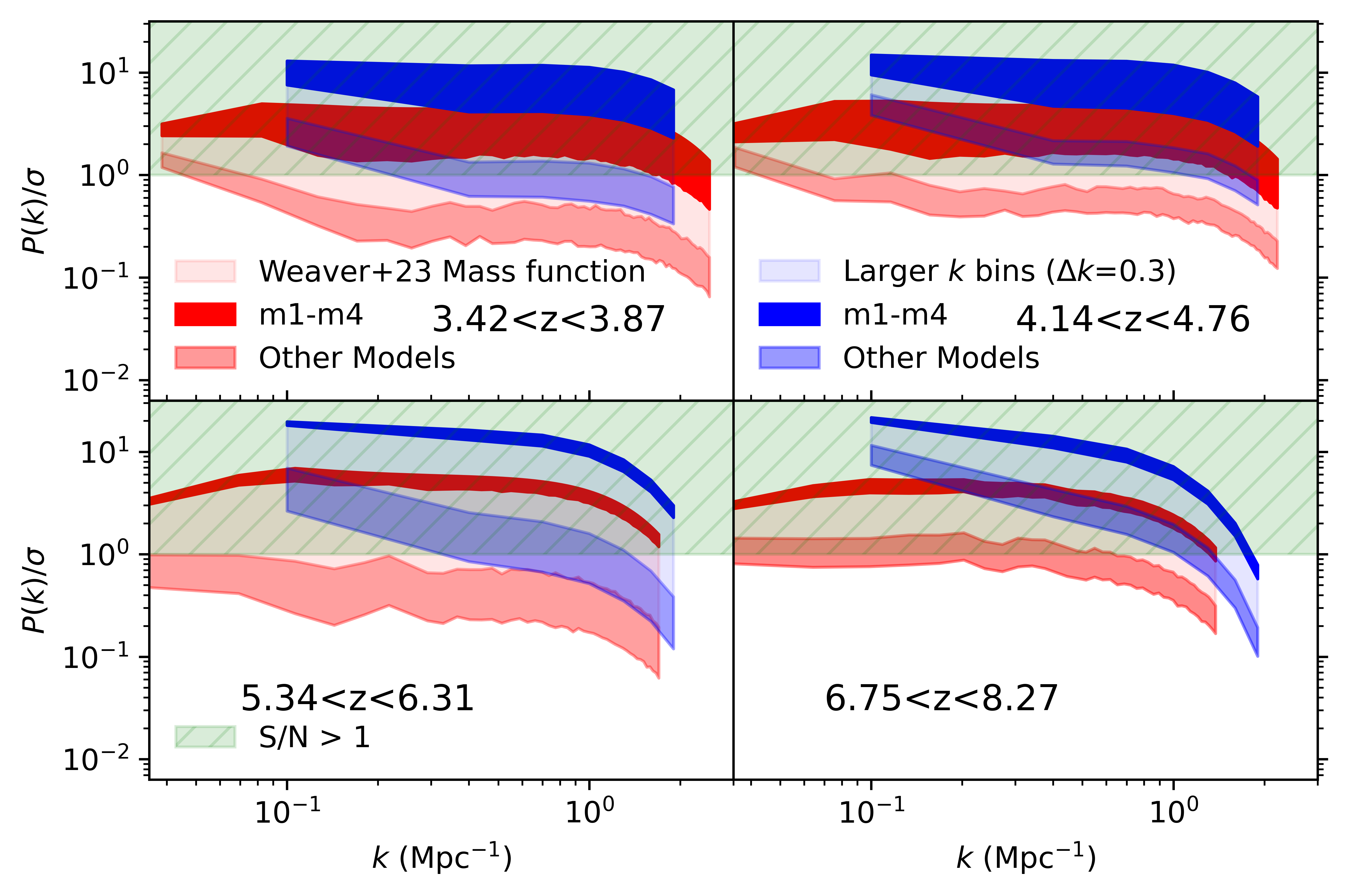}
 \captionof{figure}{As Fig. \ref{fig:FigDefaultPkDivSigma}, but for the extrapolated data based on the mass function fit with parameters given by \cite{Weaver_2022b}. The specific shaded region is for the ALPINE models m1-m4.}
 \label{fig:FigMassInterpPkDivSigma}
\end{figure*}

We also applied the same error analysis methods for the extrapolated samples, focusing on the mass function extrapolation using \cite{Weaver_2022b} to maintain consistency, resulting in Fig. \ref{fig:FigMassInterpPkDivSigma}. This figure shows that all models have greater S/N in comparison to the original versions in Fig. \ref{fig:FigDefaultPkDivSigma}. This is especially clear when using wider bins as all models give S/N$>$1 for all redshift bands. Therefore, in the scenario where these samples accurately reflect reality, observations should be possible when using wider bins for all redshift ranges as has been predicted by prior work. When using narrow $k$ bins, the ALPINE models consistently have S/N$>$1, however the other models do not breach this barrier even at the lowest redshift. We can alternatively view this as models working for no redshift bands (previous literature models) or all redshift bands (ALPINE models). This clear divide is somewhat surprising, as we would expect to recover usable results for all models at the lower redshifts, and other works such as \cite{CCAT_Prime_Collaboration_2022} expected no models to work for $6.75<z<8.27$. Regardless, in order to guarantee detections observers must use wider $k$ bins, which reduces the resolution of our results. However, even with this drawback, it is still likely that we will be able to recover usable results within our initial observation period, assuming these more complete extrapolated samples reflect reality.
\section{Discussion} \label{sec:discussion}
In this study, we demonstrate the feasibility of generating meaningful constraints on line intensity mapping (LIM) power spectra for a specific field, employing empirical data from a known sample and [CII] luminosity models based on galaxy bulk property data. In addition, it is also possible to extend this sample via plausible extrapolation methods to create reasonable limits for these power spectra, which are consistent with the power spectra of previous simulated work. This approach shows promise for making predictions in other fields, provided that we can use samples with bulk properties across sufficiently extensive and deep volumes. Moreover, we have identified several opportunities to refine our methodology, using forthcoming data and preliminary findings from the EoR-Spec DSS. These potential improvements are explored in detail below.

When solely using empirical data, the lower limits on power spectra we established in Table \ref{table:k vals} are primarily beneficial for forecasting observations with EoR-Spec. However, these can also be related to previously conducted simulations (e.g., Fig. \ref{fig:FigCompareDefaultPS}) and our analysis confirms that all previously reported power spectra exceed our absolute lower limits. Notably, some of these power spectra fall beneath our m3 and m4 models within the redshift intervals $3.42<z<3.87$ and $4.14<z<4.76$, which we attribute to inherent biases in these models discussed later.

However, the expected observations' S/N under EoR-Spec's parameters are below 1 across most $k$ (Fig. \ref{fig:FigDefaultPkDivSigma}), even with wide binning for the majority of models. Furthermore, extensive masking beyond $z>6.3$ obliterated any features of large-scale structures which existed within the COSMOS field, leaving only the most luminous star-forming galaxies (SFGs) in COSMOS 2020. As a result, for the redshift range $6.75<z<8.27$, we can only draw conclusions from the small-scale regions. Even still, the observed magnitudes may be influenced by a preferential bias towards brighter galaxies. Consequently, we find that employing this methodology in fields with significant voids is impractical.

The absolute minimum constraints we derived serve as conservative limits for potential outcomes of future observed LIM cube power spectra, and therefore are valuable for observational planning. However, to address the limitations presented by these conservative estimates, we adopted an extrapolation approach to mitigate the incompleteness in our data sets and to produce more comprehensive samples. While our specific procedure relates to the existing COSMOS 2020 catalogue, for example CANDELS lies within the COSMOS field, this methodology should be adaptable to other data sets and observational fields. Encouragingly, the results shown in Fig. \ref{fig:FigSpecificinterpPS} demonstrate that our models' shot-noise components align with the previous work, suggesting the number of galaxies extrapolated from the FARMER LP is appropriate. Despite this, only the ALPINE models m1-m4 reach the uppermost predicted limits, which suggests that these models and similar prior analyses may be overly optimistic.

Due to the extrapolation adopted here and the corresponding substantial assumptions used to extend beyond purely empirical data, the upper limits presented in Table \ref{table:compare k interp} are less stringent than our absolute lower bounds. Consequently, the shot-noise component in extrapolated power spectra is a valuable but not infallible indicator for forecasting observational results and comparing to previous work. The clustering signal is weaker relative to prior simulations, likely due to discrepancies between our clustering assumptions for VT in Sect. \ref{sec:interpolationVT} and those employed in prior work. However, because of the uncertainty surrounding the accuracy of these prior works' clustering assumptions, this is a challenging obstacle based on incomplete information. Future works will focus on refining this aspect, considering alternative weighting methodologies for VT and closely comparing our clustering assumptions with those from previous studies. Additionally, the incorporation of overdensity maps similar to those proposed by \cite{Brinch_2023} could enhance the accuracy of our predictions, using data from forthcoming analyses or initial EoR-Spec observational results to offer a more robust framework for understanding galaxy clustering within LIM.

Analysing the S/N of our extrapolated models reveals that utilising wider $k$ bins consistently achieves S/N$>$1 across all models and redshift intervals, suggesting that detecting a signal in actual observational scenarios is highly probable. However, for narrower $k$ bins, only the ALPINE-based models (m1-m4) maintain an acceptable S/N across various redshifts. In contrast, other models exhibit poor S/N at all redshifts, even with our assumption of 100\% observational efficiency. Aside from implying that the lower limits of power spectra may be challenging to distinguish when using EoR-Spec, the acceptable models are all based on ALPINE data which significantly deviate from the standard [CII]-SFR models. The power spectra of these models show a more pronounced decrease in magnitude with redshift compared to others, potentially due to them being derived from galaxies with high [CII] luminosity. This carries an underlying assumption that all galaxies exhibit substantial [CII] emission, leading to a rapid decline in signal as the galaxy count decreases. These ALPINE galaxies are also not representative of the redshift ranges $3.42<z<3.87$ and $6.75<z<8.27$, rendering the highest limit estimations for these intervals (based on model m3) questionable. While we believe that incorporating a variety of [CII] emission models is crucial, with models m3 and m4 exploring the conceptual space where there is a smaller difference in emission between [CII]-dim and [CII]-bright galaxies, their discrepancy with the existing literature is clear. Even so, models m1 and m2 still approximately align with the results from the literature's [CII]-SFR models for most redshifts, with m2 even having a similar linear log-log relationship to the literature models (albeit with stellar mass). In future work we plan to continue utilising these models while also investigating strategies to address the limitations presented by ALPINE data, such as its restricted redshift coverage, bias towards star-forming galaxies (SFGs), and UV-selection criteria. Additionally, we aim to incorporate [CII] data from higher redshifts, such as from REBELS \citep{Bouwens_2022} upon its full public release, to enhance our models' comprehensiveness and accuracy.

An inherent challenge with the extrapolation process lies in the uncertainty in determining the fit parameters for Schechter functions. While it is feasible to fit a Schechter function to the high-luminosity upper knee section after applying CANDELS extrapolation, the scarcity of data at lower masses and luminosities introduces ambiguity in identifying appropriate fits. Our approach was to select fits that minimised errors according to least-squares fitting procedures and with gradients that were in alignment with previous studies, such as those reported by \cite{Weaver_2022b}. This was to prevent extrapolating a physically impossible number of galaxies, such as cases which would add millions of galaxies per 0.1\,dex interval. In most scenarios the differences in the resulting power spectra are minimal, as illustrated in Fig. \ref{fig:FigDW}, however it would be ideal to constrain these fits using data sets which offer more comprehensive coverage of the mass and luminosity functions. Upcoming surveys such as COSMOS-Web (\citealt{Casey_2023}; \citealt{Franco_2023}; \citealt{Silverman_2023}) and other deep pencil field studies are expected to provide insights into the fainter ends of these functions, thereby serving as valuable benchmarks for calibrating Schechter curve fits. In this way we expect to enhance the reliability and accuracy of extrapolations used in future analyses.

Our foundational methodology is sound, however we must integrate additional variables to increase precision in forecasting upcoming observations, which we omitted in this work to maintain consistency with the previous literature. Line intensity mapping (LIM) inherently measures aggregate signals and thus includes interlopers within the cubes, such as CO and [OIII] line emission from galaxies at different redshifts. These should subsequently be isolated and removed using foreground signal elimination techniques, such as cross-referencing with catalogues of galaxies identified at higher resolution. It is one of EoR-Spec DSS's objectives to cross-correlate LIM data from various lines, as highlighted by \cite{Eli_Visbal_2010} and \cite{Chung_2022}. We must also account for various complex noise sources, including atmospheric noise, the influence of the instrument's scanning pattern, and the masking of specific areas. Ongoing research endeavours, such as those by Karoumpis et al. (in prep.) addressing CO foregrounds, Dev et al. (in prep.) applying frameworks such as Time Ordered Astrophysics Scalable Tools (TOAST) for atmospheric noise and scanning strategies, and \cite{Roy_2023b} examining cross-correlation's impact on S/N, are actively addressing these challenges. Each issue presents its unique set of complications, which makes gauging errors increasingly complex — for example, CO interlopers pose significant challenges at lower frequency, whereas atmospheric noise predominantly affects high-frequency channels. Fortunately, the early FYST science data will serve as a valuable resource for calibrating our error estimations. Employing techniques such as cross-correlation and atmospheric de-striping will likely alleviate major error sources. Furthermore, as there will be varying levels of sensitivity over the 2000 hours observation period of EoR-Spec \citep{CCAT_Prime_Collaboration_2022}, our methodology enables the generation of distinct predictions for various observation phases, enhancing the applicability of our forecasts.

Our investigation is tailored to the regions observed by EoR-Spec DSS, thereby restricting its scope as we require samples covering a sufficiently large sky area ($\sim$1\,$\deg^2$) with comprehensive bulk property data within the E-COSMOS or E-CDFS regions. Currently, our analysis is confined to using COSMOS 2020, as the CANDELS/GOODS-S data set \citep{Hsu_2014} lacks the necessary scale to accurately assess large-scale clustering signal. However, upcoming data from the Euclid mission \citep{Euclid_2022} will extend over 20\,$\deg^2$ in the GOODS-S field at appropriate redshifts, thereby allowing for the generation of meaningful predictions for this field. In the interim, we can also use our method to project shot-noise estimates for experiments covering smaller areas, such as TIME and CONCERTO. Additionally, we can also examine the relative impact of individual large galaxy clusters on the power spectra, using both observed and simulated clusters for this analysis. The capability of LIM to survey wider fields makes these clusters, for instance the CRLE and AZTEC-3 protoclusters (\citealt{Riechers_2010}; \citealt{Capak_2011}; \citealt{Pavesi_2018}; \citealt{Vieira_2022}), a key focus for subsequent simulations and future research. This could be achieved by incorporating simulated clusters into our maps to determine their relative impact on the power spectra. Finally, we also plan to delve into additional analytical techniques that can be applied to Fourier-transformed cubes, such as delta-variance and bi-spectra statistics. These approaches hold the potential to reveal further insights, enriching our understanding of the underlying cosmic structures.
\section{Conclusion} \label{sec:conclusion}
In this work, we introduce an empirical methodology for creating mock [CII] line intensity mapping (LIM) cubes using an existing sample, in this case: the FARMER LP sub-sample of the COSMOS 2020 galaxy catalogue. This approach enables us to make power spectra predictions for the E-COSMOS field, which will be surveyed by the EoR-Spec Deep Spectroscopic Survey (DSS) during its 2000-hour campaign commencing in 2026. These LIM cubes are tailored to the observational specifications of the EoR-Spec instrument aboard the Fred Young Submillimeter Telescope (FYST). Given the sample's incompleteness towards faint galaxies, the resulting power spectra establish lower bounds, which we quantified for $k=1\,\textrm{Mpc}^{-1}$ as $P(k)/2\pi^2=3.06\times10^7,\,1.43\times10^7,\,9.80\times10^5,\,2.77\times10^5\,$(Jy sr$^{-1})^2$ across the redshift intervals of $3.42<z<3.87$, $4.14<z<4.76$, $5.34<z<6.31$, $6.75<z<8.27$, respectively (Table \ref{table:k vals}). Furthermore, we developed techniques to extrapolate from this data set, producing results similar to previous simulations.

These simulated LIM cubes incorporate [CII] luminosity models from the previous literature as well as those derived from ALPINE data, including models that deviate from linear log-log [CII]-SFR relationships. The models were applied across four aforementioned redshift bands to create a map spanning $1.2\times1.2 \deg^2$. The key findings from our work are summarised as follows:

\begin{itemize}
\item We created four [CII] models solely using ALPINE galaxy bulk property data, which diverged from the conventional linear log-log [CII]-SFR relationship. These models produce power spectra with a range of $\sim$1\,dex, and are consistent with existing [CII] models when applied to the FARMER LP sub-sample. However, these models are less constrained at redshift ranges beyond ALPINE's redshift coverage ($3.42<z<3.87$ and $6.75<z<8.27$).
\item Our approach is capable of generating mock LIM cubes using available galaxy catalogues, such as COSMOS 2020. Given the incompleteness of these samples at the faint end of the luminosity function, the resultant power spectra serve as lower bounds, reflecting only the galaxies that have been detected. All previous power spectra from the literature exceed these minimum values for z$<$6.3. Our results are less conclusive above this point, due to the greater level of incompleteness of COSMOS 2020 at higher redshifts.
\item We evaluated the impact of incompleteness from the variance in power spectra between the original sample and prior predictions, and extrapolated from the empirical data set in FARMER LP by considering truncated mass and luminosity functions along with the detection rate ratio between the CANDELS data and COSMOS 2020. This extension from the original sample increases the magnitude of our empirically guided mock power spectra by approximately 0.5\,dex, placing them within the range of earlier simulations. Aside from providing this concordance, the extrapolation also offers a tool to estimate the signal strength for forthcoming observations. From this, we can identify potential power spectra ranges allowed by previous predictions that are only reached by specific ALPINE models, which are therefore less likely to reflect the signal strength that is to be observed.
\item In our power spectra from the extrapolations, we predicted a result of S/N$>$1 only when using wide spatial frequency bins ($\Delta k=0.3$ Mpc$^{-1}$). However, this assumes 100\% signal recovery efficiency, so future investigations will necessitate a more detailed consideration of foreground line signals and atmospheric noise.
\end{itemize}

This work provides empirical support for the previous simulation work. While the anticipated challenges in observations will be significant at higher redshift bands, we remain confident that the planned FYST/EoR-Spec surveys will be able to obtain meaningful constraints for $z<6$. Future efforts will extend this work to include realistic foregrounds and to obtain more solid constraints at $z>6,$ based on incoming JWST and Euclid data in the COSMOS field.
\begin{acknowledgements}
We wish to thank members of the CCAT Collaboration (including Dongwoo Chung and Patrick Breyesse) as well as Daniel Vieira, Cristiano Porciani, and Peter Schilke for their useful input in discussions. We also wish to thank John Weaver for his input regarding the COSMOS 2020 sample, including mass completeness calculations.
We wish to acknowledge the contributions of the individual components of the COSMOS 2020 survey (\href{https://cosmos.astro.caltech.edu/}{https://cosmos.astro.caltech.edu/}). In specific, this work is
based on observations collected at the European Southern Observatory under ESO programme ID 179.A-2005 and on data products produced by CALET and the Cambridge Astronomy Survey Unit on behalf of the UltraVISTA consortium. It is also based on data products from observations made with ESO Telescopes at the La Silla Paranal Observatory under ESO programme ID 179.A-2005 and on data products produced by CALET and the Cambridge Astronomy Survey Unit on behalf of the UltraVISTA consortium.
The CCAT project, FYST and Prime-Cam instrument have been supported by generous contributions from the Fred M. Young, Jr. Charitable Trust, Cornell University, and the Canada Foundation for Innovation and the Provinces of Ontario, Alberta, and British Columbia. The construction of the FYST telescope was supported by the Gro{\ss}ger{\"a}te-Programm of the German Science Foundation (Deutsche Forschungsgemeinschaft, DFG) under grant INST 216/733-1 FUGG, as well as funding from Universit{\"a}t zu K{\"o}ln, Universit{\"a}t Bonn and the Max Planck Institut f{\"u}r Astrophysik, Garching. The construction of EoR-Spec is supported by NSF grant AST-2009767. The construction of the 350\,GHz instrument module for Prime-Cam is supported by NSF grant AST-2117631.
We gratefully acknowledge the support provided by the Collaborative Research Center 1601 (SFB 1601 sub-projects C3, C6) funded by the Deutsche Forschungsgemeinschaft (DFG, German Research Foundation) – 500700252. In addition, Thomas Nikola gratefully acknowledges support from grants NSF AAG-1910107 and NSF ATI-2009767.
We made extensive use of the Python packages astroPy \citep{Astropy_Collaboration_2013,Astropy_Collaboration_2018,Astropy_Collaboration_2022}, numPy \citep{harris2020array}, scipy \citep{Virtanen_2020}, and matplotlib \citep{Hunter_2007}.
Many thanks to the article by Dr. David Nichols on use of colours for the colourblind: \href{https://davidmathlogic.com/colorblind/}{https://davidmathlogic.com/colorblind/}. We also give thanks to the work of Chentao Yang for correct LaTex formatting: \href{https://github.com/yangcht/AA-bibstyle-with-hyperlink}{https://github.com/yangcht/AA-bibstyle-with-hyperlink}. In addition, we made extensive use of NASA's Astrophysics Data System Bibliographic Services throughout this work.

\end{acknowledgements}

\bibliographystyle{aa}
\bibliography{sample631}

\appendix
\section{ALPINE model checks}\label{appendix:sanitycheck}
\begin{figure}[t]
 \centering
 \includegraphics[width=\linewidth]{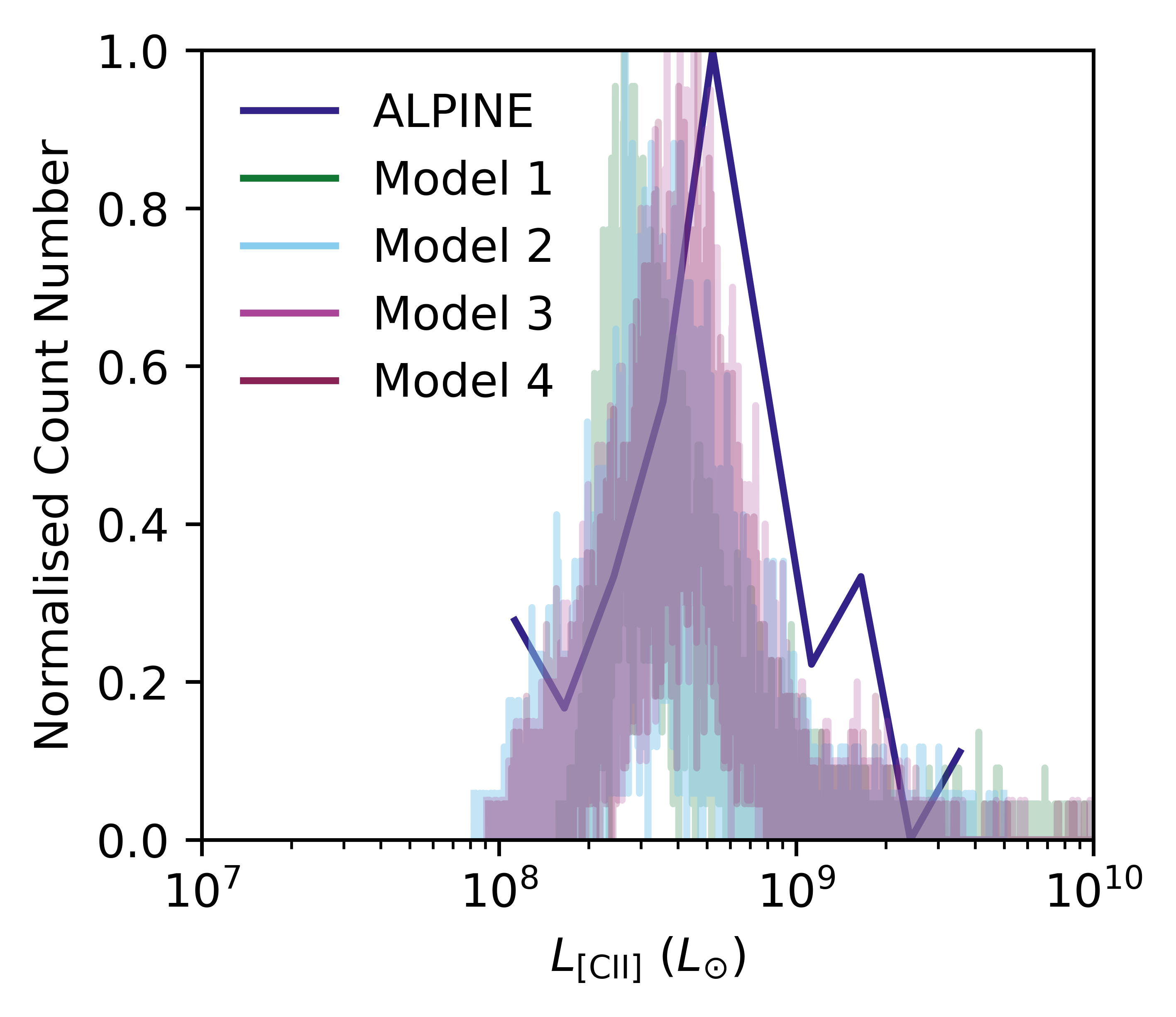}
 \captionof{figure}{Comparison of the histogram of the ALPINE [CII] luminosities compared to m1-m4 histograms when applied to FARMER LP at $4.4<z<4.65$ and $5.05<z<5.9$. As these resemble Gaussian distributions, we made fits to these histograms and compared the means and standard deviations.}
 \label{fig:FigModelHistograms}
\end{figure}

When creating the ALPINE [CII] models, we attempted to fit Eq. (\ref{eq:5}) to all combinations of bulk properties. We performed this fitting for the individual galaxies in the sample as well as binned galaxies using least-squares fitting procedures. When binning we used 5 bins with 13 galaxies apiece, with the bins sorted by [CII] luminosity, stellar mass, SFR, or metallicity. This produced a vast number of models, most of which were poor fits, so we applied multiple checks:

\begin{itemize}
  \item We calculated the reduced chi-squared statistic ($\chi^2_\nu$) for each model as described in Eq. (\ref{eq:7}), where $\nu$ is the number of degrees of freedom, $L_{[\textrm{CII}], i}(X)$ is the [CII] luminosity of each data point with bulk properties $X$, $L_{\textrm{[CII], model}}(X)$ is the [CII] luminosity given by the model when substituting in the same bulk properties $X$, and $\sigma_i^2$ is the error in [CII] luminosity for each data point. When calculating this for a binned fit, we performed error propagation to find $\sigma_i^2$, and we took $X$ as the averaged bulk properties of each galaxy within the bin. We took the number of degrees of freedom $\nu$ as the number of fit parameters subtracted from the number of data points. A model was viewed as acceptable when its $\chi^2_\nu$ is within $4\,\sigma$ of the expected value for a good model, 1. This is when we are within the limits set by Eq. (\ref{eq:8}), where we set the number of $\sigma$ $n=4$ and $\nu$ depends on the model. This has a large amount of leniency, however we use this limit to allow as many models as possible. For example, m1 has 4 fit parameters, and was calculated from binned data (5 data points), so $\nu=5-4=1$. As $n=4$, the limit $\chi^2_{\nu, \textrm{ n}\sigma\textrm{ limit}}=6.65$. This statistical concept is described in detail in the literature (for instance \citealt{Hughes}, pg. 102-107). 
\end{itemize}
\begin{equation}
\chi_{\nu}^2=\frac{1}{\nu}\sum_{i} \left[ \frac{L_{\textrm{[CII], model}}(X)-L_{[\textrm{CII}], i}(X)}{\sigma_i} \right]^2,
\label{eq:7}
\end{equation}
\begin{equation}
    \chi_{\nu, \textrm{ n}\sigma\textrm{ limit}}^2=1+n\sqrt{\frac{2}{\nu}}.
\label{eq:8}
\end{equation}
\begin{itemize}
  \item For each model, we checked if it diverged significantly from the expected range of [CII] luminosities when using outlier values, that is above $10^{13} L_{\odot}$  or below $10^4L_{\odot}$. We inserted the maximum and minimum values of galaxy bulk properties within FARMER LP ($3.42<z<8.27$) as shown in Table \ref{table:data2}. If a model failed this test it was rejected.
\end{itemize}
\begin{table}[ht!]
\caption{Maximum and minimum bulk properties within the FARMER LP sample.}
\centering
\begin{tabular}{c c c}
 \hline\hline
 Value & Max & Min \\ [0.5ex]
 \hline
 $\log_{10}\left(\frac{M_\star}{M_\odot}\right)$ & 6.600 & 12.002 \\
 $\log_{10}\left(\frac{\textrm{SFR}}{M_\odot \textrm{yr}^{-1}}\right)$ & -4.705 & 3.807  \\
 $\log_{10}\left(\frac{\textrm{sSFR}}{\textrm{yr}^{-1}}\right)$ & -14.756 & -7.337  \\
 $\frac{Z}{12+\log\textrm{(O/H)}}$ & 5.822 & 9.083  \\
 $\log_{10}\left(\frac{Z}{12+\log\textrm{(O/H)}}\right)$ & 0.765 & 0.958  \\ [1ex]
 \hline
\end{tabular}
\label{table:data2}
\end{table}
\begin{itemize}
  \item It was important to determine if each model was consistent with the known ALPINE [CII] luminosities when applied to the COSMOS 2020 data. We applied the test models to FARMER LP data in the same redshift range as ALPINE ($4.4<z<4.65$ and $5.05<z<5.9$), calculated the histograms of the resulting [CII] luminosities, and fitted a Gaussian distribution to the histograms, which we compared to the equivalent distribution of the ALPINE [CII] data. If these distributions had similar means and standard deviations ($<20$\% difference), we accepted the model. This is demonstrated for our successful models in Fig. \ref{fig:FigModelHistograms}.
\end{itemize}
\begin{itemize}
  \item Finally, we checked that each model made basic physical sense. For example, we had a seemingly acceptable test model that was a quadratic in stellar mass - that is, a model that produced high [CII] emission with very high OR very low stellar mass. As this is clearly non-physical when considering the [CII] emission of dwarf galaxies, we excluded this model.
\end{itemize}
\section{Assumptions made for constructing intensity cubes and power spectra}\label{appendix:mapmakingvar}
In our methodology, we made many assumptions about how we should construct our intensity cubes, which we discuss and justify in this section. 

\begin{figure}[t]
 \centering
 \includegraphics[width=\linewidth]{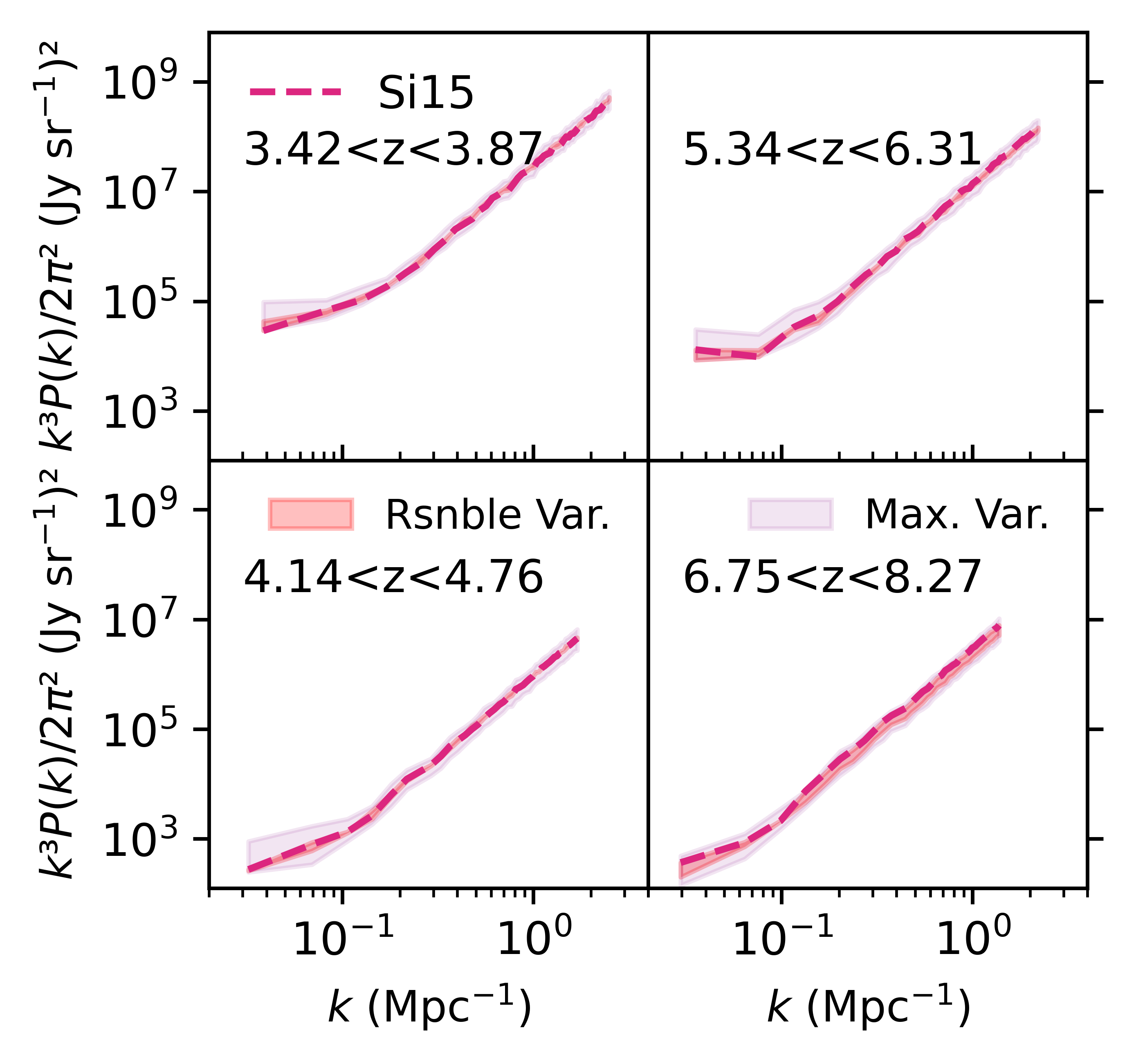}
 \captionof{figure}{Comparison of the power spectra of Si15 with samples using randomised redshifts. Reasonable variation fits randomise the sample galaxies' redshifts within one standard deviation, with maximum variation fits including as many or as few galaxies as is plausible. In most cases, the reasonable variation power spectra are nearly indistinguishable from the default power spectra. We do not include other [CII] models for clarity.}
 \label{fig:FigRedshiftVar}
\end{figure}

\begin{figure}[t]
 \centering
 \includegraphics[width=\linewidth]{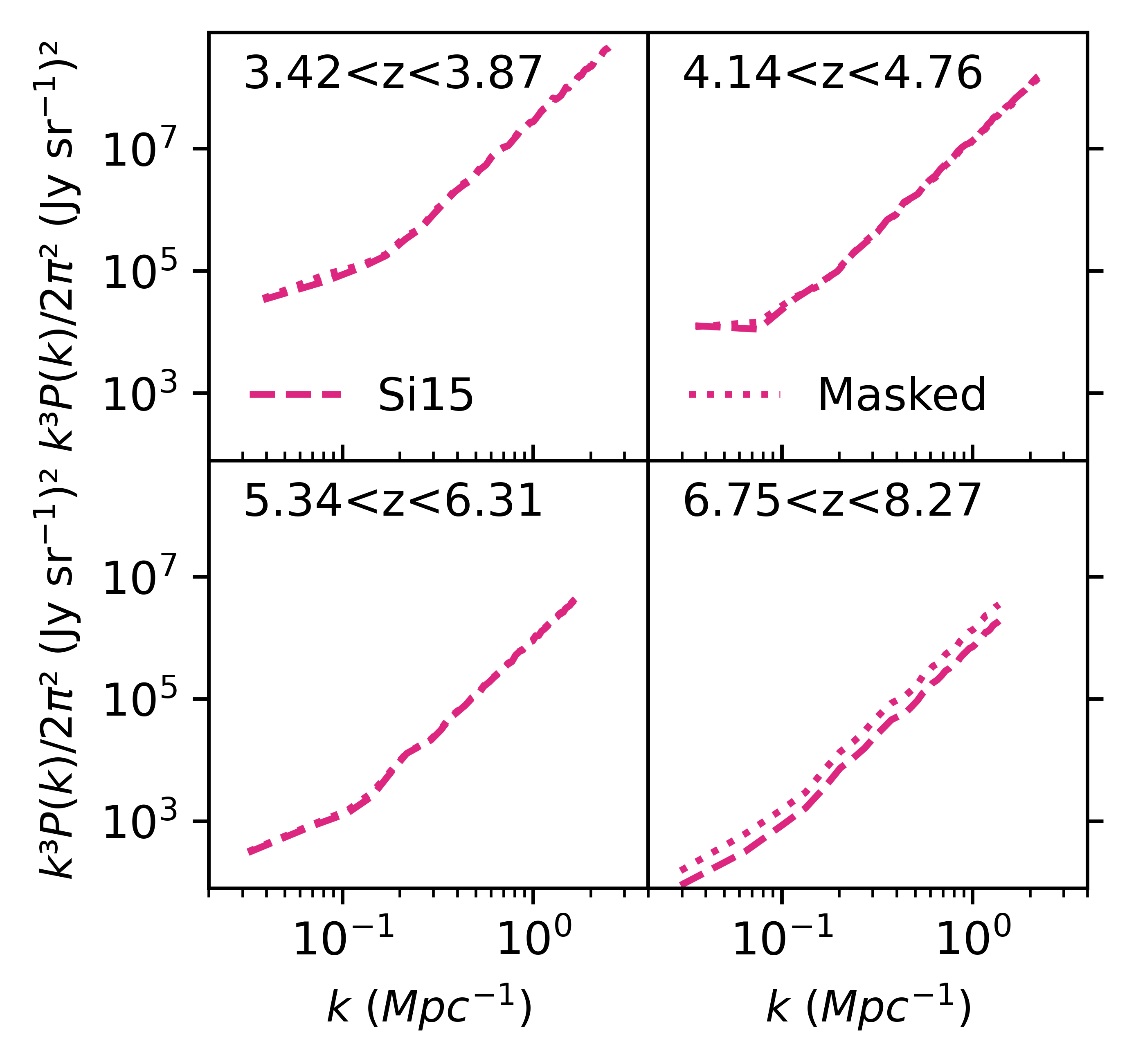}
 \captionof{figure}{Comparison of the power spectra of Si15 at all redshifts with versions where we did not perform extrapolation to add galaxies in the stellar mask. We do not show other models for clarity. As shown, the change in magnitude is negligible for clustering scales (except for $6.75<z<8.27$).}
 \label{fig:FigNoMask}
\end{figure}

A key assumption in using redshift data from FARMER LP to select which galaxies lie within each band is that the error in photometric redshift does not meaningfully impact the power spectrum. For a non-negligible proportion of galaxies, their photometric redshift errors are greater than the redshift range covered by individual frequency slices. Therefore the galaxy could actually `belong' to a different slice of the 3D intensity cube, or a galaxy may be erroneously included or excluded from the cube, thereby impacting the power spectrum. To check that the power spectra magnitude is unaffected we compared the power spectra of the default FARMER LP (when using Si15), to versions where each galaxy within the sample had its redshift increased by $\sigma_z$, redshift decreased by $\sigma_z$, or changed by a random fraction of $\pm \sigma_z$. Here $\sigma_z$ is the error calculated from the 68\% confidence limit in photometric redshift. We also used more extreme samples that tried to include or exclude as many galaxies as possible in a given redshift band (within 1\,$\sigma_z$). As shown in Fig. \ref{fig:FigRedshiftVar}, there is negligible deviation in shot-noise ($<0.1$\,dex) for the reasonable variations, with slightly greater deviation in the clustering signal. There are greater changes when using the extreme examples, with shot-noise deviation up to 0.25\,dex and clustering signal deviation up to 0.5\,dex. While these changes are more perceptible, the samples must be contorted in ways which are extremely unlikely to occur in reality. Consequently we do not believe that deviations caused by photometric redshift errors could impact our results meaningfully.
We also assumed that extrapolation within the masked regions of COSMOS 2020 was appropriate, and that by averaging 10 of these power spectra we do not introduce false structure. We verified this by comparing the power spectra of intensity cubes with and without mask extrapolation over all redshift bands using Si15 (Fig. \ref{fig:FigNoMask}). There is negligible difference between the power spectra for small $k$ because the masked areas are too small ($<1$\,arcsec$^2$) to meaningfully contribute to large-scale structure. When observing shot-noise, there is a small decrease in shot-noise magnitude when we fill the mask ($<0.1$\,dex) as the power spectra calculations when masking necessitated shrinking the volume. A decrease in volume of 10\%, corresponds to a 0.1\,dex shot-noise increase, but as the average intensity per voxel is increased by filling in the mask we instead see a much smaller increase. The only exception for this trend is $6.75<z<8.27$, as the mask covers a much wider area (50\%) and therefore we fill in far more signal. Overall, we demonstrated that leaving the stellar mask empty does not change any structural results significantly for $z<6.3$, but marginally overestimates the overall shot-noise of the power spectrum - hence, our assumption is valid. In any case, the calculated power spectra error is much larger than this discrepancy.
\begin{figure}[t]
 \centering
 \includegraphics[width=\linewidth]{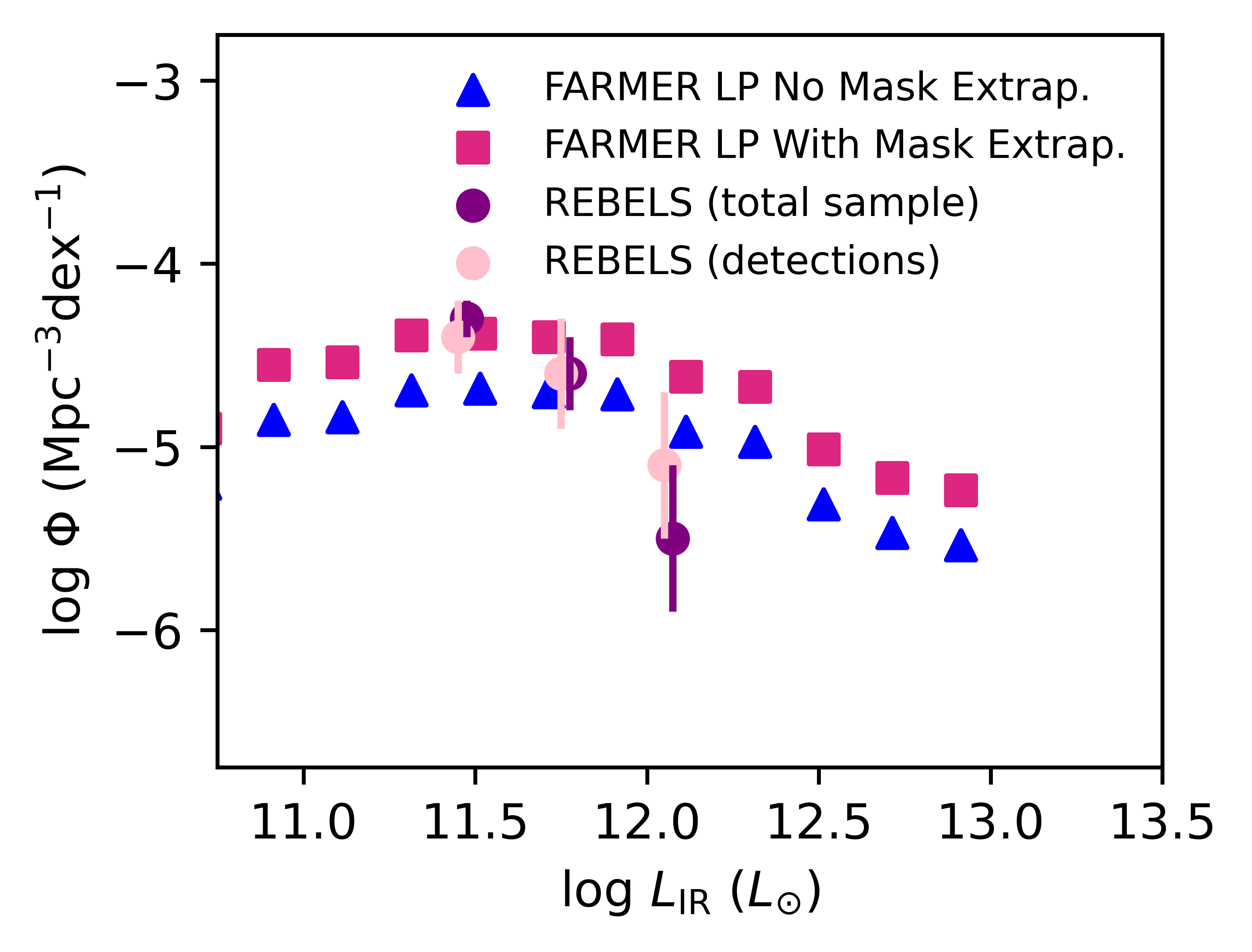}
 \captionof{figure}{Comparison of the IR luminosity functions of REBELS when using the entire sample or when using only detections, and FARMER LP with and without mask extrapolation. We do not plot the predicted ranges of Schechter functions from \cite{Barrufet_2023} due to the limited number of data points. The points of the REBELS functions are at the same luminosity, and are slightly offset in the figure to increase clarity.}
 \label{fig:FigREBELS}
\end{figure}
\begin{figure*}[t]
 \centering
 \includegraphics[width=\linewidth]{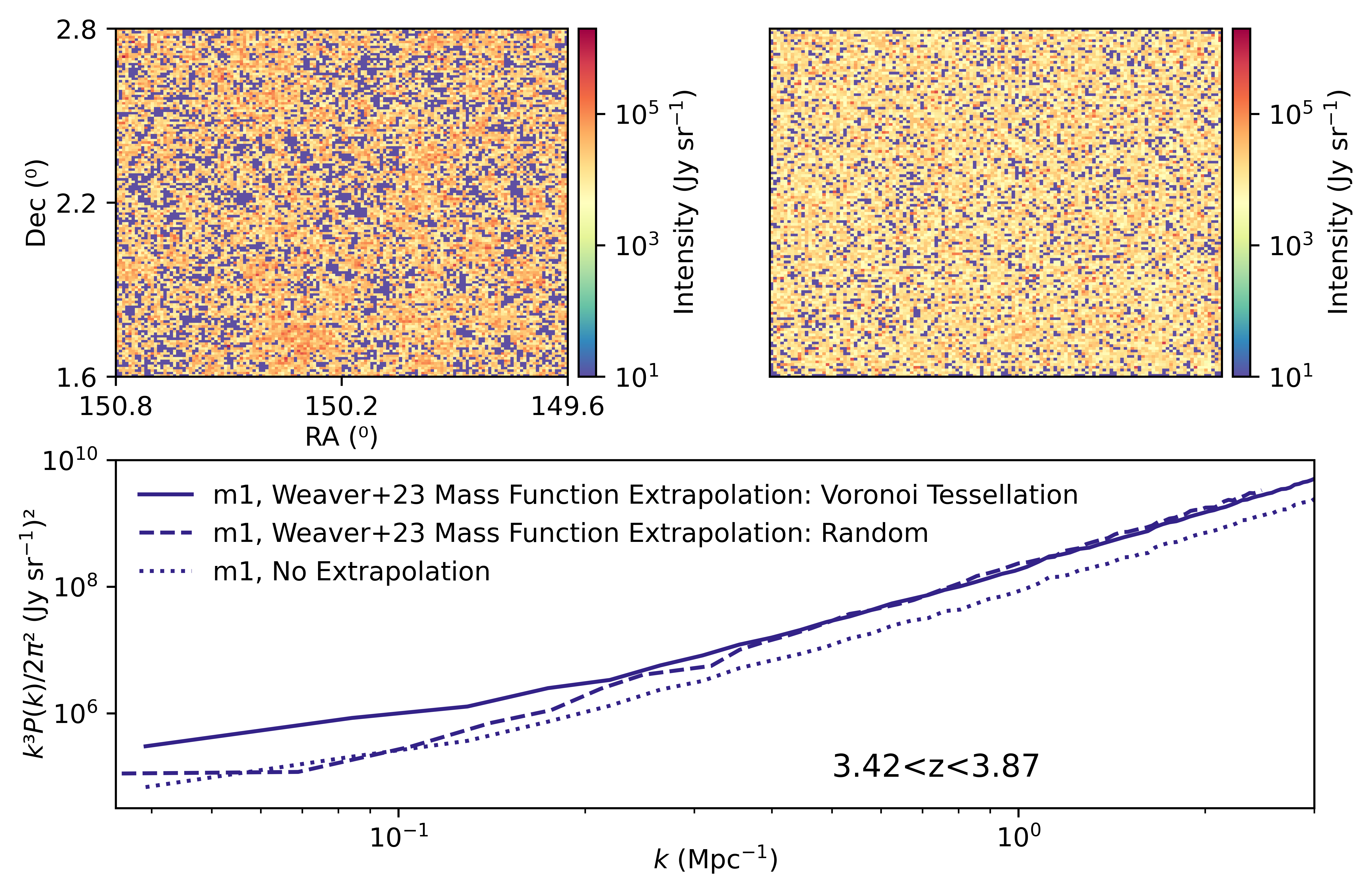}
 \captionof{figure}{Comparison of power spectra demonstrating the impact of VT on extrapolation. In the left subplot we show a cross-section of the LIM cube at $3.42<z<3.87$ for m1 when using VT to select galaxy location. In the right subplot we see the same, except for the random location selection case. In the lower subplot we compare the power spectra.}
 \label{fig:FigVoronoiVSRandom}
\end{figure*}
The only exception to the previous paragraph is the $6.75<z<8.27$ band, where we take a far greater degree of mask extrapolation and approximately double the original sample. Considering the innate bias towards the brightest galaxies at these distances in FARMER LP, this leads to results that eclipse those from equivalent simulated work (e.g. Fig. \ref{fig:FigCompareDefaultPS}). This discrepancy is significant and surprising, especially in the context of the other redshift bands where there is greater concordance. In order to investigate whether it is appropriate to apply the standard stellar mask extrapolation principles to the $6.75<z<8.27$ band, we compare the IR luminosity functions of this band and of REBELS \citep{Bouwens_2022}, a survey targeting 42 galaxies at $6.4<z<7.7$ including 15 spectroscopically confirmed [CII] sources. We use this sample with its known data as a useful point of reference, even if most of its data do not lie within the COSMOS field, with its IR luminosity function taken from \cite{Barrufet_2023}. Figure \ref{fig:FigREBELS} compares this to the IR luminosity functions of FARMER LP in the $6.75<z<8.27$ band with and without the mask extrapolation, and we find that there is strong agreement in both cases for $L_{\textrm{IR}}<10^{12}\,L\odot$. However, it is clear from a visual examination that the high-luminosity end of the FARMER LP luminosity functions would be in disagreement from any Schechter curve that could be fitted to the REBELS data. This is true even without mask interpolation. Correspondingly, we do find it appropriate to be cautious of results at these redshifts. We continued to use the same mask extrapolation methods for this band to ensure consistency with the other redshift bands. 

Finally, it is important to retain the existing structure when extrapolating to prevent changes in the clustering signal, so we used the Voronoi tessellation (VT) method as in Sect. \ref{sec:interpolationVT}. We validated this technique by comparing cubes and power spectra using m1 at $3.42<z<3.87$, when using no extrapolation, extrapolation with random locations, and extrapolation with locations determined using VT (Fig. \ref{fig:FigVoronoiVSRandom}). For both extrapolations, the shot-noise increase is similar and expected ($\sim$0.5\,dex). However the structure component is very different, with the clustering in the random location case regressing to the no extrapolation case, while the VT case shows a relatively large kick. This becomes clear when comparing the intensity cubes, as the random location signal is similar to white noise, while the VT case clearly demonstrates structure. While there is the potential concern of VT over-weighting around the existing galaxies of the sample, this is in fact ideal for us as FARMER LP is likely to include the bright centres of galaxy clusters. Despite this, we find that the clustering signal from our extrapolated samples is weaker in comparison to work from previous literature (Fig. \ref{fig:FigSpecificinterpPS}, indicating that this method will require refinement and calibration in the future.
\section{Power spectra error}\label{appendix:psnoise}
As a first estimate for calculating the error in the power spectrum, we included the thermal instrumental noise as well-understood errors in beam smoothing and sample variance, which we briefly discussed in our methodology. This is detailed further by \cite{Li_2016} and \cite{Chung_2020} where the authors calculate it analytically using Eq. (\ref{eq:14}):
\begin{equation}
\sigma(k)= \frac{P(k)+P_\textrm{n}(k)}{\sqrt{N_{\textrm{modes}}(k,\Delta k)}} \frac{1}{W(k)},
\label{eq:14}
\end{equation}
where $P(k)$ is the [CII] power spectra, $P_\textrm{n}(k)$ is the power spectra due to thermal noise, $N_{\textrm{modes}}(k,\Delta k)$ is the number of modes that are averaged over to calculate $P(k)$ (i.e. the number of points in 3D $k$ space with $k-\Delta k/2<k<k+\Delta k/2$ where $\Delta k$ is the shell width), and $W(k)$ is a power spectrum attenuation factor for large $k$ caused by the smoothing of the map via the instrumentation beam. The $\Delta k$ we use depends on our use case: for maximum precision we divide the total $k$ range by a number of intervals equal to half the number of pixels across the length of the map, resulting in $\Delta k=0.044, 0.040, 0.037, 0.035$\,Mpc$^{-1}$ from lower to higher redshift. For lower noise we use wide bins which are more appropriate for an actual instrument use case, $\Delta k=0.3$\,Mpc$^{-1}$. As the thermal noise for each pixel in the instrument is assumed to follow a purely random white noise Gaussian, the thermal noise power spectra is a constant as described in Eq. (\ref{eq:15}):
\begin{equation}
P_\textrm{n}(k)= \frac{\sigma_{\textrm{pix}}^2/t_{\textrm{pix}}}{(\textrm{Jy\,sr}^{-1})^2}\frac{V_{\textrm{vox}}}{\textrm{Mpc}^3},
\label{eq:15}
\end{equation}
where $\sigma_{\textrm{pix}}$ is the sensitivity per pixel per frequency slice, $t_{\textrm{pix}}$ is observing time, and $V_{\textrm{vox}}$ is the comoving volume covered by a voxel. $\sigma_{\textrm{pix}}/t_{\textrm{pix}}^{1/2}$, the survey sensitivity per pixel in units of Jy\,sr$^{-1}$, is described for EoR-Spec in Table 1 of \cite{Chung_2020} ($2.2\times10^4$, $1.2\times10^4$, $6.2\times10^3$, $3.9\times10^3$ for $3.42<z<3.87$, $4.14<z<4.76$, $5.34<z<6.31$, $6.75<z<8.27$ respectively), making the fractional term a constant. This assumes fixed observing conditions, notably a 45 degree observing angle with 0.4mm precipitable water vapour.

As $N_{\textrm{modes}}(k,\Delta k)$ is equivalent to the number of Fourier-transformed voxels in a thin shell at radius $k$ (with width $\Delta k$) in 3D $k$ space, we calculated the volume of this shell via Eq. (\ref{eq:16}) (\citealt{Li_2016}):
\begin{gather}
\begin{aligned}
N_{\textrm{modes}}(k,\Delta k)  & =4\pi k^2\Delta k \,n(k)  \\ & =4\pi k^2\Delta k \left(\frac{2\pi}{L_1}\frac{2\pi}{L_2}\frac{2\pi}{L_3}\right)^{-1}\frac{1}{\textrm{Mpc}^3} \\  &  =4\pi k^2\Delta k  \frac{V_{\textrm{surv}}/{8\pi^3}}{\textrm{Mpc}^3}=\frac{k^2\Delta k}{4 \pi^2 } \frac{V_{\textrm{surv}}}{\textrm{Mpc}^3}, 
\label{eq:16}
\end{aligned}
\raisetag{47pt}
\end{gather}
where $n(k)$ is the number density of voxels, $L_{1,2,3}$ are the comoving dimensions of the cube in Mpc, and $V_{\textrm{surv}}$ is the comoving volume of the whole survey in Mpc$^3$. Larger $k$ bins result in this term being larger by an approximate factor of 10, and thus give a smaller error by an approximate factor of 3. However, \cite{Chung_2020} note that we are limited by instrument resolution for high $k$ modes, which truncates $k$ values past $k_{\delta_\nu}=\pi/r_{\textrm{com}}$ (the spatial frequency corresponding to the smallest side length of a voxel, equivalent to the map pixel length). This corresponds to shells where we miss several $k$ modes (as in Fig. \ref{fig:FigPonthieuDiagram}). The altered expression is given by Eq. (\ref{eq:17}):
\begin{equation}
N_{\textrm{modes}}(k,\Delta k)=\frac{\textrm{min}({k,k_{\delta_\nu}})k\Delta k}{4 \pi^2}  \frac{V_{\textrm{surv}}}{\textrm{Mpc}^3}.
\label{eq:17}
\end{equation}
Finally the instrument attenuation factor, $W(k)$, is described as in Eq. (\ref{eq:18}):
\begin{align}
W(k)&= e^{-k^2 \sigma_\perp^2}\int_{0}^{1}e^{-k^2 (\sigma_\parallel^2-\sigma_\perp^2)\mu^2} \textrm{d}\mu,\\
\sigma_\parallel&=\frac{c}{H(z)}\frac{\Delta\nu_b(1+z)}{2.355\nu_{\textrm{obs}}},\\ 
\sigma_\perp&=\frac{D_\textrm{angular}(z)\Delta\theta_\textrm{beam}}{2.355},
\label{eq:18}
\end{align}
where $\Delta \nu_\textrm{beam}$ is the frequency width of a slice (the spectral element of a beam), $\nu_{\textrm{obs}}$ is the frequency of [CII] at a given redshift, and $D_\textrm{angular}(z)\Delta\theta_\textrm{beam}$ is the size of a voxel on map scale in Mpc. By combining all of these, we calculated $\sigma(k)$, the effective sensitivity limit $\sigma(k)/W(k)$, and the effective S/N $P(k)/\sigma(k)$.
\section{Mass and luminosity function fit parameters}\label{appendix:fitparams}
This section has the Schechter fit parameters for the mass and luminosity functions used, according to Eqs. (\ref{eq:19}) and (\ref{eq:20}) given in Sect. \ref{sec:interpolationMLF}. This information is stored in Tables \ref{table:massfitdata} and \ref{table:lumfitdata}. For the various fits we used in Fig. \ref{fig:FigDW}, we include the fit parameters in Table \ref{table:DWFig}. For ease of least-squares fitting, we used the logarithmic form of the equations (Eqs. \ref{eq:19a} and \ref{eq:19b}), so the tables store appropriate parameters in $10^{X}$ format.
\begin{gather}
  \begin{aligned}
\log_{10}(\Phi(M))\textrm{d}\log_{10}(M) & =\log_{10}(\Phi_0)-\alpha[\log_{10}(M)-\log_{10}(M_\textrm{c})]\\&-\log_{10}(\textrm{e})\left(\frac{10^{\log_{10}(M)}}{10^{\log_{10}(M_\textrm{c})}}\right),
\label{eq:19a}
\end{aligned}
\raisetag{20pt}
\end{gather}
\begin{gather}
  \begin{aligned}
\log_{10}(\Phi(L))\textrm{d}\log_{10}(L) & =\log_{10}(\Phi_0)-\alpha[\log_{10}(L)-\log_{10}(L_\textrm{c})]\\&-\log_{10}(\textrm{e})\left(\frac{10^{\log_{10}(L)}}{10^{\log_{10}(L_\textrm{c})}}\right).
\label{eq:19b}
\end{aligned}
\raisetag{20pt}
\end{gather}
\begin{table}[b]
\caption{Mass Schechter function fit parameters.}
\centering
\begin{tabular}{c c c c}
 \hline\hline
 Fit & $\Phi_0$ & $\alpha$ & $M_\textrm{c}$ \\ [0.5ex]
 \hline
 COSMOS 2020 $3.42<z<3.87$ & $10^{-3.68}$&0.46& $10^{10.83}$ \\
 COSMOS 2020 $4.14<z<4.76$& $10^{-3.70}$&0.46& $10^{10.46}$ \\
 COSMOS 2020 $5.34<z<6.31$& $10^{-4.22}$&0.46& $10^{10.14}$\\
 COSMOS 2020 $6.75<z<8.27$& $10^{-4.52}$ & 0.46 & $10^{10.18}$ \\
 FARMER LP Fit $3.42<z<3.87$ & $10^{-2.83}$&  0.35& $10^{10.40}$ \\
 FARMER LP Fit $4.14<z<4.76$& $10^{-3.67}$&  0.52& $10^{10.71}$\\
 FARMER LP Fit $5.34<z<6.31$& $10^{-3.32}$& -0.03& $10^{10.01}$ \\
 FARMER LP Fit $6.75<z<8.27$& $10^{-5.17}$&  0.46& $10^{11.39}$ \\ [1ex]
 \hline
\end{tabular}
\tablefoot{`COSMOS 2020' fit parameters were taken from \cite{Weaver_2022b}.}
\label{table:massfitdata}
\end{table}
\begin{table}[b]
\caption{Luminosity Schechter function fit parameters (as used in Fig. \ref{fig:FigDW} only).}
\centering
\begin{tabular}{c c c c c}
 \hline\hline
 Fit & $\Phi_0$ & $\alpha$ & $L_\textrm{c}$\\ [0.5ex]
 \hline 
 Actual & $10^{-2.80}$&0.26& $10^{8.94}$ \\
 Underestimate & $10^{-2.71}$&-0.003& $10^{8.79}$ \\
 Overestimate & $10^{-3.96}$&0.94& $10^{9.67}$\\
 Similar& $10^{-2.92}$ & 0.19 & $10^{8.98}$\\ [1ex]
 \hline
\end{tabular}
\tablefoot{These fit parameters were applied to the DL14 Entire [CII] luminosity function at $3.42<z<3.87$.}
\label{table:DWFig}
\end{table}
\begin{table*}[t]
\caption{Luminosity Schechter function fit parameters.}
\centering
\begin{tabular}{c c c c}
 \hline\hline
 Fit & $\Phi_0$ & $\alpha$ & $L_\textrm{c}$ \\ [0.5ex]
 \hline
 FARMER LP m1 Fit $3.42<z<3.87$ & $10^{-2.69}$&0.45& $10^{9.01}$ \\
 FARMER LP m1 Fit $4.14<z<4.76$& $10^{-2.30}$&-1.39& $10^{8.28}$ \\
 FARMER LP m1 Fit $5.34<z<6.31$& $10^{-3.06}$&0.04& $10^{8.63}$\\
 FARMER LP m1 Fit $6.75<z<8.27$& $10^{-4.88}$ & 0.60 & $10^{9.65}$ \\
 FARMER LP m2 Fit $3.42<z<3.87$ & $10^{-2.54}$&0.06& $10^{8.84}$ \\
 FARMER LP m2 Fit $4.14<z<4.76$& $10^{-2.11}$&0.37& $10^{8.52}$ \\
 FARMER LP m2 Fit $5.34<z<6.31$& $10^{-3.64}$&-0.32& $10^{8.83}$\\
 FARMER LP m2 Fit $6.75<z<8.27$& $10^{-3.84}$ & -0.71 & $10^{8.74}$ \\
 FARMER LP m3 Fit $3.42<z<3.87$ & $10^{-1.71}$&-0.73& $10^{8.41}$ \\
 FARMER LP m3 Fit $4.14<z<4.76$& $10^{-1.93}$&-0.93& $10^{8.28}$ \\
 FARMER LP m3 Fit $5.34<z<6.31$& $10^{-4.62}$&0.56& $10^{9.65}$\\
 FARMER LP m3 Fit $6.75<z<8.27$& $10^{-4.19}$ & -0.41 & $10^{8.92}$ \\
 FARMER LP m4 Fit $3.42<z<3.87$ & $10^{-1.69}$&-0.02& $10^{8.52}$ \\
 FARMER LP m4 Fit $4.14<z<4.76$& $10^{-1.84}$&-0.26& $10^{8.36}$ \\
 FARMER LP m4 Fit $5.34<z<6.31$& $10^{-2.65}$&-1.54& $10^{8.22}$\\
 FARMER LP m4 Fit $6.75<z<8.27$& $10^{-3.99}$ & -1.39 & $10^{8.58}$ \\
 FARMER LP DL14 Entire Fit $3.42<z<3.87$ & $10^{-2.80}$&0.26& $10^{8.94}$ \\
 FARMER LP DL14 Entire Fit $4.14<z<4.76$& $10^{-2.98}$&0.23& $10^{8.83}$ \\
 FARMER LP DL14 Entire Fit $5.34<z<6.31$& $10^{-3.87}$&0.65& $10^{8.96}$\\
 FARMER LP DL14 Entire Fit $6.75<z<8.27$& $10^{-4.93}$ & 0.44 & $10^{10.03}$ \\
 FARMER LP DL14 Starburst Fit $3.42<z<3.87$ & $10^{-3.02}$&0.50& $10^{9.11}$ \\
 FARMER LP DL14 Starburst Fit $4.14<z<4.76$& $10^{-2.87}$&0.15& $10^{8.84}$ \\
 FARMER LP DL14 Starburst Fit $5.34<z<6.31$& $10^{-3.87}$&0.52& $10^{9.01}$\\
 FARMER LP DL14 Starburst Fit $6.75<z<8.27$& $10^{-10.75}$ & 0.68 & $10^{17.86}$ \\
 FARMER LP Si15 Fit $3.42<z<3.87$ & $10^{-2.90}$&0.45& $10^{8.94}$ \\
 FARMER LP Si15 Fit $4.14<z<4.76$& $10^{-3.13}$&0.48& $10^{8.96}$ \\
 FARMER LP Si15 Fit $5.34<z<6.31$& $10^{-3.01}$&-0.42& $10^{8.37}$\\
 FARMER LP Si15 Fit $6.75<z<8.27$& $10^{-11.45}$ & 0.78 & $10^{17.61}$ \\
 FARMER LP Sc20 Fit $3.42<z<3.87$ & $10^{-3.40}$&0.52& $10^{9.24}$ \\
 FARMER LP Sc20 Fit $4.14<z<4.76$& $10^{-4.12}$&0.68& $10^{9.72}$ \\
 FARMER LP Sc20 Fit $5.34<z<6.31$& $10^{-4.28}$&0.58& $10^{9.10}$\\
 FARMER LP Sc20 Fit $6.75<z<8.27$& $10^{-9.41}$ & 0.52 & $10^{17.88}$ \\
 FARMER LP Ro22 Fit $3.42<z<3.87$ & $10^{-3.20}$&0.46& $10^{9.21}$ \\
 FARMER LP Ro22 Fit $4.14<z<4.76$& $10^{-3.76}$&0.59& $10^{9.52}$ \\
 FARMER LP Ro22 Fit $5.34<z<6.31$& $10^{-4.27}$&0.77& $10^{9.18}$\\
 FARMER LP Ro22 Fit $6.75<z<8.27$& $10^{-9.34}$ & 0.48 & $10^{18.52}$ \\ [1ex]
 \hline
\end{tabular}
\label{table:lumfitdata}
\end{table*}
\end{document}